\documentclass[twocolumn]{aastex6}
\usepackage{amsmath}

\newcommand{\HSthree}{HS$^{3}$}
\newcommand{\Mstar}{\ifmmode M_{\ast} \else $M_{\ast}$\fi}
\newcommand{\kms}{\ifmmode \mathrm{km~s^{-1}} \else km~s$^{-1}$\fi}
\newcommand{\smpy}{\ifmmode M_{\sun}~\mathrm{yr}^{-1} \else $M_{\sun}$~yr$^{-1}$\fi}
\newcommand{\msol}{\ifmmode M_{\sun} \else $M_{\sun}$\fi}
\newcommand{\smpc}{\ifmmode M_{\sun}~\mbox{pc}^{-2} \else $M_{\sun}$~pc$^{-2}$\fi}
\newcommand{\smkpc}{\ifmmode M_{\sun}~\mbox{yr}^{-1}~\mbox{kpc}^{-2} \else $M_{\sun}$~yr$^{-1}$~kpc$^{-2}$\fi}
\newcommand{\htwo}{\ifmmode \text{H}_{2} \else H$_{2}$\fi}
\newcommand{\ha}{\ifmmode \mbox{H}\alpha \else H$\alpha$\fi}
\newcommand{\sightwo}{\ifmmode \Sigma_{\textnormal{H}_{2}} \else $\Sigma_{\textnormal{H}_{2}}$\fi}
\newcommand{\sigmol}{\ifmmode \Sigma_{\textnormal{mol}} \else $\Sigma_{\textnormal{mol}}$\fi}
\newcommand{\sigsfr}{\ifmmode \Sigma_{\textnormal{SFR}} \else $\Sigma_{\textnormal{SFR}}$\fi}
\newcommand{\siggas}{\ifmmode \Sigma_{\textnormal{gas}} \else $\Sigma_{\textnormal{gas}}$\fi}
\newcommand{\sighi}{\ifmmode \Sigma_\text{H\textsc{i}} \else $\Sigma_\text{H\textsc{i}}$\fi}
\newcommand{\sigdust}{\ifmmode \Sigma_\text{dust} \else $\Sigma_\text{dust}$\fi}
\newcommand{\sigstar}{\ifmmode \Sigma_{*} \else $\Sigma_{*}$\fi}
\newcommand{\sigdotstar}{\ifmmode \dot{\Sigma}_{*} \else $\dot{\Sigma}_{*}$\fi}
\newcommand{\cplus}{\ifmmode \mbox{C}^{+} \else C$^{+}$\fi}
\newcommand{\CII}{\ifmmode \text{[C\textsc{ii}]} \else {\sc [C\,ii]}\fi}
\newcommand{\thirteenCII}{\ifmmode \text{[}^{13}\text{C\textsc{ii}]} \else {\sc [$^{13}$C\,ii]}\fi}
\newcommand{\twelveCII}{\ifmmode \text{[}^{12}\text{C\textsc{ii}]} \else {\sc [$^{12}$C\,ii]}\fi}
\newcommand{\CI}{\ifmmode \text{[C\textsc{i}]} \else {\sc [C\,i]}\fi}
\newcommand{\OI}{\ifmmode \text{[O\textsc{i}]} \else {\sc [O\,i]}\fi}
\newcommand{\NII}{\ifmmode \text{[N\textsc{ii}]} \else {\sc [N\,ii]}\fi}
\newcommand{\OIII}{\ifmmode \text{[O\textsc{iii}]} \else {\sc [O\,iii]}\fi}

\newcommand{\hi}{\ifmmode \text{H\textsc{i}} \else {\sc H\,i}\fi}
\newcommand{\hii}{\ifmmode \text{H\textsc{ii}} \else H\,{\sc ii}\fi}
\newcommand{\Nhi}{\ifmmode N_\text{H\textsc{i}} \else $N_\text{H\textsc{i}}$\fi}
\newcommand{\Nmol}{\ifmmode N_\text{mol} \else $N_\text{mol}$\fi}
\newcommand{\Nhtwo}{\ifmmode N_{\textnormal{H}_{2}} \else $N_{\textnormal{H}_{2}}$\fi}
\newcommand{\Mhi}{\ifmmode M_\text{H\textsc{i}} \else $M_\text{H\textsc{i}}$\fi}
\newcommand{\Mhtwo}{\ifmmode M_{\textnormal{H}_{2}} \else $M_{\textnormal{H}_{2}}$\fi}
\newcommand{\Mmol}{\ifmmode M_{\textnormal{mol}} \else $M_{\textnormal{mol}}$\fi}
\newcommand{\dGDR}{\ifmmode \delta_{\textnormal{GDR}} \else $\delta_{\textnormal{GDR}}$\fi}
\newcommand{\CO}{\ifmmode ^{12}\rm{CO} \else $^{12}$CO\fi}
\newcommand{\thirteenCO}{\ifmmode ^{13}\rm{CO} \else $^{13}$CO\fi}
\newcommand{\CeighteenO}{\ifmmode \rm{C}^{18}\rm{O} \else C$^{18}$O\fi}
\newcommand{\taudep}{\ifmmode \tau_{\textnormal{dep}} \else $\tau_{\textnormal{dep}}$\fi}
\newcommand{\taumoldep}{\ifmmode \tau^{\textnormal{mol}}_{\textnormal{dep}} \else $\tau^{\textnormal{mol}}_{\textnormal{dep}}$\fi}
\newcommand{\Av}{\ifmmode A_{V} \else $A_{V}$\fi}
\newcommand{\COtoCII}{\ifmmode \textnormal{CO}/\CII \else $\textnormal{CO}/\CII$}

\newcommand{\myemail}{katie.jameson@anu.edu.au}

\shorttitle{\CII-bright and CO-bright Gas in the SMC}
\shortauthors{Jameson et al.}

\begin{document}


\title{First Results from the {\it Herschel} and ALMA Spectroscopic Surveys of the SMC: \\ The Relationship Between \CII-bright Gas and CO-bright Gas at Low Metallicity}\thanks{{\it Herschel} is an ESA space observatory with science instruments provided by European-led Principal Investigator consortia and with important participation from NASA.}

\author{Katherine E. Jameson \altaffilmark{1,2}, Alberto D. Bolatto\altaffilmark{1}, Mark Wolfire\altaffilmark{1}, Steven R. Warren \altaffilmark{3}, Rodrigo Herrera-Camus\altaffilmark{4}, Kevin Croxall\altaffilmark{5,6}, Eric Pellegrini\altaffilmark{7}, John-David Smith \altaffilmark{8}, Monica Rubio\altaffilmark{9}, Remy Indebetouw \altaffilmark{10,11}, Frank P. Israel\altaffilmark{12}, Margaret Meixner\altaffilmark{13}, Julia Roman-Duval\altaffilmark{13}, Jacco Th. van Loon\altaffilmark{14}, Erik Muller\altaffilmark{15}, Celia Verdugo\altaffilmark{16}, Hans Zinnecker\altaffilmark{17}, Yoko Okada\altaffilmark{18}}

\email{\myemail}

\altaffiltext{1}{Astronomy Department and Laboratory for Millimeter-wave Astronomy, University of Maryland, College Park, MD 20742}
\altaffiltext{2}{Research School of Astronomy and Astrophysics, Australian National University, Canberra ACT 2611, Australia}
\altaffiltext{3}{Cray, Inc., 380 Jackson Street, Suite 210, St. Paul, MN 55101, USA}
\altaffiltext{4}{Max-Planck-Institut f\"{u}r extraterrestrische Physik, Giessenbachstr., D-85748 Garching, Germany}
\altaffiltext{5}{Department of Astronomy, The Ohio State University, 4051 McPherson Laboratory, 140 West 18th Avenue, Columbus, OH 43210, USA}
\altaffiltext{6}{Illumination Works LLC, 5650 Blazer Parkway, Suite 152, Dublin OH 43017}
\altaffiltext{7}{Zentrum f\"{u}r Astronomie, Institut f\"{u}r Theoretische Astrophysik, Universit\"{a}t Heidelberg, D-69120 Heidelberg, Germany}
\altaffiltext{8}{Department of Physics \& Astronomy, University of Toledo, 2801 W Bancroft St, Toledo, OH 43606}
\altaffiltext{9}{Departamento de Astronom\'{\i}a, Universidad de Chile, Casilla 36-D, Chile}
\altaffiltext{10}{Department of Astronomy, University of Virginia, PO Box 400325, Charlottesville, VA 22904, USA}
\altaffiltext{11}{National Radio Astronomy Observatory, 520 Edgemont Road, Charlottesville, VA 22903, USA}
\altaffiltext{12}{Sterrewacht Leiden, Leiden University, PO Box 9513, 2300 RA Leiden, The Netherlands}
\altaffiltext{13}{Space Telescope Science Institute, 3700 San Martin Dr., Baltimore, MD 21218, USA}
\altaffiltext{14}{Lennard-Jones Laboratories, Keele University, Staffordshire ST5 5BG, UK}
\altaffiltext{15}{National Astronomical Observatory of Japan, Chile Observatory, 2-21-1 Osawa, Mitaka, Tokyo 181-8588}
\altaffiltext{16}{The Joint ALMA Observatory, Alonso de Córdova 3107, Vitacura, Santiago, Chile} 
\altaffiltext{17}{SOFIA Science Center, Deutsches SOFIA Institut, NASA Ames Research Center, Moffett Field, CA, 94035, USA}
\altaffiltext{18}{I. Physikalisches Institut der Universit\"{a}t zu K\"{o}ln, Z\"{u}lpicher Stra{\ss}e 77, 50937, K\"{o}ln, Germany}















\altaffiltext{1}{National Radio Astronomy Observatory, 520 Edgemont Road, Charlottesville, VA 22903, USA}

\begin{abstract}

The Small Magellanic Cloud (SMC) provides the only laboratory to study the structure of molecular gas at high resolution and low metallicity. We present results from the $Herschel$ Spectroscopic Survey of the SMC (\HSthree), which mapped the key far-IR cooling lines \CII, \OI, \NII, and \OIII\ in five star-forming regions, and new ALMA 7m-array maps of \CO\ and \thirteenCO\ $(2-1)$ with coverage overlapping four of the five \HSthree\ regions. We detect \CII\ and \OI\ throughout all of the regions mapped. The data allow us to compare the structure of the molecular clouds and surrounding photodissociation regions using \thirteenCO, \CO, \CII, and \OI\ emission at $\lesssim{10\arcsec}$ ($<3$ pc) scales. We estimate \Av\ using far-IR thermal continuum emission from dust and find the \COtoCII\ ratios reach the Milky Way value at high \Av\ in the centers of the clouds and fall to $\sim{1/5-1/10\times}$ the Milky Way value in the outskirts, indicating the presence of translucent molecular gas not traced by bright \CO\ emission. We estimate the amount of molecular gas traced by bright \CII\ emission at low \Av\ and bright \CO\ emission at high \Av. We find that most of the molecular gas is at low \Av\ and traced by bright \CII\ emission, but that faint \CO\ emission appears to extend to where we estimate the \htwo-to-\hi\ transition occurs. By converting our \htwo\ gas estimates to a CO-to-\htwo\ conversion factor ($X_{CO}$), we show that $X_{CO}$ is primarily a function of \Av, consistent with simulations and models of low metallicity molecular clouds.

\end{abstract}




\section{Introduction}

Molecular clouds are the sites of the first stages of star formation. The structure of molecular clouds and the transition from atomic to molecular gas can affect what fraction of the gas participates in star formation. The effects of metallicity on the structure and properties of molecular clouds \citep{rub93,bol08,hey09,hug10,sch12}, and the resulting effects on star formation, are not well understood due to the difficulty in observing \htwo\ and the molecular-to-atomic transition at low metallicity. Without knowledge of these effects, simulations of molecular clouds and star formation at low metallicity are largely unconstrained. Galaxy evolution simulations, particularly at the very early times when metallicities are low, rely on an accurate understanding of the fraction of gas available for star formation. At a metallicity of $Z\sim{1/5}~Z_{\sun}$ \citep{duf84,kur99,pag03} and a distance of $D\approx63$~kpc, the Small Magellanic Cloud (SMC) provides an ideal laboratory to study the effects of low metallicity on the molecular gas and the molecular to atomic transition. 

The transition from atomic to molecular gas occurs at the outer edges of the molecular cloud, where the shielding is lower and molecules are more easily dissociated. These edges are referred to as photodissociation regions (PDRs). Studying the molecular gas structure requires understanding the distribution of \htwo\ from the dense cloud cores to the diffuse outer layers of the clouds. The most common tracer of molecular gas is \CO. At low metallicity, the dissociating far-UV (FUV) radiation field strengths are higher because there is less dust to shield the molecular gas. The \htwo\ gas, however, is expected to be more prevalent than CO due to the ability of \htwo\ to more effectively self-shield against dissociating FUV photons. While more prevalent, the lowest energy line transition arising directly from \htwo\ has a temperature-equivalent energy of $E_{u}/k=510$ K and critical density of $n_{crit,H}\sim{1000}$ cm$^{-2}$, which will trace only warm ($T\gtrsim{100}$ K) molecular gas.  Both observations and modeling suggest that $\sim{}30\% - 50\%$ of the H$_{2}$ in the Solar Neighborhood resides in a ``CO-faint'' phase \citep[e.g.,][]{gre05, wol10, pla11}. Studies of the SMC suggest this phase to encompass $80\% - 90\%$ of all the H$_{2}$ \citep{isr97, pak98, ler07, ler11, bol11}, likely dominating the molecular reservoir available to star formation. 

In regions where CO is photo-dissociated, the carbon is present as neutral carbon, C$^{0}$, and singly ionized carbon, C$^{+}$. Given the CO dissociation energy of 10.6 eV and the C ionization potential of 11.3 eV, a large fraction of the carbon will be ionized throughout the interstellar medium (ISM). The \CII\ 158 \micron\ line (arising from the $^{2}P^{0}_{3/2}\rightarrow^{2}P^{0}_{1/2}$ fine structure transition), with an energy above ground of $\Delta E/k=91$ K, originates from the ``CO-faint'' \htwo\ gas as well as the neutral atomic and ionized gas. The \CII\ line thus offers the potential to estimate the amount of molecular gas not traced by bright CO emission, particularly in low metallicity environments where a significant fraction of the \htwo\ may not be traced by bright CO emission: after removing the contributions to \CII\ from atomic and ionized gas, the remaining emission can be attributed to molecular gas. To then convert the \CII\ emission to a molecular gas column density requires some knowledge of the conditions of the gas (namely volume density and temperature, which determine the \CII\ excitation). Early \CII\ observations of low metallicity environments from the {\em Kuiper Airborne Observatory} have shown bright emission and high \CII/CO ratios that are best explained by a significant amount of \htwo\ not traced by CO emission in star-forming regions of the Magellanic Clouds \citep{pog95,isr96,isr11} and IC 10 \citep{mad97}. Even at higher metallicity in the Milky Way, spectral decomposition of the \CII\ line using the GOT C$+$ survey shows that molecular gas not associated with bright CO emission (called ``CO-dark'' or ``CO-faint'' molecular gas) accounts for $\sim{30\%}$ of the total molecular mass \citep{pin13,lan14}.   

To estimate the total amount of molecular gas we need both \CII\ and CO observations: the \CII\ emission traces the molecular gas at in the outer parts of the cloud within the PDR, while the CO emission traces the remaining molecular gas at in the denser, inner regions of the cloud. One way to trace the depth probed along the line of sight is to use the visual extinction due to dust, \Av. Low \Av\ indicates a lower column of dust and gas associated with diffuse gas and the PDR region in the outskirts of molecular clouds. Higher \Av\ indicates a higher column of dust and gas and the transition into the denser regions of molecular clouds. In terms of \Av, \CII\ will trace the molecular gas at low \Av\ and CO will trace the molecular gas at high \Av. The existing CO data have only traced the high \Av\ molecular gas. The Magellanic Clouds have been studied extensively in CO with earliest surveys completed using the Columbia 1.2m \citep{coh88,rub91}. Since then, many higher resolution surveys of the SMC have taken place using Nanten \citep{miz01}, the Swedish-ESO Submillimetre Telescope (SEST) \citep{isr93,rub93}, and Mopra \citep{mul10}. Their typical angular resolution of $\sim{30\arcsec}$ ($\sim{10}$ pc), however, makes it difficult to use them to study individual star-forming regions. 

In this study, we present new $Herschel$ far-infrared line observations, including Photoconductor Array Camera and Spectrometer (PACS) \CII\ and \OI\ observations, from the $Herschel$ Spectroscopic Survey of the SMC (HS$^{3}$), together with new ALMA {\em Morita-san} Compact Array (ACA) \CO, \thirteenCO, and \CeighteenO\ observations of the Southwest Bar of the SMC, all at a resolution of $\sim{5-10}\arcsec$ ($\sim{1.5-3}$ pc). The ACA resolution is similar to that of the PACS spectroscopy, which allows us to produce estimates of molecular gas from \CII\ and CO at comparable resolutions and investigate how the ``\CII-bright'' molecular gas relates to the ``CO-bright'' molecular gas at low metallicity.

In Section \ref{section:observations} we describe the details of the HS$^{3}$ and ALMA SMC observations and data reduction, as well as ancillary data used for this study. We present the main results of the two surveys in Section \ref{section:results}. Our methodology to estimate molecular gas using \CII\ and \CO\ emission is described in Section \ref{section:estimating_h2}. We discuss the results of our new molecular gas estimates in Section \ref{section:discussion}, including a comparison to previous dust-based estimates and converting our estimates to CO-to-\htwo\ conversion factor values to compare to models and simulations of molecular clouds at low metallicity. Finally, Section \ref{section:conclusions} summarizes our work and outlines the main conclusions of this study.

\input{table1.tex}
\input{table2.tex}

\section{Observations}
\label{section:observations}

\subsection{The Herschel Spectroscopic Survey of the SMC}
\label{subsection:HS3_data}

The $Herschel$ \citep{pil10} Spectroscopic Survey of the SMC (HS$^{3}$) maps the key far-infrared (far-IR) lines of \CII\ 158 \micron, \OI\ 63 \micron, \OIII\ 88 \micron, and \NII\ 122 \micron\ with the PACS spectrometer \citep{pog10} and obtain Spectral and Photometric Imaging Receiver \citep[SPIRE,][]{gri10} Fourier Transform Spectrometer (FTS) observations (that include \NII\ 205 \micron) in five regions across the SMC with varying star formation activity and ISM conditions. These targets were covered using strips oriented to span the range from the predominantly molecular to the presumably atomic regime. The strips are fully sampled in \CII\ and \OI, while only a few pointings were observed for \NII\ and \OIII.

The HS$^{3}$ targeted regions with a range of star formation activity, overlapping with the $Spitzer$ Spectroscopic Survey of the SMC (S$^{4}$MC; \citealt{san12}) whenever possible, and spanning a range of ``CO-faint'' molecular gas fraction using dust-based molecular gas estimates \citep{bol11} going from the peaks out to the more diffuse gas. The main survey covers 5 star-forming areas which we refer to as `N83' (also includes N84), `SWBarN' (covers N27), `SWBarS' (covers N13), `N22' (also includes N25, N26, H36, and H35) and a smaller square region called `SWDarkPK' that covers a region with a dust-based peak in the molecular gas without any associated CO emission as seen in the NANTEN \CO\ map \citep{miz01}. The `N' numbered regions refer to \hii\ regions from the catalog by \citet{hen56}, and the `H' numbered regions are from the catalog of \ha\ structures by \citet{dav76}. 

The \CII\ and \OI\ maps are strips that encompass the peaks in CO, star formation, and ``CO-faint'' \htwo\ as traced by dust.  Using the PACS spectrometer $47\arcsec\times47\arcsec$ field of view, the strips were sampled using rasters with sizes $33\arcsec\times11$ (\CII) and $24\arcsec\times15$ (\OI) by $23.5\arcsec\times3$. Both \OIII\ and \NII\ observations were targeted at the location of the main ionizing source in each region and sampled with $23.5\arcsec\times2$ by $23.5\arcsec\times2$ raster. The PACS maps used the unchopped scan mode with a common absolute reference position placed south of the SMC ``Wing'' and observed at least once every two hours. The PACS spectrometer has a beam FWHM of $\theta\sim9.5\arcsec$ at the wavelength for \OI\ (63 \micron), \OIII\ (88 \micron), $\theta\sim10\arcsec$ at \NII\ (122 \micron), and $\theta\sim12\arcsec$ at \CII\ (158 \micron), which have corresponding spectral resolutions of $\sim100$, 120, 320, and 230 km s$^{-1}$ \citep{pog10}. Tables \ref{table:HS3} and \ref{table:HS3_NII_OIII} list the positions and uncertainties for all the PACS spectroscopy line images. The SPIRE FTS observations were ``intermediate sampling'' single-pointing (with a $2\arcmin$ circular field of view) at high resolution at the star-forming peak, which is typically close to the peak in \CO, for the N83, SWBarN, SWBarS, N22 regions. In addition to the main survey regions FTS observations, one single-pointing covered the brightest \hii\ region N66, which has PACS \CII\ and \OI\ observations as part of Guaranteed Time Key Project SHINING and is included in the $Herschel$ Dwarf Galaxy Survey (DGS; \citealt{mad13}). 

\subsubsection{Data Reduction}

PACS spectral observations were obtained in the Un-Chopped mapping mode and reduced using the Herschel Interactive Processing Environment (HIPE) version 12.0.2765 \citep{ott10}.  Reductions applied the standard spectral response functions, flat field corrections, and flagged instrument artifacts and bad pixels  \citep[see][]{pog10,cro12}.  The dark current, determined from each individual observation, was subtracted during processing as it was not removed via chopping.  {\it Herschel's} baseline exhibits significant baseline drifts and distinctive instrumental transients are common occurrences.  These instabilities result in a variable non-astrophysical continuum, which is dominated by emission from {\it Herschel} itself.  

Transient signals are strongly correlated with motions of the PACS grating and of {\it Herschel}. Using fits of the \citet{dra07} dust model to spectral energy distributions of galaxies in the KINGFISH sample we estimate the expected astrophysical continuum is less than 2\% of the spectral continuum detected at \CII\ 158$\mu$m.  Given that the other spectral lines were located farther from the peak of the dust-continuum than the \CII\ line, we assume that thermal dust emission is undetected in the PACS spectra. Thus, the continuum adjacent to the expected locations of the observed fine-structure lines should be constant and is used to correct for transients. This has significantly improved our ability to detect line emission. 

The averages of the clean off-observations obtained were subtracted from observations to correct for the thermal background contributed by {\it Herschel}.  Subsequently, all spectra within a given spatial element were combined.  Final spectral cubes with 2.06\arcsec\ spatial pixels were created by combining individual pointings using the Drizzle algorithm implemented in HIPE.  In-flight flux calibrations\footnote{Calibration Version 65} were applied to the data.  These calibrations resulted in absolute flux uncertainties on the order of 15\% with relative flux uncertainties between each {\it Herschel} pointing on the order of $\sim$10\%. 

The long and short wavelength SPIRE-FTS arrays (FWHMs of $34\arcsec$ and $19\arcsec$, respectively) are arranged in concentric circles and are dithered 4 times to provide complete coverage of the mapped region. The FTS data reduction started with level 0.5 data, which was temperature drift corrected, detector clipped, and time shift corrected using HIPE (version 11). A semi-extended-source correction \citep{wu13} was applied to the individual bolometer (level 1) data before mapping. Spectral cubes were produced using the corrected bolometer fluxes. 

\input{table3.tex}
\input{table3b.tex}

\subsection{ALMA Survey of the Southwest Bar}
\label{subsection:ALMA_data}

We mapped four regions in the Southwest Bar of the SMC in \CO, \thirteenCO, and \CeighteenO\ $(2-1)$ using Band 6 of the ALMA Atacama Compact Array (ACA; 7m-array consisting of 11 antennas) and Total Power array (TP; 12m single-dish) during Cycle 2. Three of these regions were previously mapped in \CO\ and \thirteenCO\ $(2-1)$ using the Swedish ESO-Submillimeter Telescope (SEST) \citep{rub93a,rub93,rub96}, but at a resolution of $22\arcsec$. The ACA maps were observed using a mosaic with 22.1\arcsec\ spacing of 47 pointings for the N22, SWBarS, and SWBarN regions and 52 pointings for the SWDarkPK with 25.5 s integration time per pointing. Both \CO\ and \thirteenCO\ were observed with 117.2 MHz (152 km s$^{-1}$) bandwidth and 121.15 kHz (0.2 km s$^{-1}$) spectral resolution. We chose a somewhat broader bandwidth for \CeighteenO\ of 468.8 MHz (642 km s$^{-1}$) and corresponding 0.24 MHz (0.12 km s$^{-1}$) spectral resolution, and used the fourth spectral window for continuum (1875.0 MHz bandwidth, 7.81 MHz resolution). During the early ALMA cycles the fast-mapping capabilities of the array were fairly limited, and so we decided to cover only half of the strips mapped by HS$^{3}$. The coverage of the maps overlaps approximately with the main CO emission known to be present in the strips, except for SWDarkPK where the PACS map is small and we covered its entire area.

\begin{figure*}[t]
\epsscale{1.0}
\plotone{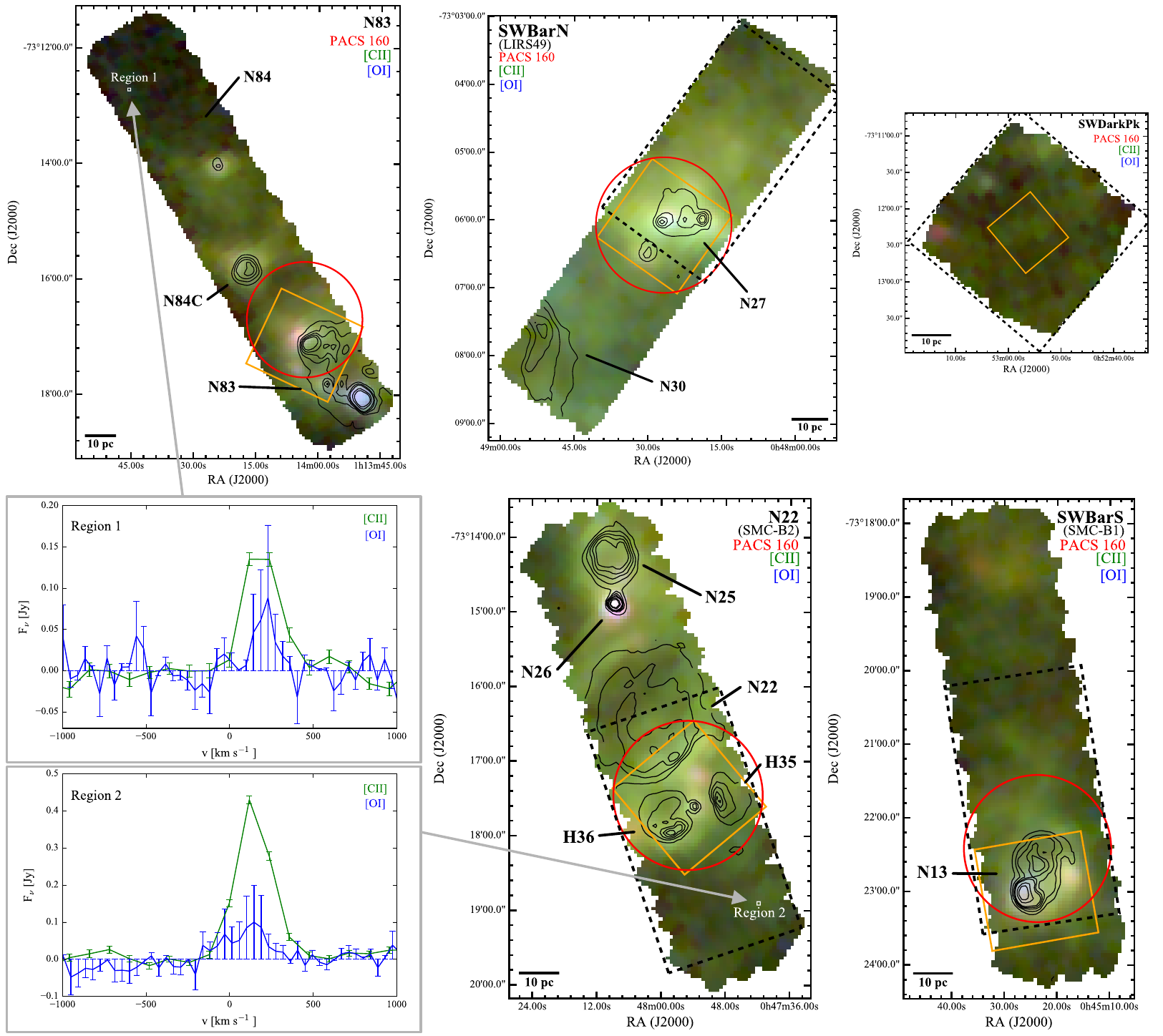}
\caption{RGB composites of the five HS$^{3}$ regions. The \CII\ (green) and \OI\ (blue) are on the same intensity scale from 0 to $3\times10^{-7}$ W m$^{-2}$ sr$^{-1}$, whereas the PACS 160 \micron\ image (red) is shown on a scale from 0 to $2\times10^{-5}$ W m$^{-2}$ sr$^{-1}$. All images are displayed using a logarithmic stretch. The black contours show MCELS \ha\ intensity \citep{smi99} at linear intervals (1, 2, 3, 4, 5, 10, 15 10$^{-14}$ ergs cm$^{-2}$ s$^{-1}$) to show the location of massive star formation throughout the regions with the region designations based on \ha\ from \citet{hen56} and \citet{dav76} indicated. The names of overlapping regions from \citet{rub93a,rub93,rub96} SEST surveys are listed in parentheses. The orange squares show the coverage of the \OIII\ observations (\NII\ has approximately the same coverage) and the red circles show the area covered by the FTS observations. The black dashed line rectangles show the approximate coverage of the ALMA maps. The two inset spectra, labeled Region 1 and 2, show spectral extractions from the PACS cube in some of the faintest regions covered by the strips. We clearly detect \CII\ emission throughout the faint areas and, somewhat unexpectedly, also \OI\ 63~$\mu$m.
\label{fig:HS3_RGB}}
\end{figure*}

We used the Common Astronomy Software Applications (CASA; \citealt{mcm07}) package to reduce, combine, and image the data. The ACA data were calibrated with the pipeline using CASA version 4.2.2, and no modifications were made to the calibration script. We used the calibrated delivered TP data, which were manually calibrated using CASA version 4.5.0 (described in the official CASA guide) with the exception of the SWBarN \CO\ spectral window (SPW 17). For the SWBarN SPW 17 we modified the baseline subtraction in the calibration script to avoid channels with line emission (the delivered calibration script included all channels when fitting the baseline). We created the reduced measurement set using the script provided with the delivered ACA data and use the imaged TP SPWs as part of the delivered data. 

We cleaned each spectral window of the ACA data and imaged it using {\sc clean} and then used {\sc feather} to combine the ACA images with the corresponding TP image re-gridded to match the ACA data (using CASA version 4.7.0). We used Briggs weighting with a robust parameter of 0.5 and cleaned to $\sim{2.5}\times{\rm{RMS}}$ found away from strong emission in the dirty data cube. As there was no noticeable continuum emission, the effect of continuum subtraction was negligible and we did not include any continuum subtraction for the final imaged cubes. Due to the short integration times and the arrangement of the 7m-array, we used conservative masks for cleaning to reduce the effects of the poor $u-v$ coverage of the ACA-only data. The data were imaged at 0.3 km s$^{-1}$ spectral resolution with synthesized beam sizes of $\sim{7\arcsec} \times 5.5\arcsec$ (2.1 pc $\times$ 1.7 pc) for \CO\ $(2-1)$. The combination of the ACA and TP data make the observations sensitive to all spatial scales. The previously published single-dish SEST \CO\ $(2-1)$ data from \citet{rub93a} overlaps with the SWBarN region, which they refer to as LIRS49. They found a peak temperature of 2.56 K in a 43\arcsec\ beam at $\alpha(\rm{B}1950)=00h46m33s$, $\delta(\rm{B}1950)=-73d22m00s$ and we find a peak temperature of 2.55 K in the same aperture in the SWBarN ACA+TP data when convolved to 43\arcsec\ resolution. The positions, beam sizes, and sensitivities of the observations are listed in Tables \ref{table:ALMA} and \ref{table:ALMA_cont}.

\begin{figure*}[t]
\epsscale{1.0}
\plotone{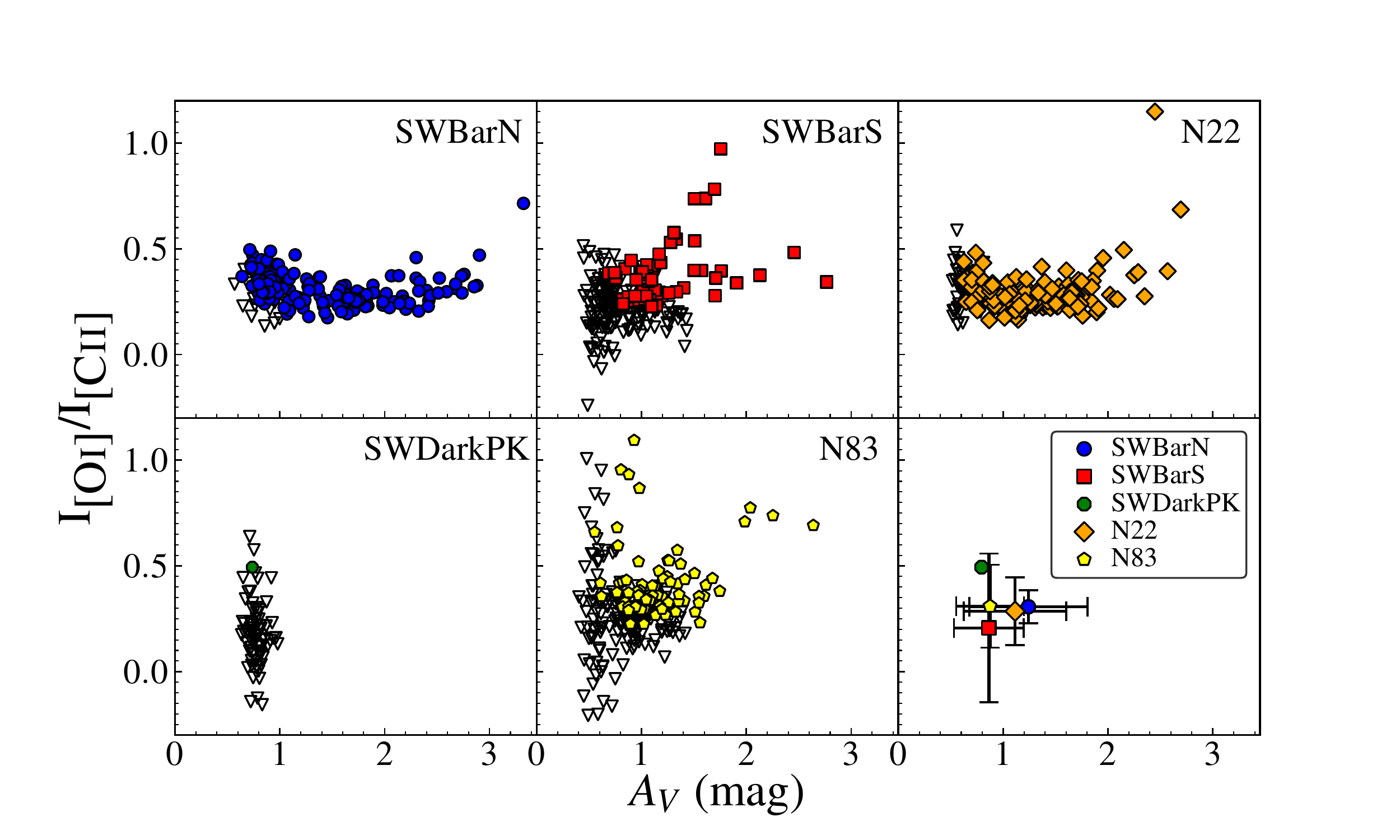}
\caption{The ratio of the integrated intensity of \OI\ to that of \CII\ as a function of $A_{V}$ for the HS$^{3}$ regions. The colored symbols show independent measurements detected at $>3\sigma$ in both \OI\ and \CII\ with the downward pointing triangles indicating upper limits (where $I_{\OI}<3\sigma$). The bottom right panel shows the average values with the error bars showing $1\sigma$ on mean for each of the regions except for SWDarkPK, for which we show the one measurement $>3\sigma$. The mean line ratios were calculated to include the upper limits (``left-censored'' data) using the cenfit routine in the R package NADA \citep{lee17,hel05}. The \OI/\CII\ ratio is mostly constant across the regions, independent of $A_{V}$, and has a typical value of $\sim{0.3}$. 
\label{fig:OI_CII}}
\end{figure*}

\subsection{\hi\ Data}
\label{subsection:HI_data}

The neutral atomic gas data come from 21 cm line observations of \hi. We use the \hi\ map from \citet{sta99} that  combined Australian Telescope Compact Array (ATCA) and Parkes 64m radio telescope data. The interferometric ATCA data set the map resolution at $1.6\arcmin$ ($r\sim30$ pc in the SMC), but the data are sensitive to all size scales due to the combination of interferometric and single-dish data. The observed brightness temperature of the 21 cm line emission is converted to \hi\ column density (\Nhi) assuming optically thin emission. The observed brightness temperature of the 21 cm line emission is converted to \hi\ column density (\Nhi) using: 
\[ \Nhi = 1.823\times{10^{18}}\frac{\mbox{ cm}^{-2}}{\mbox{K km s}^{-1}} \int T_{B}(v)~\mathrm{d}v\,. \]
The SMC map has an RMS column density of $5.0\times10^{19}$ cm$^{-2}$. While most of the \hi\ emission is likely optical thin, some fraction will be optically thick and the optically thin assumption will cause us to underestimate \Nhi. \citet{sta99} produced a statistical correction to account for optically thick \hi\ line emission in the SMC, however the correction is based only on 13 \hi\ absorption measurements with only two in the Southwest Bar. We choose not to apply the correction since it has little effect on our \htwo\ estimate from \CII\ (see Section \ref{subsection:CII_from_HI}).

\begin{figure*}[t]
\epsscale{1.2}
\plotone{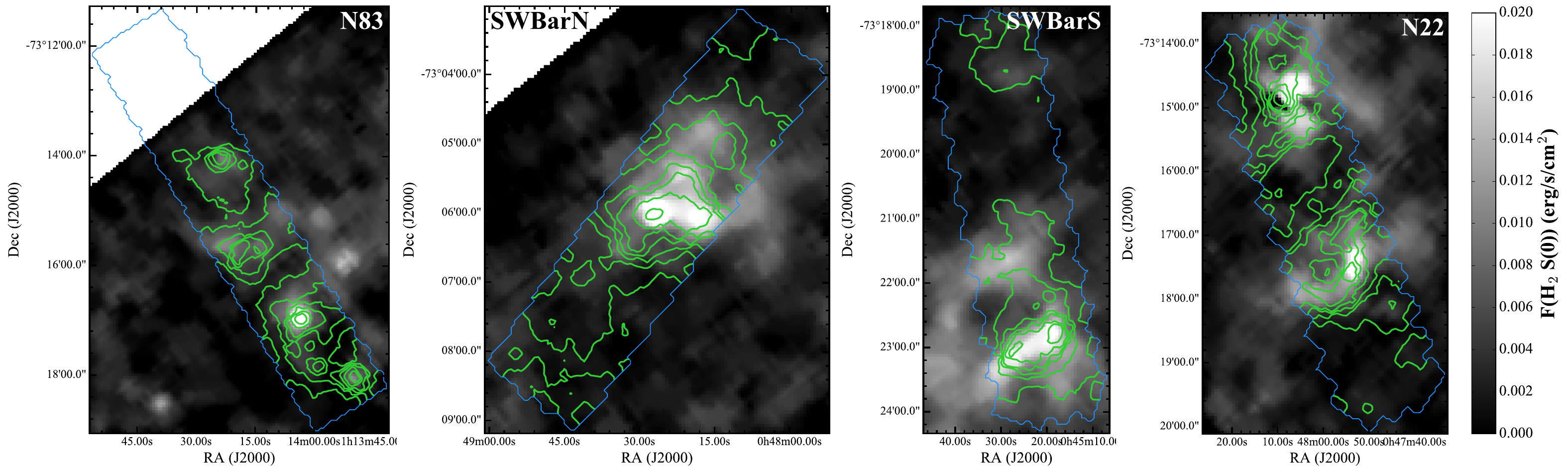}
\caption{Images of the \htwo\ S(0) 28.2 \micron\ line from S$^{4}$MC \citep{san12} resampled to match the HS$^{3}$ \CII\ images. Contours show the \CII\ integrated intensity at levels of 0.3, 0.5, 0.7, 0.9, 1.2, 1.5, 2.0, $3.0\times{10}^{-7}$ W m$^{-2}$ sr$^{-1}$, with the blue line showing the \CII\ map coverage. The excellent correspondence between the structures provides evidence that \CII\ is tracing the molecular gas in the PDRs.
\label{fig:H2_S0}}
\end{figure*}

\subsection{Additional Data}
\label{subsection:add_data}

We use mid-infrared $Spitzer$ IRAC and MIPS data from the SMC-SAGE \citep{gor11} and S$^{3}$MC \citep{bol07} surveys and spectroscopic IRS data, particularly \htwo\ rotational lines, from the  S$^{4}$MC \citep{san12} survey. The maps of the \htwo\ rotational line images were produced by fitting and removing the baseline near the line and then calculating the total line intensity. We also use a velocity-resolved \CII\ spectrum from the GREAT heterodyne instrument \citep{hey12} on board the Stratospheric Observatory for Infrared Astronomy (SOFIA) \citep{tem14} from the SMC survey presented in R. Herrera-Camus et al. 2017 (in preparation).

Since the ALMA Survey focuses on the Southwest Bar of the SMC, there is no comparable map of CO from ALMA for N83. However, there are new APEX\footnote{This publication is based on data acquired with the Atacama Pathfinder Experiment (APEX). APEX is a collaboration between the Max-Planck-Institut fur Radioastronomie, the European Southern Observatory, and the Onsala Space Observatory.} maps of \CO\ $(2-1)$ that overlap the N83 HS$^{3}$ region (PI: Rubio). We use the APEX data for the N83 region to be able to make similar comparisons to the HS3$^{3}$ data, but note the lower resolution ($\sim{25\arcsec}$) limits the analysis. To do this, we take the additional step of convolving and re-gridding the $Herschel$ spectroscopic maps (\OI, \CII) to match that of the APEX \CO\ $(2-1)$ map. 

\subsection{Total-infrared}
\label{subsection:TIR}

We determine the total-infrared (TIR) intensity (from $3~\micron-1100~\micron$) using the $Spitzer$ 24 \micron\ and 70 \micron\ (no $Herschel$ 70 \micron\ map exists) from SMC-SAGE \citep{gor11} combined with $Herschel$ 100 \micron, 160 \micron, and 250 \micron\ images from HERITAGE \citep{mei13}. All of the images are convolved to the lowest resolution of the $Spitzer$ 70 \micron\ image ($\sim{18\arcsec}$) using the convolution kernels from \citet{ani11}. The total-infrared intensity is calculated following the prescription by \citet{gal13}:
\begin{equation}
S_{\rm{TIR}} = \sum{c_{i}S_{i}},
\end{equation}
all in units of W kpc$^{-2}$, where the coefficients ($c_{i}$) are 2.013, 0.508, 0.393, 0.599, and 0.680 for 24 \micron, 70 \micron, 100 \micron, 160 \micron, and 250 \micron, respectively.

\subsection{Estimating $A_{V}$}
\label{subsection:Av}

We investigate the structure of the photodissociation region and molecular cloud by using the visual extinction ($A_{V}$) as an indicator of the total column through the cloud, and as a  to gauge the depth within the cloud associated with the observations. To match the high resolution of the \CII, \OI, and ALMA CO data, we use the optical depth at 160 $\mu$m ($\tau_{160}$) and the HERITAGE 160 \micron\ map of the SMC \citep{mei13} as the basis for producing a map of $A_{V}$. \citet{lee15} fit a modified blackbody with $\beta=1.5$ to the SMC HERITAGE 100 \micron, 160 \micron, 250 \micron, and 350 \micron\ data for the SMC. We re-sample their map of fitted dust temperatures at the lower resolution of the 350 \micron\ $Herschel$ map ($\sim{30\arcsec}$) to the higher resolution 160 \micron\ map ($\sim{12\arcsec}$) in order to estimate $\tau_{160}$ at a resolution comparable to the \CII\ and ALMA CO maps. We convert from $\tau_{160}$ to $A_{V}$ using $A_{V}\sim{2200\tau_{160}}$ from \citet{lee15}, which is based on measurements in the Milky Way and provides similar $A_{V}$ values as those found using UV/optical and NIR color excess methods (see Figure 1 in \citealt{lee15}). We stress that these values of $A_{V}$ are estimates that include uncertainties associated with the assumptions made in the dust modeling (e.g. the assumption of a single dust temperature) and the conversion from $\tau_{160}$ to $A_{V}$. While the extinction at $\sim{11}$ eV (ionization potential of carbon) would be a more relevant quantity to our study of \CII\ and CO emission, the conversion from $\tau_{160}$ to $A_{\rm{11eV}}$ is highly uncertain.

\section{Results}
\label{section:results}

We present the high resolution imaging ($\sim{10\arcsec}~\sim{3}$ pc) of a suite of far-IR cooling lines from the $Herschel$ Spectroscopic Survey of the SMC (HS$^{3}$) and CO from the ALMA ACA in the SMC. \CII\ and \OI\ lines were detected in all of the regions targeted, and \CII\ is detected throughout all of the regions. The ALMA ACA+TP data shows clear detections of \CO\ and \thirteenCO\ ($2-1$) emission in all of the regions, but \CeighteenO\ is not detected. In this section we discuss the comparison of the \OI, \CII, and \CO\ emission.

\subsection{\CII\ and \OI}
\label{subsection:CII_OI}

The \CII\ 158 \micron\ line dominates the cooling of the warm ($T\sim{100}$ K) neutral gas because of the high carbon abundance, its lower ionization potential of 11.26 eV, and an energy equivalent temperature of 92 K. Ionized carbon, C$^{+}$, exists throughout most phases of the ISM except in the dense molecular gas where most of the carbon is locked in CO. The \OI\ 63 \micron\ line also contributes to the gas cooling, but with an energy equivalent temperature of 228 K and a high critical density ($\sim{10^{5}}$ cm$^{-3}$), this line dominates over the \CII\ emission only in the densest gas. Indeed, the bright ``blue'' knots of \OI\ emission in Fig. \ref{fig:HS3_RGB} are coincident with bright knots of \ha\ emission (in black contours) associated with very recent massive star formation in dense and presumably warm structures bathed by intense radiation. Oxygen can remain neutral in regions with ionized hydrogen and \ha\ emission due to its slightly higher ionization potential of 13.62 eV. In warm PDRs subjected to radiation fields larger than $10^3$ Habings and due to the difference in critical densities, the \OI\ to \CII\ ratio is a good indicator of density, but in colder gas and particularly below $n\lesssim10^4$\,cm$^{-3}$ it is mostly sensitive to temperature and the incident radiation field (e.g., \citealt{kau99}). 

Figure \ref{fig:HS3_RGB} shows the \OI\ and \CII\ integrated intensity images in combination with the 160 \micron\ PACS image showing dust continuum emission. The differences in the local star formation, shown by \ha\ in black contours, produce different structures and varying intensities of \CII, \OI, and dust emission. In many faint \CII\ regions in the diffuse gas the \OI\ line is detected (see inset spectra in Figure \ref{fig:HS3_RGB}). In principle it is possible for \OI\ 63 \micron\ to be a very important coolant for the warm neutral phase (WNM) of the ISM \citep{wol95,wol03}. Despite the high critical density of this transition, the high temperature of the WNM excites \OI\ making it an efficient coolant even in $n\sim1$\,cm$^{-3}$ gas. 

\begin{figure}[t]
\epsscale{1.1}
\plotone{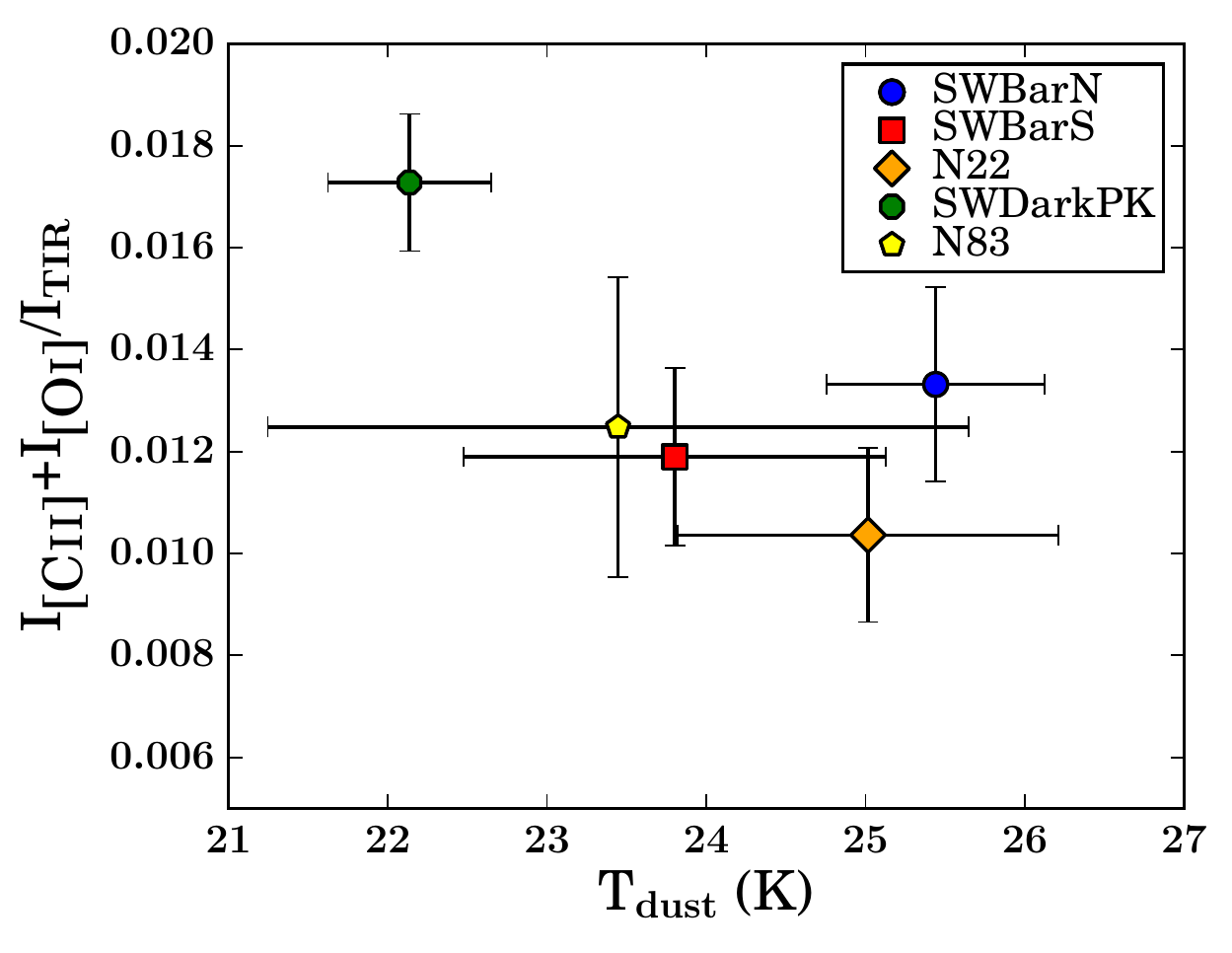}
\caption{The mean ratios of the \CII\ and \OI\ intensities to the TIR, an indicator of the photoelectric heating efficiency, as a function of mean dust temperature (T$_{\rm{dust}}$) from \citet{lee15}. The error bars show $1\sigma$ on the mean.
\label{fig:PE_G0}}
\end{figure}

\begin{figure*}[t]
\epsscale{1.2}
\plotone{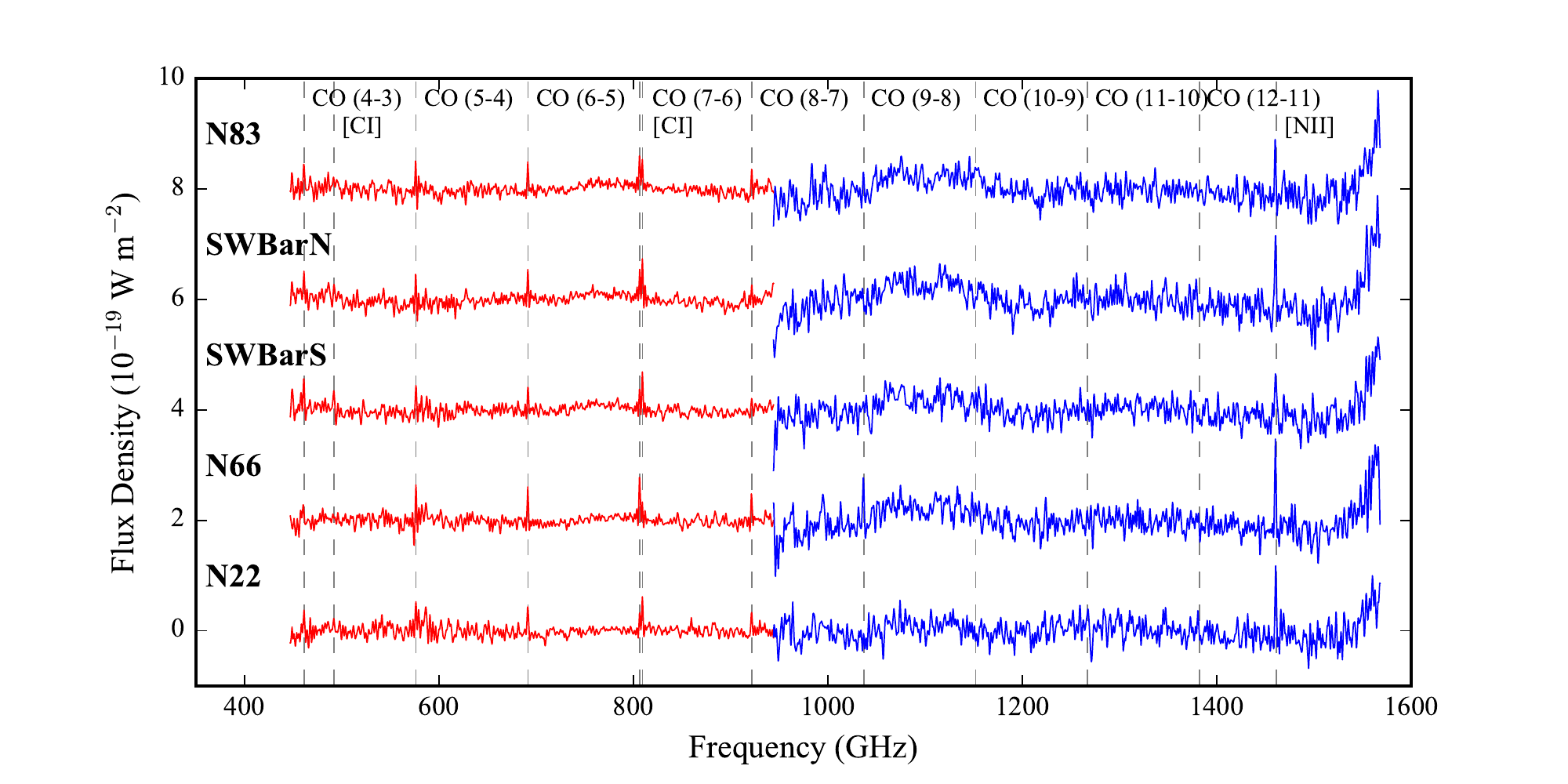}
\caption{FTS SLW (red) and SSW (blue) spectra averaged over all the bolometers for HS$^{3}$ regions SWBarN, SWBarS, N22, and N83, plus N66 (the target of PACS observations from the $Herschel$ GTKP SHINING project) with each spectra being offset by $2\times{10^{-19}}$ W m$^{-2}$ sr$^{-1}$. The SLW and SSW spectra have a second-order polynomial fit and subtracted to remove the baseline. The grey dashed lines indicate the positions of possible spectral lines. There are no detections of the CO ladder in the SSW except for CO (9--8) in N66, which also displays CO (8--7) in the SLW, showing that gas associated with the molecular complex in this giant \hii\ region is warm and highly excited.
\label{fig:FTS_spec}}
\end{figure*}

\subsubsection{\OI\ self-absorption}
\label{subsubsection:OI_abs}
A challenge with interpreting velocity unresolved observations of \OI\ 63 \micron, such as ours ($\Delta \rm{v}\sim{100}$ km s$^{-1}$), is the potential effect of self-absorption or absorption from cold gas along the line of sight, a phenomenon originally identified through anomalous \OI\ 145 \micron\ to 63 \micron\ integrated line ratios. Indeed some Milky Way massive star-forming regions show significant self-absorption and absorption by foreground cold clouds containing O$^{0}$ in velocity-resolved observations of \OI\ \citep{pog96,leu15}. It is unknown how widespread this phenomenon is in the SMC, where 145 \micron\ observations do not exist and velocity-resolved observations are very limited.

In the Milky Way, heavy \OI\ self-absorption is usually accompanied by \CII\ absorption \citep[e.g.,][]{leu15}. There is no indication of absorption in recent \CII\ velocity-resolved profiles (\citealt{req16}; R. Herrera-Camus et al. 2017, in preparation), and no clear evidence of self-absorption in \OI\ velocity-resolved profiles (Okada et al. 2017, in preparation) in the star-forming regions N25 (located in the north end of the HS$^{3}$ ``N22'' region), N66, and N88 in the SMC. Given the high radiation fields and low \Av\ throughout much of the SMC, this suggests that in the SMC there is a dearth of high \Av\ cold material that may absorb \OI\ along the line of sight, while absorption contamination is likely more common in the Milky Way \citep[e.g.,][]{leu15}. As mentioned above, another indicator of optical depth or absorption in the \OI\ 63 \micron\ line is anomalously high \OI\ 145 \micron\ to 63 \micron\ ratios \citep{sta83}. While there are no $Herschel$ PACS observations of \OI\ 145 \micron\ in the SMC, in the LMC three regions were observed in \OI\ 145 \micron\ to 63 \micron\ by \citep{cor15} and 30 Doradus by \citep{che16}. Only two of these regions, N159 \citep[the site of the brightest CO emission in the LMC;][]{isr93} and 30 Doradus (one of the most active star-forming regions in the Local Group), have a high 145 \micron\ to 63 \micron\ ratios with a ratio of 0.11 found in N159 and $>0.1$ in 30 Doradus, which is not much higher than the theoretical limit of 0.1 for the expected ratio for optically thin emission for $T>300$ K \citep{tie85}. If the line were self-absorbed in our observations, we would expect to see lower \OI/\CII\ ratios in high density regions at higher \Av, but we see the opposite. We conclude that it is unlikely that the \OI\ 63 \micron\ line is significantly affected by absorption in our SMC observations.

\subsubsection{\OI-to-\CII\ Ratio}
\label{subsubsection:OI_CII_ratio}
What is the origin of the observed \CII\ emission?
Figure \ref{fig:OI_CII} shows the integrated intensity ratio of \OI\ to \CII. The observed ratio is approximately constant with a value of $\OI/\CII\sim{0.3}$. One of the main departures from the mostly flat trend in $\OI/\CII$ with \Av\ are a cluster of higher ratio values in the SWBarS, which are found in the \hii\ region N13 indicating the presence of warm, dense gas. This is also the typical value observed in the disks of the KINGFISH sample of nearby galaxies \citep{her15}. Using the \CII\ and \OI\ cooling curves calculated for diffuse gas under SMC conditions (Wolfire et al. 2017, in preparation), a ratio $\OI/\CII\sim{0.3}$ is indicative of densities of $\sim{10^{2}-10^{3}}$ cm$^{-3}$, which are consistent with dense CNM and/or molecular gas. Figure \ref{fig:H2_S0} shows the similarity between the mid-infrared \htwo\ S(0) quadrupole rotational line at 28.8 \micron\ and the \CII\ emission. This clearly demonstrates that molecular gas is associated with the \CII-emitting material, strongly suggesting that most of the \CII\ emission in our mapped regions has a PDR origin and it arises from the surfaces of molecular clouds.

\subsubsection{Photoelectric Heating Efficiency at Low Metallicity}

The dominant heating source is the photoelectric effect where a dust grain absorbs a FUV photon and ejects an electron that heats the gas through collisions. The \CII\ and \OI\ far-IR (FIR) line emission dominates the cooling of the diffuse atomic and molecular gas, as well as PDRs. By combining the \CII\ and \OI\ line emission, we can account for most of the gas cooling that is attributed to gas heated by the photoelectric effect. Taking the ratio of \CII\ and \OI\ intensities to the TIR intensity indicates the fraction of the power absorbed by the grains that goes into heating the gas through the photoelectric effect. Figure \ref{fig:PE_G0} shows the mean $\CII+\OI/$TIR ratios for each of the HS$^{3}$ regions as a function of the mean dust temperature from \citet{lee15}. The ratio of $\CII+\OI/$TIR ranges from $\sim{0.01-0.018}$ in the SMC. This is similar to the LMC, where \citet{rub09} find that the \CII\ in emission from BICE observations accounts for $\sim{1\%}$ of the TIR emission. These ratios are on the high end of the range observed for the Milky Way and galactic nuclei of $0.1-1\%$ using KAO observations \citep{sta91}, $<0.1-1\%$ for the nearby KINGFISH galaxies using $Herschel$ observations \citep{smi17},  normal galaxies using $ISO$ observations \citep{mal01}, and M31 using $Herschel$ observations \citep{kap15}. We also see a trend of decreasing ratios with high dust temperatures, also observed by \citet{mal01}, \citet{cro12}, and \citet{kap15}, which is commonly attributed to the increased grain charging at warmer dust temperatures that increases the energy threshold for the photoelectric effect ejection of electrons and decreases the energy, and therefore amount of gas heating, per ejected electron. 

\subsection{\NII\ and Contribution from Ionized gas}
\label{subsection:NII}

The HS$^{3}$ data set includes sparsely sampled FTS spectra for the SWBarN, SWBarS, N22, and N83 regions, as well as a pointing towards the most active star-forming region in the SMC, the giant \hii\ region N66. In Figure \ref{fig:FTS_spec} we show the long wavelength array (SLW) spectra in red and the short wavelength array (SSL) in blue with the positions of the \CO, \CI, and \NII\ lines indicated. We see clear detections of the lower rotational transitions of \CO\ and the \NII\ 205 \micron\ lines, as well as weak detections of \CI.

We do not detect the \NII\ 122 \micron\ line in any of the regions, whereas the \NII\ 205 \micron\ line is detected in all regions in the FTS spectra (see Figure \ref{fig:FTS_spec}). Because the ionization potential of nitrogen of 14.5 eV is higher than hydrogen, ionized nitrogen traces the ionized gas. The \NII\ 122 \micron\ and 205 \micron\ lines result from the fine-structure splitting of the ground state of ionized nitrogen and are primarily excited by collisions with electrons. 

The critical densities of the 122 \micron\ and 205 \micron\ lines differ and the ratio can be used to estimate the electron density ($n_{\rm{e}}$). The \NII\ 122 \micron\ line has a higher critical density for collisions with electrons ($n_{\rm{e}}\sim{300}$ cm$^{-3}$) compared to the 205 \micron\ line ($n_{\rm{e}}\sim{40}$ cm$^{-3}$), which has a critical density similar to that for exciting the \CII\ 158 \micron\ line with collisions with electrons. Thus the ratio of \CII/\NII\ 205 \micron\ in ionized gas is independent of density, and it depends only on the relative abundances of the ions which are likely similar to the elemental abundances. 

The sensitivity of the \NII\ 122 \micron\ observations is $\sim{3\times10^{-10}}$ W m$^{-2}$ sr$^{-1}$, while the range of detected \NII\ 205 \micron\ intensities is $\simå{2-5\times10^{-10}}$ W m$^{-2}$ sr$^{-1}$. Based on these measurements, the \NII\ 122/205 ratio is $\lesssim{1}$ in regions where the \NII\ 205 \micron\ line is detected. Using the electron collision strengths from \citet{tay11}, this translates to an upper limit to the electron density of $n_{e}\lesssim{20}$ cm$^{-3}$ since this is approximately the \NII\ 122/205 ratio diagnostic lower density limit. In other words, our measurements are consistent with a relatively low density for the ionized material, somewhat lower than the mean ionized gas density observed in the KINGFISH sample of galaxy disks of $n_e\sim30$\,cm$^{-3}$ \citep{her16}. 

\begin{figure}[t]
\epsscale{1.2}
\plotone{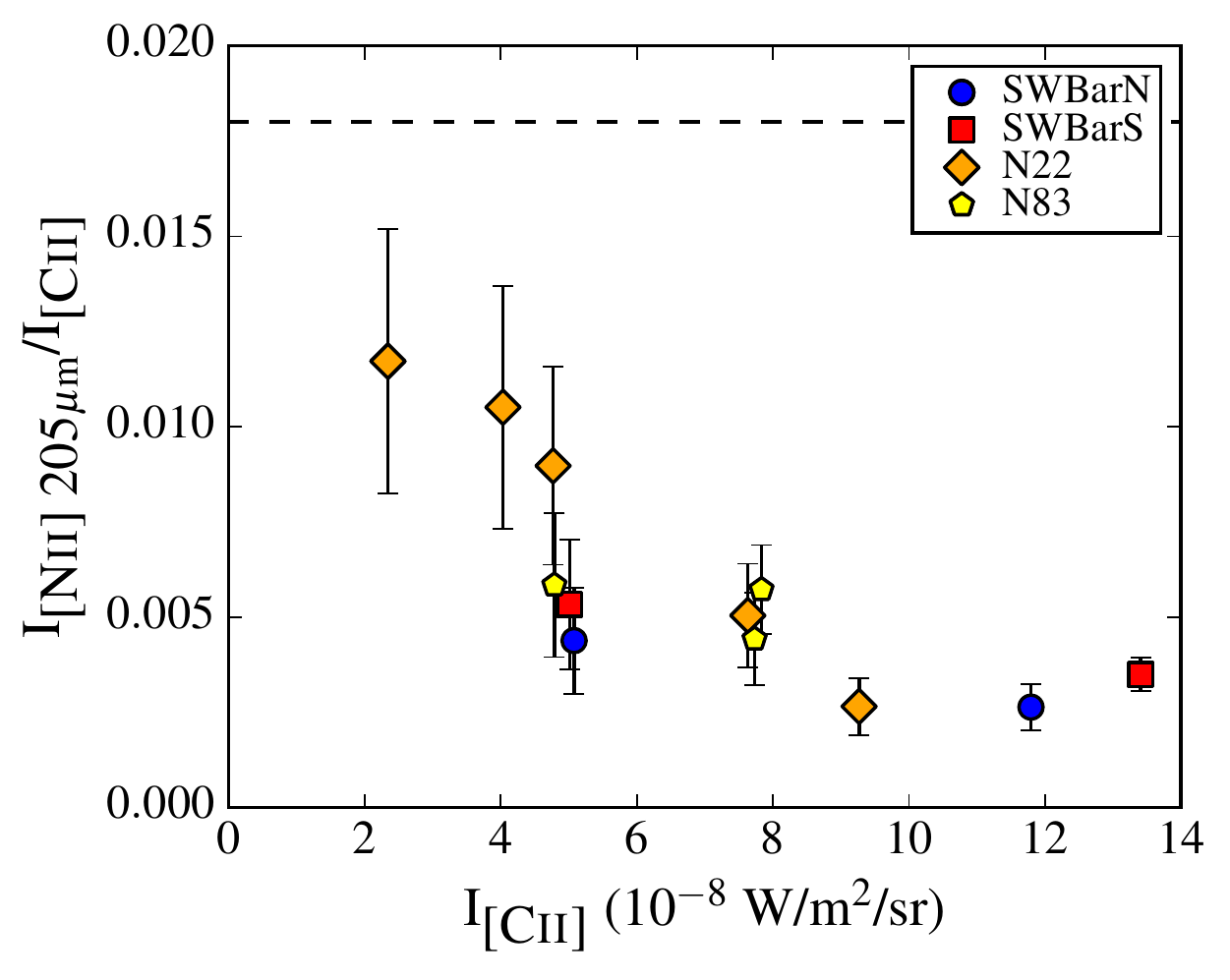}
\caption{Ratio of integrated intensities of \NII\ 205 \micron\ (I$_{\NII~\rm{205}\mu\rm{m}}$) to \CII\ (I$_{\CII}$) as a function of the \CII\ intensity for FTS bolometer measurement of \NII\ 205 \micron\ that is $>3\sigma$ for each of the regions. The error bars show the $1\sigma$ uncertainty on the line ratio, which is dominated by the uncertainty in the \NII\ 205 \micron\ flux (the uncertainty in I$_{\CII}$ is smaller than the symbols). The FTS pointings for each of the regions targeted the star-forming peak, which tends to coincide with the peak CO emission. The ratios are low and naturally peak towards higher values at lower \CII\ intensities. The black dashed line shows the lowest observed ratio of $\NII~205~\rm{\micron}/\CII\sim{0.018}$ in the KINGFISH sample \citep{cro17}.  
\label{fig:NII_CII}}
\end{figure}

In Figure \ref{fig:NII_CII}, we show the \NII\ 205 \micron/\CII\ ratio for all of the pointings where \NII\ 205 \micron\ is detected at $>3\sigma$, and where the \CII\ intensity found for the central FTS bolometer position after convolving the map to the FTS resolution ($\sim{17\arcsec}$). We see that the \NII\ 205 \micron\ emission ranges from $0.2\%-1.2\%$ of the \CII\ emission. For carbon emission arising from ionized gas with a ratio ${\rm C^+/N^+\approx C/N}$ similar to Galactic we would expect a \NII\ 205 \micron/\CII$\sim0.2$, mostly independent of density due to the similarity in the critical densities \citep{tay08,tay11}. Because the \NII\ emission can only arise from ionized gas, the fact that we measure over $\sim20$ times fainter \NII\ relative to \CII\ suggests that the contribution of ionized gas to \CII\ is at most 5\%. 

These ratios represent the maximum ratios throughout the regions since the FTS observations targeted the bright CO emission, which tend to be near \hii\ regions, and as such will have the highest fraction of \CII\ emission arising from ionized gas. The observed \NII\ 205\micron/\CII\ ratios are lower than the typical values found in the KINGFISH survey of 0.057 \citep{cro17}. Similarly, \citet{cor15} find depressed \NII\ 122 \micron/\CII\ in dwarf galaxies, suggesting that ionized gas only produces a small fraction of the \CII\ emission in low metallicity environments.

\subsection{\OIII\ and Highly Ionized Material}
\label{subsection:OIII}

The \OIII\ 88 \micron\ line traces ionized gas as the second ionization potential of oxygen is $\sim{35}$ eV, much higher than hydrogen. H$^3$S obtained \OIII\ observations toward the dominant HII region in each strip. We observed bright \OIII\ emission from all of these pointings. \citet{van10} also detected \OIII\ towards a subset of their sample of compact sources in the SMC using ISO spectra. Figure \ref{fig:OIII_CII} shows the \OIII/\CII\ ratios for the SMC regions, which reach as high as $\OIII/\CII\sim{4}$. Observations of the \OIII/\CII\ ratio for higher metallicity galaxies based on ISO data presented by \citet{bra08} found lower ratios that ranged from $\OIII/\CII\sim{0.1-1.5}$. The higher values in the SMC are similar to the ratios found for dwarf galaxies observed with $Herschel$ PACS as part of the Dwarf Galaxy Survey (DGS) with a median and range of $\OIII/\CII=2.0^{+13.0}_{-0.52}$ \citep{cor15}. These comparisons to measurements in other galaxies should be tempered somewhat by the fact that the SMC observations are pointings toward \hii\ regions obtained at high spatial resolution, while the comparison work typically samples larger scales and therefore a mix of ionized and neutral material. 

\input{table4.tex}

While the critical density for \NII\ 122 \micron\ line is similar to \OIII, the \OIII\ ionization potential is much higher as is the energy above ground for excitation. \citet{cor15} suggest that the hard radiation fields found at lower metallicity in dwarf galaxies could explain the high \OIII\ emission and high \OIII\ to \NII\ 122 \micron\ ratios with a median ratio of 86 found for the DGS sample. Note that we failed to detect \NII\ 122 \micron\ emission toward these same pointings. A combination of hard radiation fields and low density ionized gas may explain the high \OIII\ to \NII\ 122 \micron\ ratios present in the SMC, with the a lower limit  $\OIII/\NII$ 122 \micron\ $\sim{25}$ (calculated using $3\sigma$ of the \OIII\ intensity and the $1\sigma$ sensitivity of the \NII\ intensity). We note that the O/N abundance ratio in the SMC is similar to that found in the Solar neighborhood \citep{rus92} and it is unlikely that a difference in abundance ratios would explain the relatively high \OIII\ emission. Understanding the ISM conditions that produce the \OIII\ line emission in low metallicity environments is critical for interpreting new and future observations of FIR cooling lines in high redshift galaxies using ALMA \citep[e.g.,][]{ino16}.

\subsection{High Resolution Molecular Gas: \CO\ and \thirteenCO}
\label{subsection:12CO_13CO}

We mapped and detected \CO\ and \thirteenCO\ $(2-1)$ in all of the regions targeted by ALMA, shown in Figure \ref{fig:ALMA_CO}. We do not detect \CeighteenO\ with our current observations, which had the minimum integration time allowed per pointing to increase the coverage of the mosaics. Despite the lower metallicity and less dust-shielding, \CO\ and  \thirteenCO\ form and emit brightly in small clumps. Our high resolution ALMA ACA data show that the bright CO emission is found in small structures, which we will quantify in a forthcoming paper (Jameson et al. 2017, in preparation), which were unresolved by previous observations. The clumpy nature of the CO emission at low metallicity has also been observed in the N83C star-forming region in the wing of the SMC \citep{mur17}, the dwarf galaxies WLM \citep{rub15} and NGC 6822 \citep{sch17}, and in 30 Doradus of the LMC \citep{ind13}. 

For most of the regions, a large fraction of the flux is recovered by the high resolution ACA maps. In the SWBarS and SWBarN regions, the ACA flux represents $\sim{60\%}$ of the flux in the combined ACA+TP \CO\ maps. In the SWDarkPK, nearly $100\%$ of the flux is from the high resolution imaging, whereas in the N22  region most of the emission is diffuse with only $\sim{30\%}$ of the flux found at high resolution. The higher fraction of diffuse \CO\ emission in N22, less in SWBarS and SWBarN, and none in SWDarkPK is likely due to their varying evolutionary stages: N22 is the most evolved region with multiple large \hii\ regions around the CO emission, SWBarS and SWBarN both are actively forming stars and have one prominent \hii\ region, and SWDarkPK has no signs of active star formation. The higher UV fields likely increases how deep the PDR extends into the molecular cloud and the amount of diffuse CO emission associated with the PDR. In the more evolved regions (particularly N22), the densest peaks of molecular gas may have already been dispersed by star formation leaving less clumps of molecular gas and increasing the fraction of diffuse CO emission.

Figure \ref{fig:12COto13CO} shows the comparison between the \CO\ and \thirteenCO\ $(2-1)$ emission. We find average \CO/\thirteenCO\ $(2-1)$ ratios of $\sim{5-7.5}$ (in units of K km s$^{-1}$). These ratios are consistent with previous measurements in the SMC toward emission peaks \citep{isr03} and in nearby galaxies \citep[e.g.,][]{pag01,kri10}. In the Milky Way the ratios are similar to what we obtain for the SMC, with an average of $\sim{5}$ \citep{sol79} in the inner Galaxy, and somewhat higher ratios of $\sim{7}$ for large parts of the plane \citep{pol88} and in the outer Galaxy clouds \citep{bra95}.

\subsection{Estimating the Optical Depth of \CO}
\label{subsection:CO_opticaldepth}

The linear trend with no turnover observed between \CO\ and \thirteenCO\ indicates that while the \CO\ $(2-1)$ transition is optically thick where there is \thirteenCO, \thirteenCO\ likely remains optically thin for these observations. The \thirteenCO-to-\CO\ ratio gives the optical depth of \thirteenCO\ $(2-1)$, and from that we can estimate the optical depth of the \CO\ $(2-1)$ emission. The Rayleigh-Jeans radiation temperature of the CO line emission is 
\begin{equation}
T_{R} = J_{R}(T_{ex})(1-e^{-\tau})
\label{eqn:Tr}
\end{equation} 
where $T_{ex}$ is the excitation temperature and $\tau$ is the optical depth of the line. The observed intensity is, 
\begin{equation}
J_{R}(T_{ex}) = \frac{h\nu}{k}\left( \frac{1}{e^{(h\nu/kT_{ex})}-1} - \frac{1}{e^{(h\nu/kT_{bg})}-1} \right)
\label{eqn:Jr}
\end{equation}
where $T_{bg}$ is the background temperature (taken to be the Cosmic Microwave background of 2.73 K). If we assume both lines share the same excitation temperature, then the ratio of the line brightness temperatures for the two isotopic species can be used to estimate the optical depth of the more abundant species. This is strictly correct only in the high-density regime ($n\gg n_{cr}$, where $n_{cr}\sim10^4$\,cm$^{-3}$ is the critical density of the \thirteenCO\ $2-1$ transition that is assumed to be optically thin), where the level populations will follow a Boltzmann distribution at the kinetic temperature of the gas. 
For \CO\ and \thirteenCO\ $(2-1)$, 
\begin{equation}
\frac{T_{R,\rm{\CO}(2-1)}}{T_{R,\rm{\thirteenCO}(2-1)}} = \frac{1-e^{-\tau_{\rm{\CO}}}}{1-e^{-\tau_{\rm{\CO}}/X}}
\end{equation}
where $X$ is the abundance ratio of $\CO/\thirteenCO$. Assuming that $\tau_{\rm{\CO}}\gg1$, we can solve for the optical depth of \CO:
\begin{equation}
\tau_{\rm{\CO}} = -X\ln{(1-T_{R,\rm{\thirteenCO}(2-1)}/T_{R,\rm{\CO}(2-1)})}
\end{equation} 
We adopt an abundance ratio of $X=70$, which is appropriate for the Milky Way \citep{wil94}, since it is intermediate between the two values found in the SMC using radiative transfer modeling of the \CO\ and \thirteenCO\ lines from \citet{nik07}. Using this isotopic abundance ratio, the average $\CO/\thirteenCO$ values (where $I_{\thirteenCO}>3\sigma$) indicate optical depths of $\tau_{\rm{\CO}} = 7.8$, 13.8, 12.5, and 19.0 for the SWBarN, SWBarS, N22, and SWDarkPK regions, respectively. In practice \CO\ is likely to be more highly excited than \thirteenCO\ due to radiative trapping, which would result in smaller \thirteenCO/\CO\ ratios and somewhat underestimating the optical depth by this method. At these size scales ($\sim{2}$ pc), the beam may include high and low density gas with the low density gas more likely to have more excited \CO, and also decrease the \thirteenCO/\CO\ ratio. However, the density in the \CII\ emitting gas (see Section \ref{subsubsection:method2} and Figure \ref{fig:PDRT}) is already high and there is unlikely to be a strong contribution from \CO\ emitting low density gas.

\begin{figure}[t]
\epsscale{1.2}
\plotone{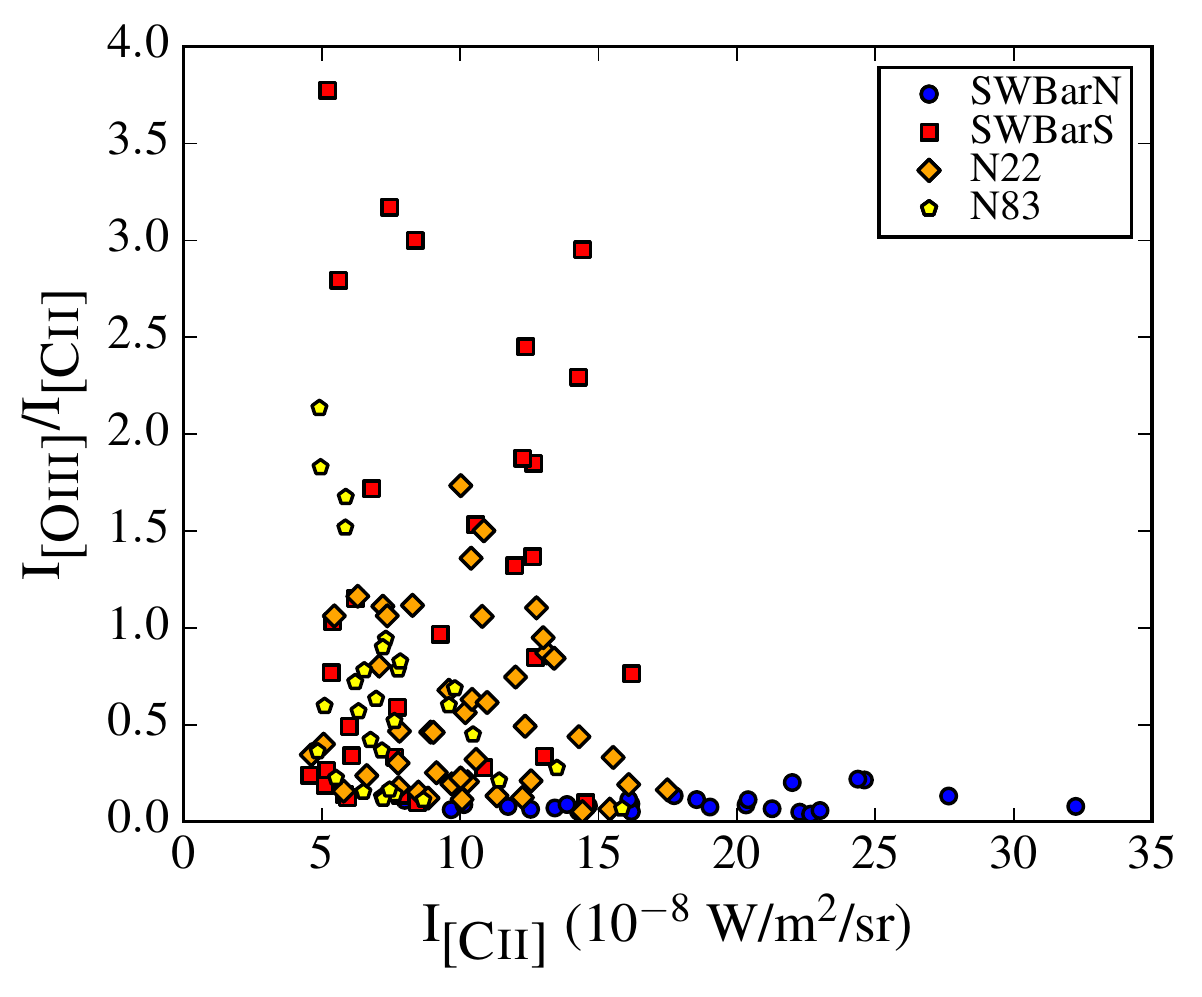}
\caption{Ratio of integrated intensities of \OIII\ 88 \micron\ (I$_{\OIII}$) to \CII\ (I$_{\CII}$) as a function of the \CII\ intensity where \OIII\ is detected at $>3\sigma$ for each of the regions. The \OIII\ maps cover the main \hii\ regions in the HS$^{3}$ regions. We see high I$_{\OIII}$/I$_{\CII}$ ratios in all of the regions except SWBarN where there is no large \hii\ region and likely less ionized gas.  
\label{fig:OIII_CII}}
\end{figure}

\begin{figure*}[t]
\epsscale{1.2}
\plotone{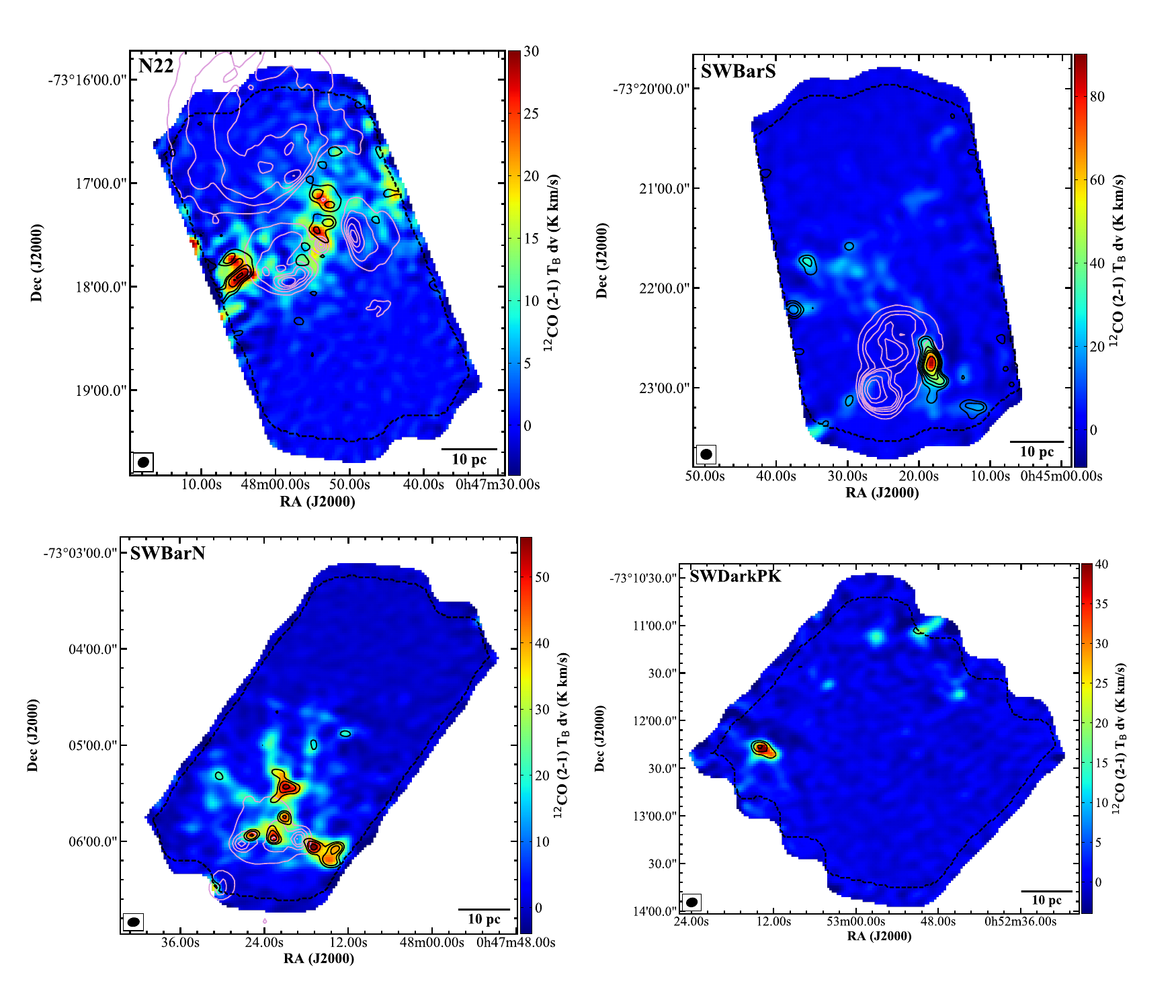}
\caption{Integrated intensity maps of ALMA ACA+TP \CO\ $(2-1)$ with black contours showing ALMA ACA \thirteenCO\ $(2-1)$ at levels of 2.5, 4, 6, 8, 10, 15 K km s$^{-1}$ and the light purple contours showing the \ha\ contours shown in Figure \ref{fig:HS3_RGB}. The black dashed line shows the coverage of the \thirteenCO\ $(2-1)$ image. The sensitivities of the maps are listed in Table \ref{table:ALMA_prop}. 
\label{fig:ALMA_CO}}
\end{figure*}

\begin{figure}[t]
\epsscale{1.2}
\plotone{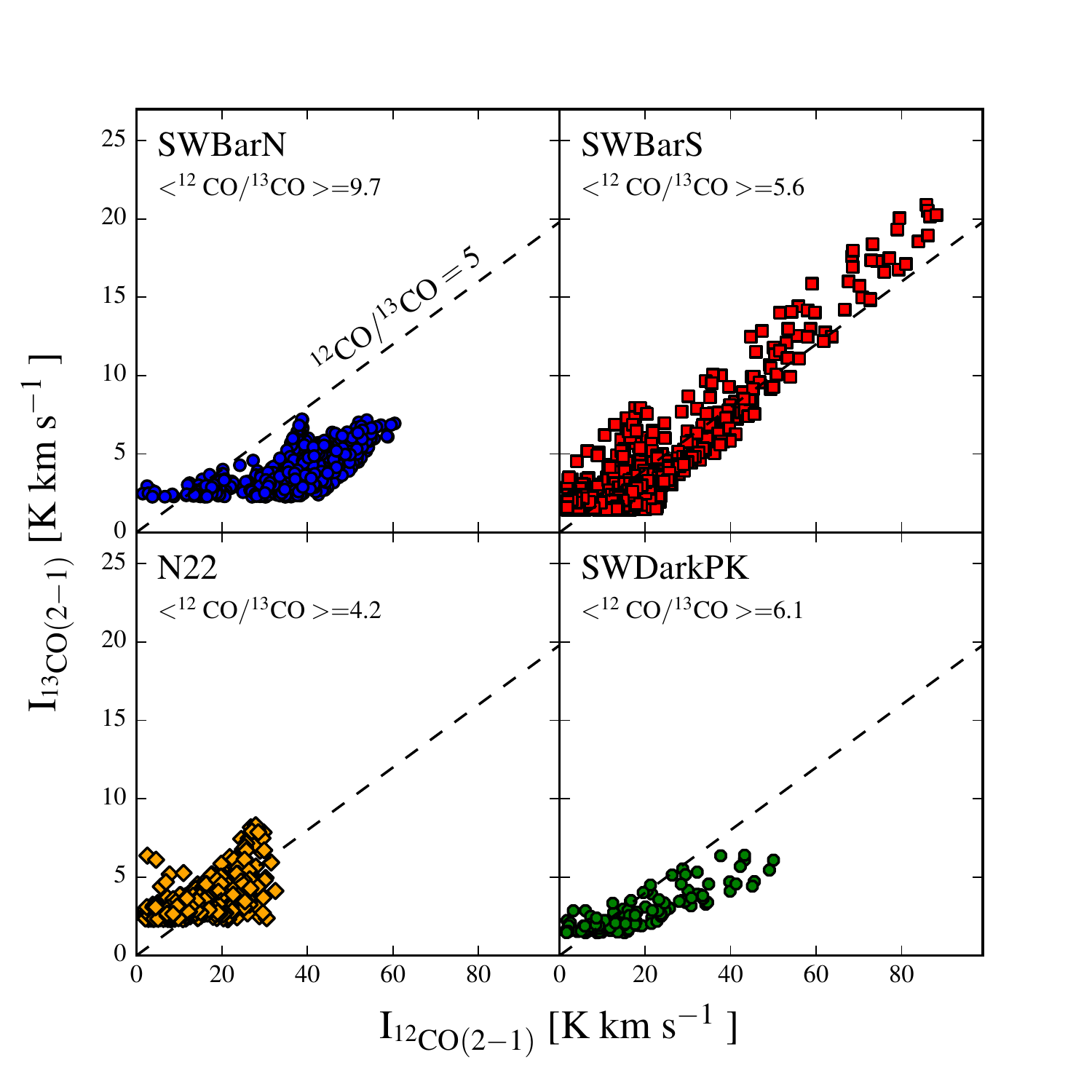}
\caption{Ratio of integrated intensity of \CO\ $(2-1)$ to \thirteenCO\ where both are detected at $>3\sigma$. The average \CO/\thirteenCO\ ($\langle\CO/\thirteenCO\rangle$) are listed for each region. The dashed line shows a typical ratio found in the Milky Way of $\CO/\thirteenCO=5$. 
\label{fig:12COto13CO}}
\end{figure}

\subsection{Estimating $X_{\text{CO}}$ Using \thirteenCO}
\label{subsection:XCO_13CO}

The \CO\ emission is likely to be optically thick under most conditions found in a molecular cloud, and indeed we estimate high average optical depths of the \CO\ $(2-1)$ line emission. Assuming, however, that \thirteenCO\ is optically thin is reasonable throughout much of a cloud, given the high isotopic ratio and the lower C abundance in the SMC. We can then use the \thirteenCO\ intensity to estimate a molecular gas column density, and from that infer a \CO-to-\htwo\ conversion factor ($X_{\text{CO}}$). 

We use equations \ref{eqn:Tr} and \ref{eqn:Jr} and the assumption that the excitation temperature is the same for both the \CO\ and \thirteenCO\ emission to calculate the optical depth of the \thirteenCO\ emission. The total column density of \thirteenCO\ ($N_{\thirteenCO}$) as a function of the excitation temperature ($T_{ex}$) and optical depth of \thirteenCO\ ($\tau_{\thirteenCO}$) for the $J=2\rightarrow1$ transition is given by \citep{gar91,bou97}:
\begin{equation}
N_{\thirteenCO} = 1.12\times10^{14}\frac{(T_{ex}+0.88)e^{5.29/T_{ex}}\int{\tau_{\thirteenCO}dv}}{1-e^{-10.6/T_{ex}}}
\end{equation}
We then make the approximation that the integral of $\tau_{\thirteenCO}$ is taken to be the line center optical depth ($\tau_{\thirteenCO,0}$) multiplied by the \thirteenCO\ full width at half maximum ($\Delta \rm{v}_{\thirteenCO}$) \citep{dic78}. To convert from $N_{\thirteenCO}$ to \Nhtwo, we scale by the $^{12}\textrm{CO}/^{13}\textrm{CO}$ abundance of 70 \citep{nik07} and the $^{12}\textrm{C}/\textrm{H}$ abundance of $2.8\times10^{-5}$ (for a justification see Section \ref{section:estimating_h2}). \textbf{This results in $\htwo/^{13}\textrm{CO}=1.25\times10^{6}$.}

\begin{figure}[t]
\epsscale{1.2}
\plotone{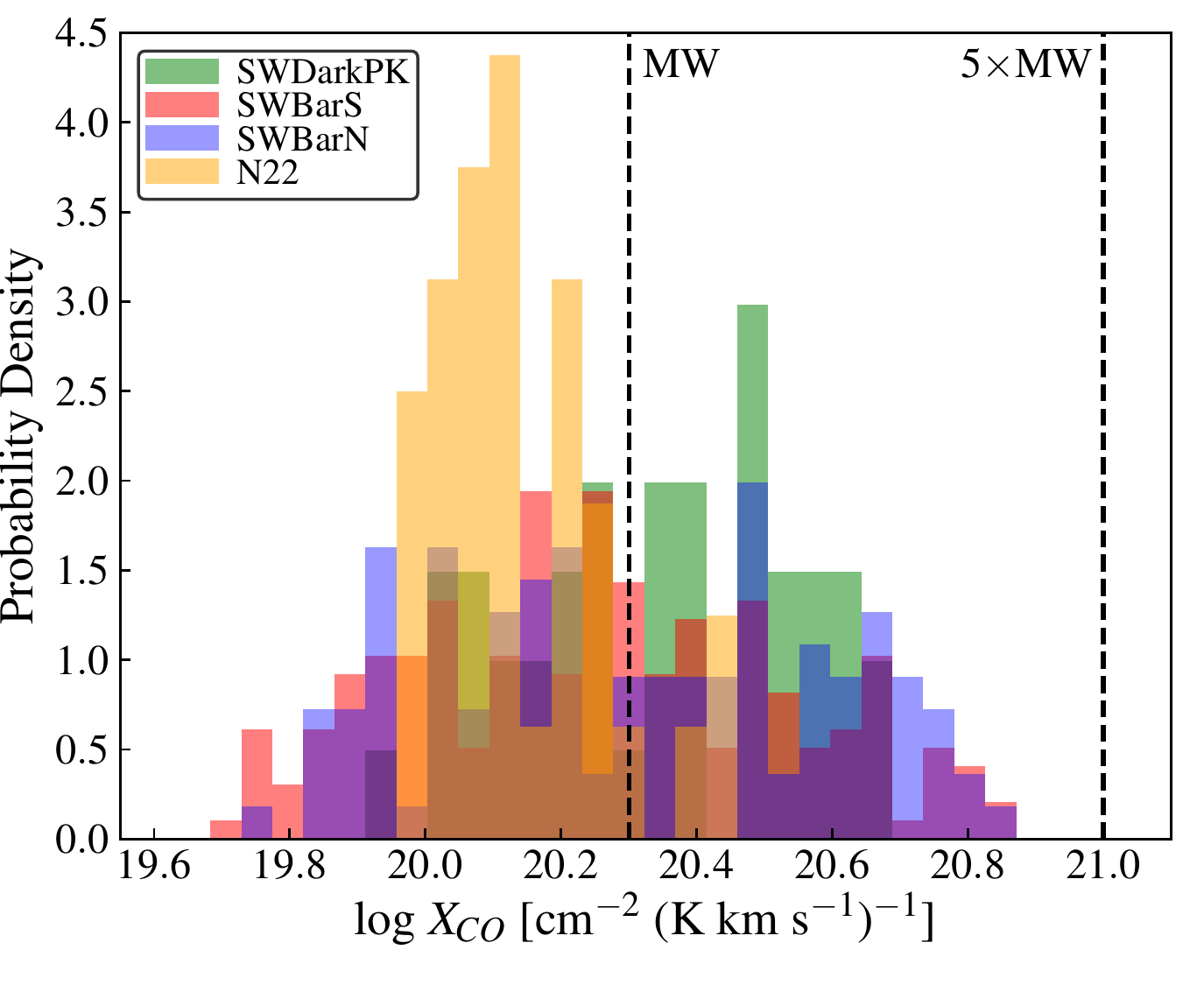}
\caption{Histogram of the probability density of $X_{\text{CO}}$ for areas in each region with I$_{^{13}\rm{CO}}>3\sigma$ and at $\Av>1$ with \Nhtwo\ estimated using \thirteenCO. The vertical dashed black line shows the Galactic value of $X_{\text{CO}}=2\times10^{20}$ cm$^{-2}$ (K km s$^{-1}$)$^{-1}$ and five times the Galactic value. We see that our estimates of $X_{\text{CO}}$ in the high-\Av\ regions peak close to the Galactic value.
\label{fig:XCO_13CO}}
\end{figure}

Figure \ref{fig:XCO_13CO} shows the distribution of $X_{\text{CO}}$ for lines of sight with I$_{^{13}\rm{CO}}>3\sigma$ and $A_{V}>1$ where we expect the molecular gas to be primarily traced by \CO\ emission. We estimate the $\Av>1$ by converting $N_{\thirteenCO}$ to \Av\ using the empirically determined relationship $\Av \approx 4\times10^{-16}N_{\thirteenCO}$ cm mag$^{-1}$ from \citet{dic78} for Milky Way dark clouds, which should be appropriate if the \thirteenCO\ abundance scales the same as the dust abundance.  The distributions have median values of $1.3-2.3\times10^{20}$ cm$^{-2}$ (K km s$^{-1}$)$^{-1}$, which are consistent with a Milky Way $X_{\text{CO}}$.

Estimating $X_{\text{CO}}$ using \thirteenCO\ involves a number of assumptions that are not well constrained. We assume that the \thirteenCO\ and \CO\ lines share the same excitation temperature, that this temperature can be inferred from the brightness of the \CO\ emission at our spatial resolution (typically $\sim2.3$~pc, see Table \ref{table:ALMA_cont}), and that the gas is in LTE. We do not account for potential beam dilution, which could decrease our measurements of $T_{ex}$ from \CO\ and cause us to overestimate $\tau_{\thirteenCO}$ and $N_{\thirteenCO}$ and overestimate $X_{\text{CO}}$. \citet{won17} make similar estimates of the molecular gas mass using \CO\ and \thirteenCO\ ALMA observations of molecular clouds in the LMC, which suffer from similar uncertainties, and find evidence that their $N_{\htwo}$ estimates were biased towards higher values. Conversely, the assumption of LTE could potentially cause $N_{\thirteenCO}$ to be underestimated by a factor of $\sim{1.3-2.5}$ based on the comparison to simulations of molecular clouds by \citet{pad00} for the typical $N_{\thirteenCO}$ in our regions ($\sim{10^{15}}$ cm$^{-2}$). Both the $^{12}\textrm{CO}/^{13}\textrm{CO}$ and $^{12}\textrm{C}/\textrm{H}$, which we assume also traces $^{12}\textrm{CO}/\textrm{H}$, are uncertain in the SMC. The existing $^{12}\textrm{C}/\textrm{H}$ measurements vary by $\pm{30\%}$. \citet{nik07} constrains the $^{12}\textrm{CO}/^{13}\textrm{CO}$ abundance ratio to $\pm\sim{30\%}$, but recent radiative transfer modeling by \citet{req16} find $^{12}\textrm{CO}/^{13}\textrm{CO}=50$, which is at the lower end of their range found by \citep{nik07}. If we are overestimating the abundance of \thirteenCO, which could be by up to a factor of 2 based on the metallicity scaling of the Milky Way measurement of $\htwo/^{13}\textrm{CO}$ from \citet{dic78}, then we would also underestimate $X_{\text{CO}}$ by the same amount.  It is unclear how all of these uncertainties would balance out in these specific regions. \textbf{Ultimately, the data are consistent with a conversion factor similar to that of the Milky Way, but we caution the reader that there is a large degree of uncertainty on this estimate.}    

\subsection{Relationship between \CO\ and \CII}
\label{subsection:CO_CII}

\begin{figure*}[t]
\epsscale{1.1}
\plotone{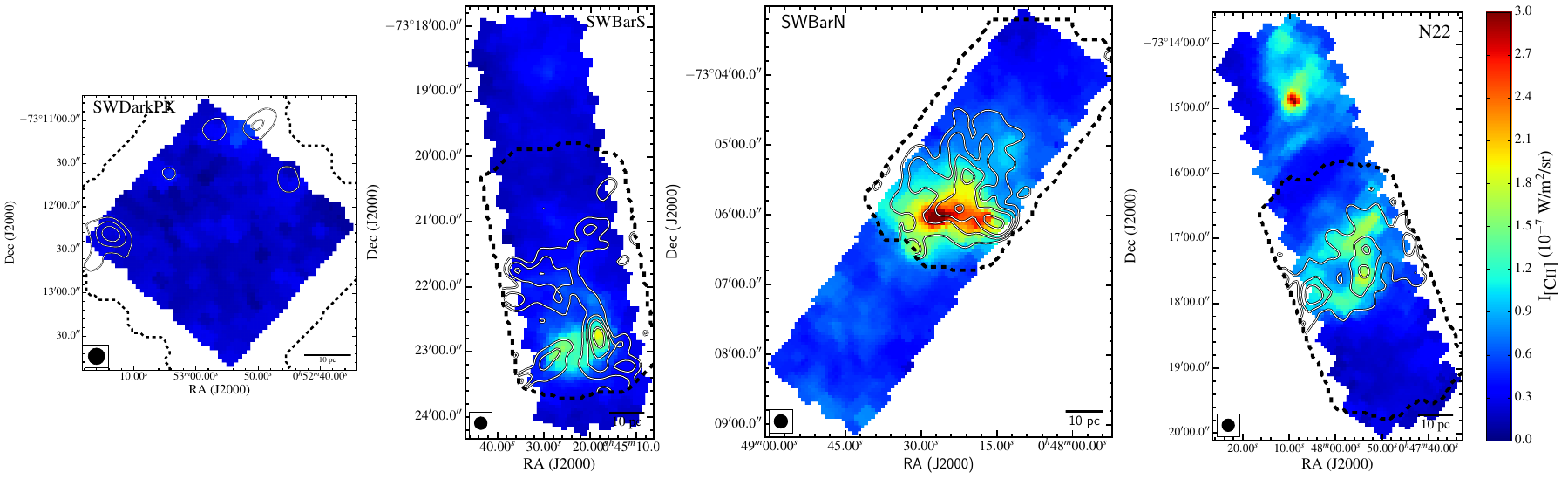}
\caption{Maps of the \CII\ integrated intensity with black contours showing the integrated intensity of ALMA ACA+TP \CO\ $(2-1)$ map convolved to the \CII\ resolution ($\theta=12\arcsec$) at levels of 0.3, 1, 2, 4, $6\times10^{-10}$ W m$^{-2}$ sr$^{-1}$ (2, 8, 16, 32, 48 K km s$^{-1}$). The dashed black lines show the coverage of the ACA maps. There is generally good agreement between the peaks in the \CII\ and the \CO\ emission and the faint \CII\ emission tends to trace the faint \CO\ structure, strongly suggesting that most of the \CII\ is originating from PDRs and is associated with molecular gas.
\label{fig:CII_CO_images}}
\end{figure*}

\CII\ emission can arise from neutral atomic gas, molecular gas, and ionized gas. Our comparison of \CII\ to \NII\ shows that only a small fraction of the \CII\ emission arises from ionized gas. While \CII\ emission is an important coolant of the cold neutral medium (\citealt{dal76}; which we explore in Section \ref{subsection:CII_from_HI}), PDRs are expected to produce bright \CII\ emission \citep[][and references therein]{hol99}. The carbon will exist as CO in the dense parts of the molecular gas, but at the edges of the cloud the lack of shielding will dissociate CO and much of the carbon will exist as C$^{+}$ while there can still be molecular (\htwo) gas \citep[e.g.,][]{isr96,bol99,wol10}. This leads to a layer of molecular gas surrounding the molecular cloud associated with \CII\ but no \CO\ emission. Studying the \CO\ to \CII\ ratio can determine whether the \CII\ emission is primarily associated with the PDR and provide insight into the structure of the molecular cloud and its associated faint CO gas.  

What is the origin of the \CII\ emission in our SMC regions? Figure \ref{fig:CII_CO_images} shows the \CII\ integrated intensity maps with the ALMA \CO\ $(2-1)$ integrated intensity shown with contours. There is a striking similarity between the structure, at both high and low intensity, but the \CII\ extends throughout the mapped region whereas the CO emission is more localized. The similarity in integrated intensity emission structure suggests that the \CII\ emission predominantly traces the molecular gas. This is reinforced by the velocity-resolved observations. Recent observations of the \CII\ 158 \micron\ line using the GREAT instrument on SOFIA for individual pointings throughout some of the HS$^{3}$ regions show that the velocity ranges covered by the \CII\ and \CO\ $(2-1)$ emission are similar, although in a FWHM sense the \CII\ line can be up to $\sim{50\%}$ wider than the \CO\ line (R. Herrera-Camus et al. 2017, in preparation). Other SOFIA GREAT observations of SMC star-forming regions N66, N25/26 (covered in the HS$^{3}$ `N22' region), and N88 show that the \CII\ profile is similar to the \CO\ profile and with the \CII\ profile up to $\sim{50\%}$ wider \citep{req16}. SOFIA GREAT observations of \CII\ in M101 and NGC6946 also show that the velocity profile of \CII\ has a better correspondence with \CO\ than \hi\ \citep{deb16}. We show one example GREAT \CII\ spectrum for one pointing in the SWBarN region compared with the beam-matched ALMA \CO\ $(2-1)$ profile and the \hi\ emission (at a resolution $\sim{1\arcmin}$) toward the same point in Figure \ref{fig:SOFIA_spec}. As mentioned above, the CO and \CII\ cover a very similar velocity range, which is much more restricted than that of the \hi\ emission. This strongly suggests that toward these molecular regions \CII\ is dominated by emission arising from molecular gas rather than atomic gas. 

Furthermore, there is a clear similarity between the \CII\ map and the map of the \htwo\ rotational line emission at $28.2$ \micron\ (see Figure \ref{fig:H2_S0}). This line requires warm gas ($T\gtrsim100$~K) to be excited, so it is a poor tracer of the bulk of the molecular mass. But the spatial coincidence between these two transitions together with the \CII-\CO\ agreement above provide strong evidence that a majority of the \CII\ emission arises from molecular gas. 

\subsection{\CO-to-\CII\ Ratio}

In Figure \ref{fig:CO_CII_OICII} we show the ratio of $\CO/\CII$ emission as a function of $A_{V}$ estimated from dust emission (see \S\ref{subsection:Av}), which indicates the total column density through the molecular cloud along the line-of-sight. The ratios are typically much lower at low $A_{V}$ and increase with $A_{V}$. At higher $A_{V}$ and deeper within the cloud, there is enough shielding from FUV radiation for significant amounts of \CO\ to exist, with an increasing fraction of the carbon in the form of CO as opposed to C or C$^{+}$. The values of this ratio in the outskirts of the clouds are typically $\sim{1/5}$ the fiducial value for the Milky Way of $\CO/\CII\sim{1/4400}$ \citep{sta91}, which is consistent with the recent global measurements at low metallicity from \citep{acc17}. Towards the centers of the clouds, at high $A_V$, the ratio tends to reach the Milky Way value or higher as in the case of the SWBarN and SWBarS regions which show compact, bright peaks in the \CO\ and \thirteenCO\ emission. The lower \CO-to-\CII\ ratios at lower $A_{V}$ suggest that there is a reservoir of \htwo\ gas that is not traced by bright CO emission.

The low range of $A_{V}$ in the SWDarkPK region and the high \CO-to-\CII\ values seen at low $A_{V}$ (e.g., in the SWBarS region) are probably in part due to the geometry of the sources, as well as physical differences in density and radiation field (see \S\ref{subsubsection:method2}). Limitations in how we estimate $A_{V}$ likely also play a role. In order to derive a dust temperature from the dust SED we need to include the longer wavelength data, which have poorer resolution. To obtain our $A_V$ estimate we have thus applied the dust temperature fit on larger ($\sim{40\arcsec}$) scales, which smooth out any dust temperature variations on small scales, particularly those likely present towards smaller CO clouds or cores. Associated with this, the simple one dust temperature modified blackbody fit is also biased toward the higher dust temperatures that will dominate the emission. A high dust temperature results in underestimating $\tau_{160}$ and $A_{V}$. This combination of inability to resolve small structures and bias toward underestimating $A_V$ should be taken into account when interpreting Figure \ref{fig:CO_CII_OICII}.

\section{Estimating \htwo\ using \CII\ and \CO}
\label{section:estimating_h2}

At low $A_{V}$, in the outer layers of the molecular cloud, the molecular gas will be associated with \CII\ emission. As the shielding increases deeper within the cloud at high $A_{V}$, the molecular gas will be traced by CO emission. Between the layers traced by \CII\ and \CO, there will be some amount of the molecular gas traced primarily by \CI, but the amount of gas associated with this layer is uncertain. Observationally, existing data suggest that the \CI\ traces a small fraction of the total molecular gas in the SMC \citep{req16,pin17}. The theoretical work by \citet{nor16} and simulations by \citet{glo15} also suggest that most of the neutral carbon will be mixed with the CO and the amount of molecular gas traced only by \CI\ is always much less than that traced by \CII\ or CO. Assuming the amount of gas traced only by \CI\ is negligible, we can estimate the total amount of molecular gas by evaluating the \CII\ emission coming from \htwo\ gas at low $A_{V}$, and combining it with the \htwo\ traced by CO emission at high $A_{V}$: 
\begin{equation}
\label{eq:Nh2}
N_{\textnormal{\htwo}} = N_{\textnormal{\htwo,\CII}} + X_{\textnormal{CO}}\,I_{\textnormal{CO}}, 
\end{equation}
To convert the \CII\ intensity to a column of \htwo\ gas we have to estimate and remove any possible contribution to the \CII\ emission from ionized and atomic gas:
\begin{equation}
\label{eq:CII_component}
I_{\textnormal{\CII,mol}} =  I_{\textnormal{\CII}} - I_{\textnormal{\CII,ionized}} - I_{\textnormal{\CII,atomic}}, 
\end{equation}
leaving only \CII\ emission arising from molecular hydrogen. This methodology is based on many previous studies that estimate molecular gas based on \CII\ emission, with the most recent from \citet{lan14} in the Milky Way, \citet{oka15} in the LMC, and \citet{req16} in the SMC. However, those studies relied primarily on \CII, \CI, and \CO, lacking any observations that would give them an indication of the temperature and/or density of the \CII-emitting gas. Our \OI\ observations allow us to use the \OI/\CII\ line ratios to constrain the conditions of the \CII-emitting gas and better estimate the amount of molecular gas.

The integrated \CII\ line intensity ($I_{\textnormal{\CII}}$) from collisional excitation assuming optically thin emission is
\begin{equation}
\label{eq:I_CII}
I_{\textnormal{\CII}} = 2.3\times10^{-24} \left[ \frac{2e^{-91.2/T}}{1+2e^{-91.2/T}+n_{\textnormal{crit}}/n} \right]N_{\textnormal{C}^{+}},
\end{equation}
where $I_{\textnormal{\CII}}$ is in units of W m$^{-2}$ sr$^{-1}$, $T$ is the kinetic gas temperature in K, $n$ is the volume density of the collisional partner (H, \htwo, or $e^{-}$) in cm$^{-3}$, $N_{\textnormal{C}^{+}}$ is the column density of C$^{+}$ in cm$^{-2}$, and $n_{\textnormal{crit}}$ is the critical density for collisions with a given partner in cm$^{-3}$ \citep{cra85}. We assume a carbon abundance ($(\textnormal{C}/\rm{H})_{\rm{SMC}}$) that is the Milky Way abundance scaled by the metallicity of the SMC, such that $(\textnormal{C}/\rm{H})_{\rm{SMC}} = Z'_{\textnormal{SMC}}\ (\rm{C}/\rm{H})_{\rm{MW}}=2.8\times10^{-5}$, taking $Z'_{\textnormal{SMC}}=0.2$ and $(\textnormal{C}/\rm{H})_{\rm{MW}}=1.4\times10^{-4}$ \citep{sof97}, which agrees with the available measurements in the SMC \citep{kur99,tch15}. 

\subsection{$I_{\text{\CII}}$ Contribution from Ionized Gas}
\label{subsection:CII_from_ionized}

We estimated a very low contribution to the total \CII\ emission from ionized gas based on our \NII\ 205 \micron\ measurements (\S\ref{subsection:NII}). Here we also estimate the possible contribution using the narrow-band \ha\ data for the SMC to estimate the electron volume density ($n_{\rm{e}}$) and column density ($N_{\rm{e}}$), and assuming an ionized gas temperature to calculate $I_{\textnormal{\CII}}$ using Equation \ref{eq:I_CII}. The \ha\ observations provide the emission measure (EM), which can be used to estimate $n_{\rm{e}}$:
\begin{equation}
\label{eq:EM}
\text{EM} = \int n_{\rm{e}}^{2} dl = 2.75 \left( \frac{T}{10^{4} \text{K}} \right)^{0.9} \left( \frac{I_{\text{\ha}}}{1 \text{R}} \right) \text{pc cm}^{-6},
\end{equation}
which we then solve for $n_{\rm{e}}$ and get
\begin{equation}
\label{eq:ne_fromHa}
n_{\rm{e}} = \left( \frac{2.75}{l} \left( \frac{T}{10^{4} \text{K}} \right)^{0.9} \left( \frac{I_{\text{\ha}}}{1 \text{R}} \right) \right)^{1/2} \text{cm}^{-3}.
\end{equation}
where $I_{\text{\ha}}$ is the \ha\ intensity (1 R = $10^{6}~(4\pi)^{-1}$ photons cm$^{-2}$ s$^{-1}$ sr$^{-1}$ $= 2.409 \times 10^{-7}$ ergs s$^{-1}$ cm$^{-2}$ sr$^{-1}$ at $\lambda=6563$ \AA), $T$ is the electron temperature, and $l$ is the length over which $n_{e}^{2}$ is integrated \citep{rey91}. This procedure is uncertain and known to be biased toward overestimating $n_e$ since the highest densities dominate the EM, but it provides an independent check on the conclusions in \S\ref{subsection:NII}, and it also gives us the ability to correct the \CII\ maps at higher resolution.  

We assume $T_{\rm{e}}=8000$ K, appropriate for Galactic \hii\ regions and the warm ionized medium (WIM). While the temperature of \hii\ regions is a strong function of metallicity and is likely higher in the SMC (\citealt{kur98} measures $T_{\rm{e}}\approx12000$ K in N66) \citet{gol12} calculates the \CII\ critical density for collisions with electrons only up to 8000 K, and reports $n_{\textnormal{crit}, \rm{e}}=44$ cm$^{-3}$. The \CII\ emission from ionized gas is not expected to dominate the total \CII\ emission (see Section \ref{subsection:NII}), so the effect of assuming a lower temperature will be minimal. The most uncertain quantities are the assumed lengths: the distance $l$ over which $n_{\rm{e}}^{2}$ is integrated, and the distance used to convert the volume density to a column density. For simplicity, we assume both to be 20 pc, which is approximately the diameter of the largest \hii\ region (N22) found within the survey regions and likely to be the size scale that produces the majority of the measured integrated $n_{\rm{e}}^{2}$. The $n_{\rm{e}}$ we estimate from the \ha\ emission is $\sim{20-25}$ cm$^{-3}$, which is consistent with the upper limit on $n_{\rm{e}}$ determined from the \NII\ 122 \micron/205 \micron\ values (see Section \ref{subsection:NII}).

The overall change after correcting for the contribution from ionized gas is small. This is consistent with the finding of \citet{cro17} that the neutral fraction of gas traced by \CII\ increases with decreasing metallicity in the Beyond the Peak sub-sample of KINGFISH galaxies. We list the average percentages of \CII\ intensity estimated to originate from the ionized gas in Table \ref{table:h2_estimates}. The amount of emission from ionized gas can be as high as $\sim{50\%}$, but this is only found within the \hii\ region where the \CII\ surface brightness is low. The effect of correcting for the ionized gas mainly affects the structure of \CII\ emission in and near the \hii\ regions where we expect little to no molecular gas.

\subsection{$I_{\text{\CII}}$ Contribution from Atomic Gas}
\label{subsection:CII_from_HI}

We estimate the amount of \CII\ emission from \hi\ by using the typical temperature and density for the conditions of the gas and assume that the column density of ionized carbon scales with the carbon abundance such that $N_{\textnormal{C}^{+}}=(\textnormal{C}^{+}/\rm{H})_{\rm{SMC}}\Nhi$. The \CII\ emission from atomic hydrogen gas arises from a combination of the warm neutral medium (WNM), with typical temperatures of $6000-12000$ K, and the denser cold neutral medium (CNM), which has temperatures of $50-120$ K. Some or all of the CNM component may be directly associated with the molecular gas as a shielding layer surrounding and mixed with the outer part of the molecular cloud in the photodissociation region. Given the differences in conditions, we make separate estimates for the \CII\ emission coming from the WNM and CNM and remove both contributions from $I_{\CII}$.

To estimate the possible \hi\ associated with the \CII\ emission, we want to select only the \hi\ with velocities in the observed range of the \CII\ emission. However, the spectral resolution of the PACS spectrometer is $\sim{240}$ km s$^{-1}$, which encompasses almost all of the \hi\ emission and is a much larger range of velocities than the $\sim{15-50}$ km s$^{-1}$ widths of the \CO\ line observations. We use the velocity-resolved observations of \CII\ from the SOFIA GREAT receiver (R. Herrera-Camus et al. 2017, in preparation) as guidance to estimate the relationship between the velocity profile of the \CII\ emission and the \CO\ throughout all our regions. 

\begin{figure}[t]
\epsscale{1.1}
\plotone{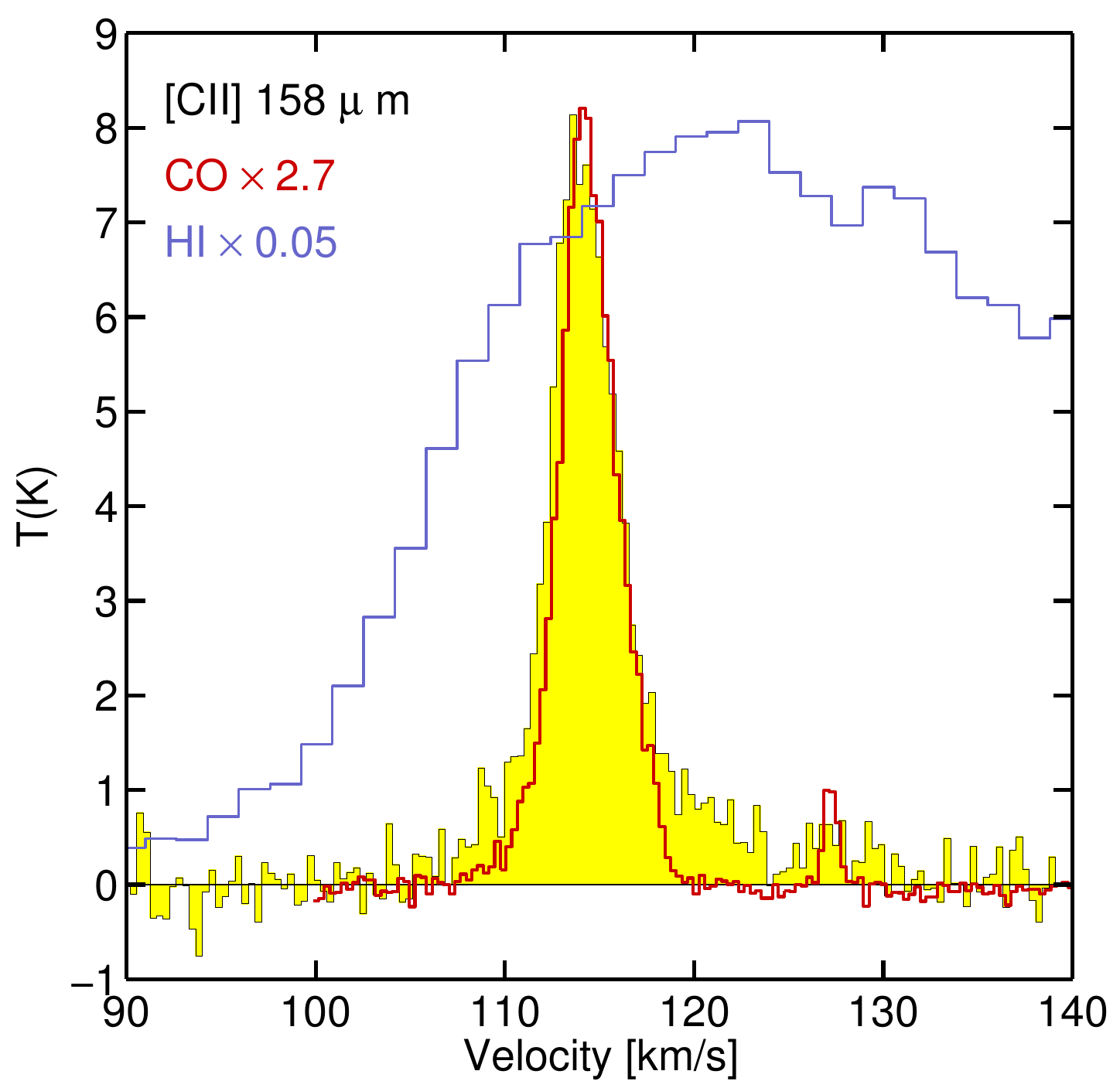}
\caption{Spectrum of \CII\ from one pointing observed with SOFIA GREAT (R. Herrera-Camus et al. 2017, in preparation; region R8) with the yellow filled histogram showing the data in 0.3 km s$^{-1}$ bins with scaled spectra of the ALMA \CO\ $(2-1)$ in red and \hi\ in purple for the same area in the SWBarN region with velocity in the LSRK reference frame. This illustrates that the velocity profile of the \CII\ emission is most similar to the \CO\ and not the \hi, suggesting that a significant fraction of the emission comes from molecular gas. It also shows that the \CII\ emission profile can have wider wings at low or no \CO\ intensity, in this case by $5-10$ \kms, which makes the total extent of the \CII\ profile wider (by at most $\sim{50\%}$) than the \CO\ line profile and has wider .
\label{fig:SOFIA_spec}}
\end{figure}

\begin{figure*}[t]
\epsscale{1.0}
\plotone{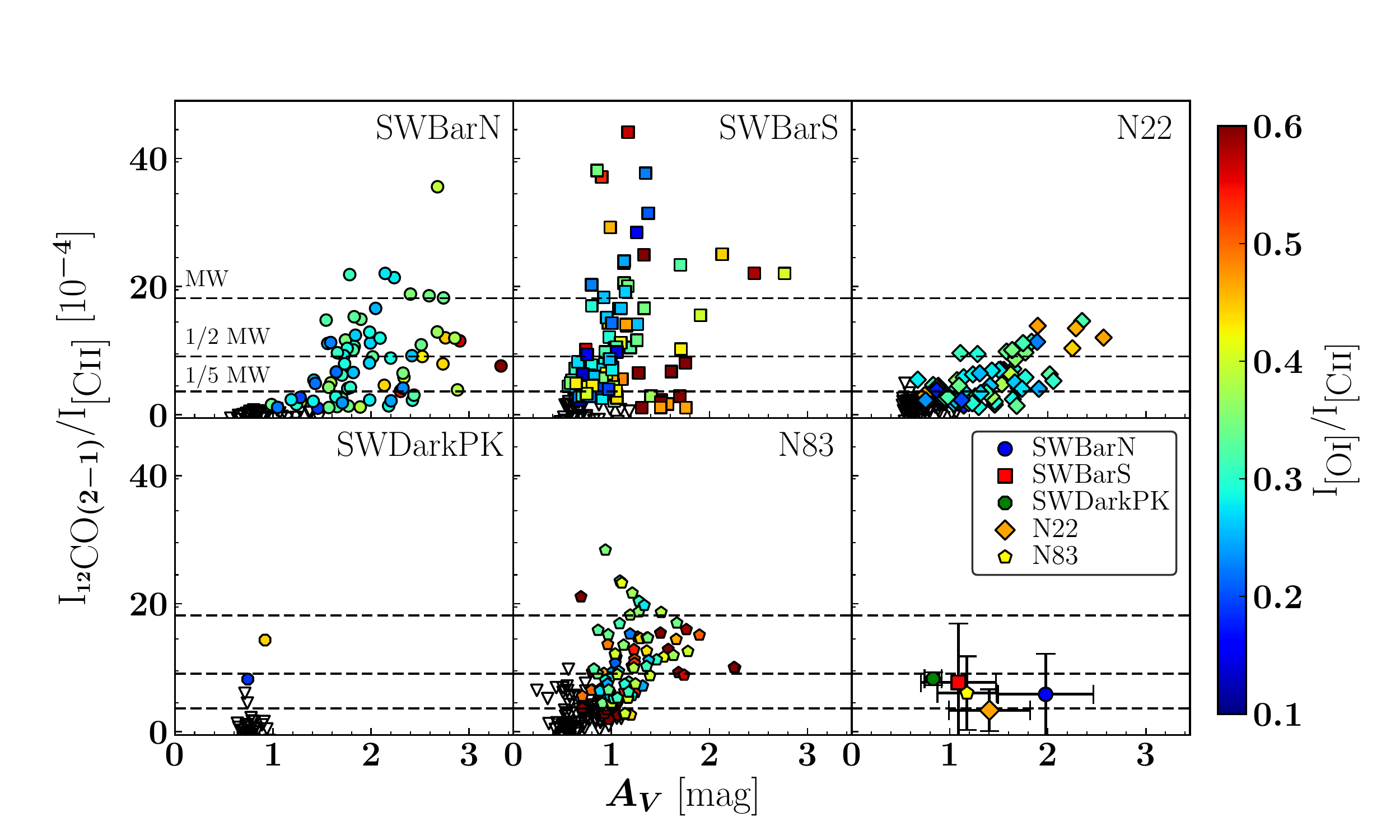}
\caption{The ratio of the integrated intensity of ALMA ACA+TP \CO\ $(2-1)$ to that of \CII\ as a function of $A_{V}$ for the HS$^{3}$ regions. The colored symbols show independent measurements detected at $>3\sigma$ in both \CO\ and \CII\; the downward pointing triangles indicate upper limits (where $I_{\textnormal{CO}}<3\sigma$). The over plotted dashed lines show scalings of the canonical \CII\ to \CO\ $(1-0)$ ratio translated to $(2-1)$ assuming thermalized emission for the Milky Way of $\sim{1/4400}$ \citep{sta91}. The bottom left panel shows the mean \CO\ $(2-1)$/\CII\ ratios for each region, which were calculated to include the upper limits (``left-censored'' data) using the cenfit routine in the R package NADA \citep{lee17,hel05}. We see a trend of \CO\ $(2-1)$/\CII\ starting at values much lower than the Milky Way value at low $A_{V}$ and increasing \CO\ $(2-1)$/\CII\ towards higher $A_{V}$. The \OI/\CII\ ratios also show an increase towards higher $A_{V}$, which is consistent with increasing density producing more \OI\ emission and/or less \CII.
\label{fig:CO_CII_OICII}}
\vspace{0.2cm}
\end{figure*}

Figure \ref{fig:SOFIA_spec} shows an example \CII\ spectrum from a pointing in the SWBarN region with the ALMA ACA \CO\ $(2-1)$ and \hi\ line profiles for the same region. The \CII\ line is clearly resolved, and shows a profile similar to the \CO\ and much narrower than the \hi, but the full extent of \CII\ profile is $\sim{50}\%$ wider than the full extent of the \CO\ line profile due to more extended wings with there is low or no \CO\ emission. \citet{req16} presents new of velocity-resolved \CII\ GREAT observations of the SMC star-forming regions N66, N25/N26, and N88 that also show similarity between \CII\ and \CO\ line profiles and find the \CII\ emission to be at most $50\%$ wider than \CO, and significantly narrower than that of \hi. Similarly, comparison of the dynamics of \CII, \hi, and CO on larger scales show that the \CII\ line widths agree better with the CO than \hi\ \citep{deb16}. The width of the \hi\ line profile is due to emission from multiple kinematic components that are likely spatially separate, and the narrow profiles of the \CO\ and \CII\ appear to originate from only one of those kinematic components. To select the \hi\ associated with the strong \CII\ emission and the molecular cloud, we created integrated intensity maps of \hi\ for each of the ALMA regions using a velocity range that is centered on the \CO\ line velocity, but $\pm25\%$ wider than the range of velocities with observed \CO\ throughout the region (not for individual lines-of-sight) to account for the possibility of $\sim{50\%}$ wider extent of the \CII\ line profile (mostly found in the wings of the line). We use these \hi\ maps to estimate the amount of \CII\ emission associated with neutral atomic gas. 

To estimate the \CII\ emission attributable to atomic gas we assume a single temperature and density for each of the WNM and CNM. The \hi\ emission will be dominated by the WNM component, which we assume to have a temperature of $\approx{8000}$ K \citep{wol03} and a typical density range of $0.1-1$ cm$^{-3}$ \citep{vel12}. We set $n_{\rm{WNM}}=1$ cm$^{-3}$ for our estimate to ensure we do not underestimate the \CII\ contribution from the WNM. The C$^{+}$ excitation critical density for $T\approx{8000}$ K of the WNM from hydrogen atoms is $n_{\text{crit}}(H^{0})=1600$ cm$^{-3}$  \citep{gol12}. For the CNM, we assume $n_{\rm{CNM}}=100$ cm$^{-3}$ and $T=40$ K based on the values found from \hi\ absorption line study by \citet{dic00}. For collisions with electrons, we take $n_{\textnormal{crit}, e}=6$ cm$^{-3}$. We calculate the critical density for collisions with hydrogen atoms for the CNM temperature using $n_{\text{crit}} = A_{ul}/R_{ul}$, where $A_{ul}$ is the spontaneous emission rate coefficient and $R_{ul}$ the collisional de-excitation rate coefficient. Using $A_{ul}=2.36\times10^{-6}$ s$^{-1}$ and the fit to $R_{ul}$ for 20 K $\leq T^{\text{kin}} \leq 2000$ K from \citet{bar05}, we find $n_{\text{crit}}(\rm{H}^{0})\approx3000$ cm$^{-3}$. 

We calculate $I_{\CII}$ using Equation \ref{eq:I_CII} for collisions with both hydrogen atoms and electrons in the WNM and CNM and collisions with \htwo\ for the CNM and add them to produce our estimate. We take an ionization fraction in neutral gas to be $n_{e^{-}}/n_{\rm{H}}=4.3\times10^{-4}$ \citep{dra11}. In order to calculate this we need the fraction of WNM and CNM in our lines of sight, which is unknown. We assume that $50\%$ of \Nhi\ associated with the CO emission is associated with the CNM and the other $50\%$ with the WNM when calculating $I_{\CII}$. \citet{dic00} find the fraction of cool \hi\ to be $\sim{15\%}$ of the total \hi\ in the SMC along a few lines of sight, however locally it may be much higher. Assuming a higher fraction of CNM will generate a higher estimate of the associated \CII\ emission, decreasing the likelihood of underestimating the \CII\ emission from \hi. Table \ref{table:h2_estimates} shows the average amount of total \CII\ emission for each region. We find that \hi\ typically only accounts for $\lesssim{5\%}$. When we apply to statistical correction for optically thick \hi\ from \citet{sta99} to \Nhi\ used to estimate the \CII\ emission from \hi, the fraction of the total \CII\ emission coming from \hi\ only slightly increases (e.g., the average increases from $2\%$ to $3\%$ for the SWBarN region). The SWDarkPK region has the highest fraction of \CII\ emission coming from \hi, likely because it is not actively forming stars and is more atomic-dominated (as also suggested by the lower CO emission indicating less molecular gas at $A_V>1$ than the other regions). These low fractions of \CII\ emission from atomic gas are consistent with the theoretical study by \citet{nor16}, who find that most of the C$^{+}$ column would be associated with \htwo\ gas.

\subsection{Converting $I_{\text{\CII}}$ to \Nhtwo}
\label{subsection:estimating_h2_cii}

After removing the contributions from ionized and neutral atomic gas, we assume the remainder of the \CII\ emission originates from molecular gas. We estimate the amount of \htwo\ gas from the remaining \CII\ emission in two ways:
\begin{enumerate}
\item{Assuming a fixed $T$ and $n$ representative of PDR regions and converting $I_{\CII}$ to \Nhtwo using Equation \ref{eq:I_CII}.}
\item{Using PDR models tuned to the SMC conditions (see Appendix \ref{appendix:smc_pdr_models}) to estimate $n$ and the FUV radiation field intensity from the observed combination of \CII, \OI, and FIR continuum, determining \Nhtwo\ from the cooling curve.}
\end{enumerate}
We explain both methods in the following sections. 

\subsubsection{Method 1: Fixed $T,~n$}
\label{subsection:method1}

Using Equation \ref{eq:I_CII}, we can solve for the \htwo\ column density given the \CII\ intensity:
\begin{equation}
\Nhtwo = \frac{4.35\times10^{23}}{(\textnormal{C}/\rm{H})_{\textnormal{SMC}}} \left[ \frac{1+2e^{-91.2/T}+n_{\textnormal{crit}}(\htwo)/n}{2e^{-91.2/T}} \right]I_{\textnormal{\CII,\htwo}}.
\end{equation}
To estimate \Nhtwo, we need only assume a temperature and volume density and calculate the appropriate critical density for collisions with \htwo\ using the latest collisional de-excitation rate fits by \citet{wie14} as a function of temperature, which includes \htwo\ spin effects in LTE:
\begin{equation}
n_{\text{crit}}=\frac{A_{ul}}{R_{ul}}= \frac{2.36\times10^{4}}{4.55+1.6e^{(-100.0~\text{K}/T)}}
\end{equation}
Our assumption is that all of the \CII\ emission associated with \htwo\ comes from the PDRs where the gas is warm and moderately dense. We can then make the simplifying assumption that this region has a single average temperature of $T=90$ K and density of $n=4000$ cm$^{-3}$ (consistent with the PDR modeling results in \S \ref{subsubsection:method2}). These are chosen as similar to the excitation temperature and critical density of the transition, and thus at about the point in which the gas emits most efficiently. For reference, the exact critical density for collisions with \htwo\ at $T=90$\,K is $n_{\text{crit}}(\htwo)=4648$ cm$^{-3}$. This choice of temperature and density results in a conservative estimate of the amount of molecular gas associated with the \CII\ emission. Assuming a lower temperature or density would decrease the amount of \CII\ emission per \htwo\ molecule and increase the amount of molecular gas needed to explain the observed emission, while increasing the temperature or density will not change our results significantly. The benefit of this methodology is its simplicity and straightforward calculation. The drawback is that it over simplifies the conditions of the \CII\ emitting gas by assuming only one constant temperature and volume density throughout each region and across all regions.

\begin{figure*}[t]
\plotone{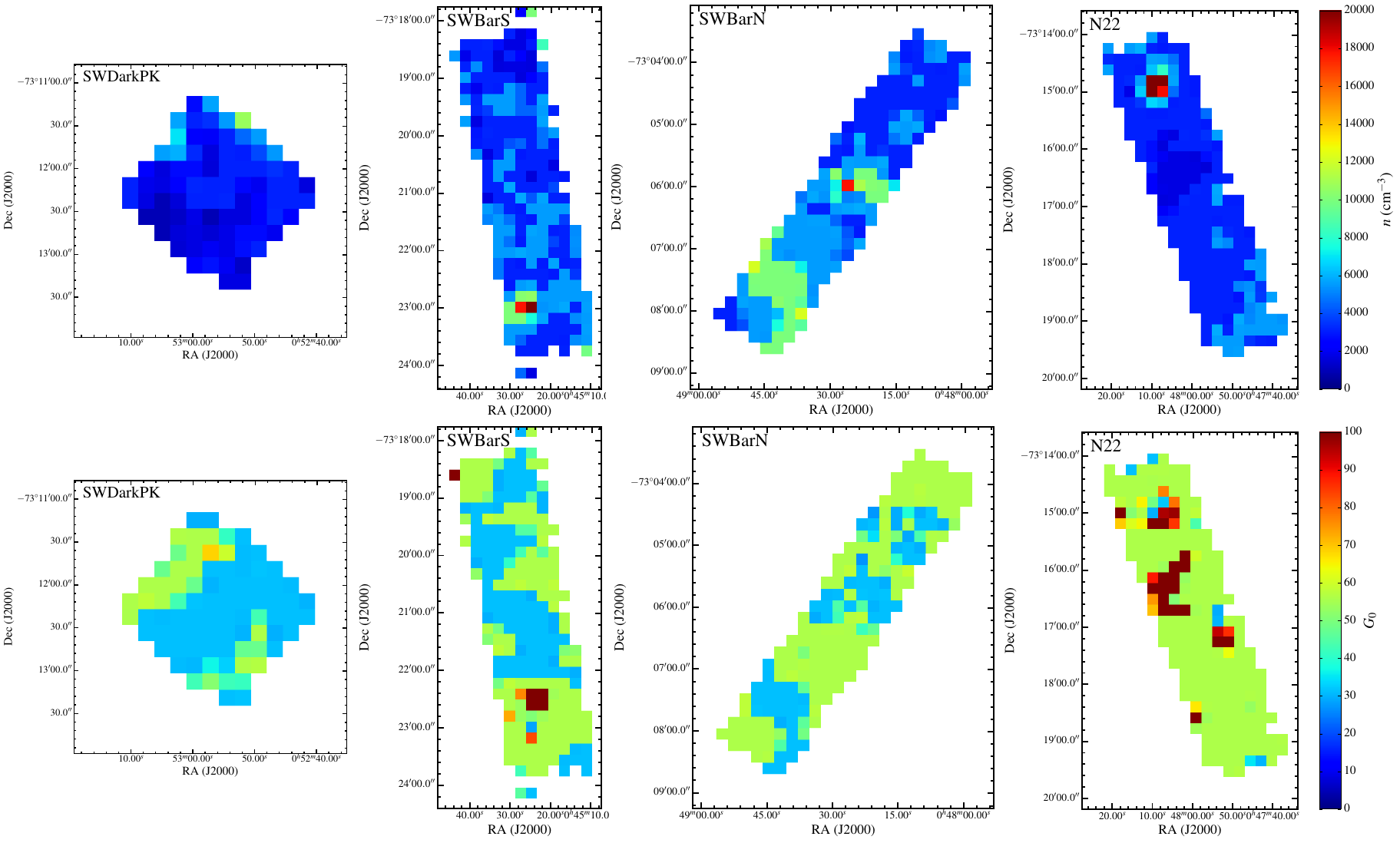}
\caption{Density and radiation field for each of the regions with HS$^{3}$ and ALMA data, obtained from physical PDR modeling developed for the conditions in the SMC (abundances, cosmic ray rates, dust to gas ratio, etc; see see Appendix \ref{appendix:smc_pdr_models}). The top row shows volume density ($n$). The bottom row shows FUV radiation field strength ($G_{0}$). 
\label{fig:PDRT}}
\end{figure*}

\subsubsection{Method 2: PDR Model}
\label{subsubsection:method2}

We can constrain the physical conditions of the gas in the PDRs by modeling the \CII, \OI, and total FIR emission from the $Spitzer$ and $Herschel$ imaging. We use new PDR models generated using the elemental abundances and grain properties for SMC conditions that are presented in Appendix \ref{appendix:smc_pdr_models}. The models solve the equilibrium chemistry, thermal balance, and radiation transfer through a PDR layer characterized by a constant H nucleus volume density ($n$ in units of cm$^{-3}$) and incident far-ultraviolet (FUV; 6 eV $<$ E $<$ 13.6 eV) radiation field ($G_{0}$ in units of the local Galactic interstellar FUV field found by Habing, $1.6 \times 10^{-3}$ erg cm$^{-2}$ s$^{-1}$). The models are based on \citet{kau06} but are updated for gas and grain surface chemistry as described in \citet{wol10} and \citet{hol12}. We also modify the grain properties and abundances as appropriate for the SMC (see Appendix \ref{appendix:smc_pdr_models}) which makes it tailored to modeling the emission from the HS$^{3}$ regions.

We use the \CII, \OI, and total-infrared (TIR) intensity images as inputs to determine the nearest volume density and FUV radiation field strength in the model grids using a version of the PDR Toolbox \citep{pou08} that is updated with the SMC models. While the PDR Toolbox and models call for far-infrared (FIR) intensity as the input, the term FIR is meant to represent all of the FUV and optical emission absorbed by dust and re-radiated in the infrared, which makes the current definition of TIR the most appropriate to compare to the models. We convolve the \CII\ and \OI\ images to the resolution of the TIR. We run the PDR Toolbox for the intensities at each matched pixel in the images and produce maps of $n$ and $G_{0}$. The images are then sampled with $\sim{1}$ pixel per beam, matching the \CII\ beam-sampled images. While the limiting resolution of the TIR map is lower ($\sim{18\arcsec}$) than the \CII\ ($\sim{12\arcsec}$), it is not significantly lower and we expect $n$ and $G_{0}$ to vary smoothly. The initial images of $n$ and $G_{0}$ show artifacts from the FIR image (due mostly to the strong beam patterns from the point sources). We take one final step and mask out regions in the images with unphysical values of $n$ or $G_{0}$ that are due to the artifacts and use a linear interpolation to replace the masked values. The results of model fits are shown in Figure \ref{fig:PDRT}. The PDR models suggest that the typical volume densities are $n\sim{10^{3}-10^{4}}$ cm$^{3}$, which is consistent with most of the gas being molecular in order to reproduce the observed far-infrared emission.

Using the maps of $n$ and $G_{0}$, we convert the \CII\ emission to \Nhtwo. We use \CII\ cooling rates as a function of $n$ that are adapted from \citet{wol03} to account for the conditions of the SMC, which we show in Figure \ref{fig:CII_cooling} (Wolfire et al. 2017, in preparation). We use the same abundances as for the PDR model. The \CII\ rates are calculated for a fiducial FUV radiation field strength of $G_{0} = 5.5$, which is approximately the average in diffuse gas (outside of the regions we mapped) in the SMC Southwest Bar based on the maps from \citet{san10} and the cooling rates were primarily intended for use in diffuse gas. While the \CII\ cooling rate is only determined for a single $G_{0}$, the effect of different $G_{0}$ is to change the photoelectric heating rate, which to first order is proportional to $G_{0}$ (\citealt{bak94}, equation 42). If \CII\ dominates the cooling, then the \CII\ line intensity is proportional to $G_{0}$. The appropriate \CII\ cooling rate per \htwo\ molecule for each pixel is determined by taking the \CII\ cooling rate appropriate for the specific value of $n$ and then scaling it linearly by the ratio of estimated $G_{0}$ to the fiducial $G_{0}$ for diffuse gas. We then use this local \CII\ cooling rate to calculate a \htwo\ column density and produce the final map. The benefit of this method is that it can account for the variations in density and radiation field in the \CII-emitting gas throughout the regions. The weakness is that the gas conditions are estimated using ``single component'' models based on plane-parallel PDRs that do not account for complex geometries, or for distributions of $n$ or $G_0$ within a spatial resolution element. This is alleviated by the fact that our physical resolution is $4$~pc, thus we expect the conditions in a beam to be relatively uniform and simple ``single component'' models to be more applicable than they would be if applied to data on much larger scales. 

\input{table5.tex}

\subsection{Converting $I_{\rm CO}$ to \Nhtwo}
\label{subsection:h2_from_CO}

Once enough shielding from dissociating radiation is built up at high $A_{V}$, carbon will mostly exist in the form of CO. The exact amount of \htwo\ traced by the \CO\ emission (and not \CII) cannot be easily calculated theoretically and relies on empirical calibrations of the conversion factor from \CO\ intensity to \Nhtwo, $X_{\text{CO}}$, typically calibrated in terms of \CO\ $(1-0)$ emission. The exact definition of $X_{\text{CO}}$ varies depending on the study, where it can refer to the total amount of \htwo\ gas is associated with the CO emission including the \htwo\ associated with \CII\ or \CI\ (e.g., unresolved CO observations) or it can refer to the molecular gas only associated with gas where the dominant form of carbon is CO. For this work, we use the latter as our definition since our observations resolve the CO emission and we are separately accounting for the \htwo\ gas where carbon is mostly \CII. To convert from \CO\ $(2-1)$ to the equivalent \CO\ $(1-0)$ integrated intensity we assume thermalized emission, which is supported by SEST observations in this region of the SMC that find \CO\ $(1-0)$/\CO\ $(2-1)$ $\sim{1}$ \citep{rub93a}. This is equivalent to assuming constant integrated intensity in K\,km\,s$^{-1}$ units, or to divide the 2--1 emission in units of W\,m$^{-2}$\,Hz$^{-1}$ by $J_u^{3}=8$. Numerical simulations of low metallicity molecular clouds show that the conversion factor reaches Galactic, high-metallicity values in the CO-bright regions \citep{she11,szu16}, suggesting that regardless of metallicity, CO traces a similar amount of \htwo\ at high $A_V$. Given the simulation results and our estimates of $X_{\text{CO}}$ (see Section \ref{subsection:XCO_13CO}), we use a Galactic CO-to-\htwo\ conversion factor $X_{\text{CO}}=2\times10^{20}$ cm$^{-2}$ (K km$^{-1}$ s$^{-1})^{-1}$ \citep{bol13}, to estimate the amount of molecular gas present in regions at high $A_V$ where most of the carbon is in the form of CO. This is consistent with an underlying picture where most of the difference in the conversion factor observed in low metallicity systems is due to the shrinking of the high $A_V$ CO-emitting cores and the growth of the outer layer of \htwo\ coextensive with C$^+$ \citep[e.g.,][]{wol10,bol13}.

\subsection{\Nhtwo\ Estimates}
\label{subsection:total_h2}

Overall, the two different methods to estimate \Nhtwo\ based on \CII\ emission produce similar results. Table \ref{table:h2_estimates} shows that average column densities and total \htwo\ masses for both methods are different by at most a factor of $\sim{1.5}$. The total mass estimates from Method 1, assuming $T=90$ K and $n=4000$ cm$^{-3}$, are higher than the estimates from Method 2, using the SMC PDR models and Wolfire et al. (2017, in preparation) \CII\ cooling rates, except for the SWDarkPK region. The fixed temperature and density produce higher estimates because they fail to account for the presence of higher density regions that have a higher emission of \CII\ per unit \htwo\ mass, as we show in the example in Figure \ref{fig:H2_compare}. The opposite is true for the SWDarkPK where the densities from the PDR modeling are somewhat lower than those in other regions, and the assumed $n=4000$ cm$^{-3}$ in Method 1 appears to be too high. This is consistent with the observed lower level of CO emission in that very quiescent region, which is suggestive of less dense molecular gas. 

The Method 1 results produce structure that is similar to the \CII\ emission, with the peak in the \Nhtwo\ occurring at the peaks in \CII\ emission, which are nearly coincident with the peaks in the CO emission. When the variations in density and radiation field strength are accounted for in the estimate in Method 2, then the peaks in the \Nhtwo\ maps occur around the peak in the CO emission (see Figure \ref{fig:H2_compare}), consistent with our understanding that \CII\ emission should trace the \htwo\ gas in the photodissociation regions surrounding the dense molecular gas traced by bright CO.  We prefer the \Nhtwo\ estimates from Method 2 because it incorporates additional information, and use them for the remaining analysis and discussion. 

\begin{figure}[t]
\epsscale{1.2}
\plotone{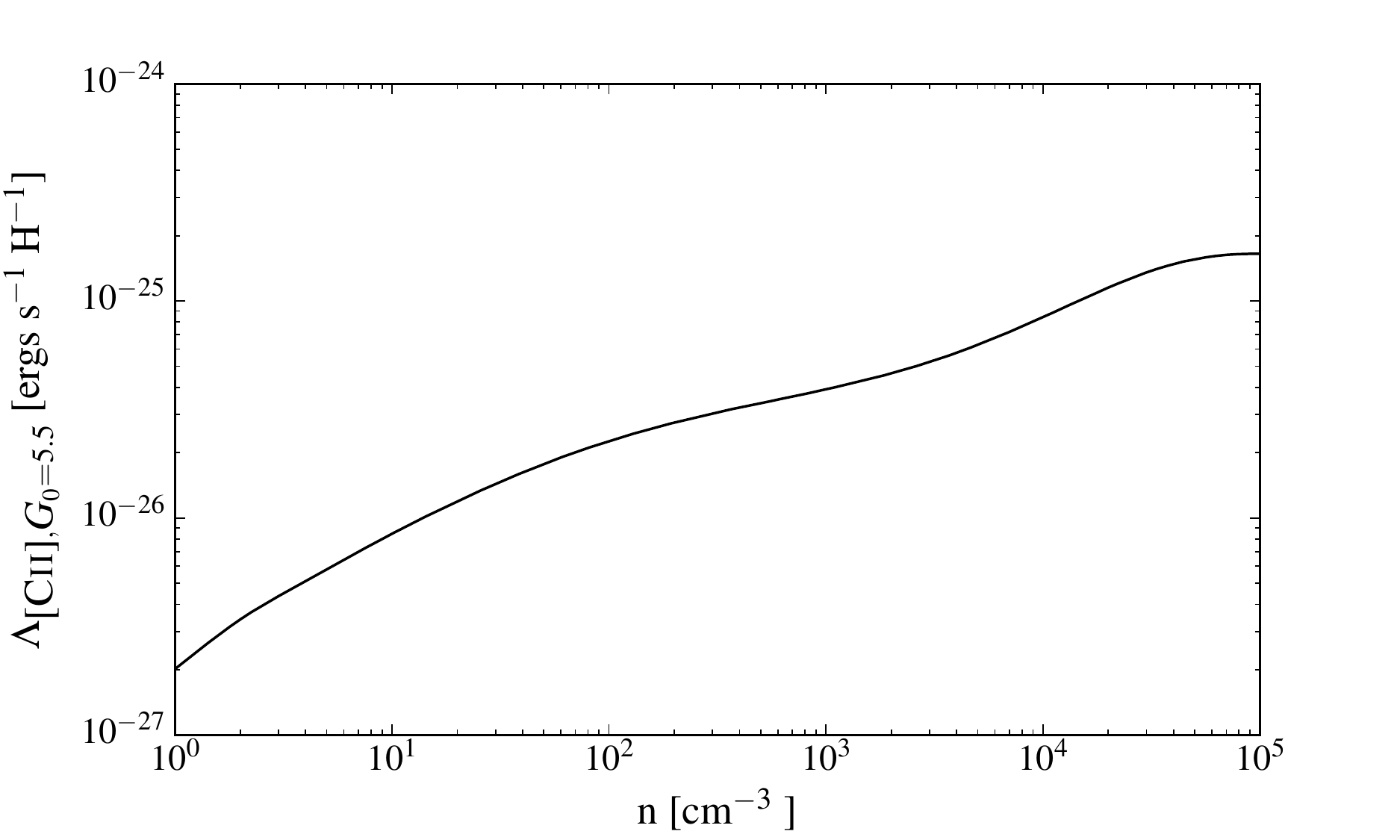}
\caption{\CII\ cooling rate ($\Lambda_{\textnormal{\CII}}$) as a function of density for the conditions in the SMC from Wolfire et al. (2017, in preparation) used to convert the \CII\ emission to a column density of \htwo. The cooling curve is calculated assuming $G_{0}=5.5$.
\label{fig:CII_cooling}}
\end{figure}

Figure \ref{fig:H2} shows the \htwo\ maps derived from \CII\ alone, from CO alone, and the total corresponding to their sum for each region. The \htwo\ estimated from \CII\ is more extended than that derived from CO and wraps around it, which agrees with the idea that the \CII\ should be primarily tracing the \htwo\ in the low $A_{V}$ portion of the molecular cloud. The CO emission traces between $\sim{5\%-60\%}$ of the total molecular gas in this methodology, with a typical percentage of only $20\%$. Note, however, that there are no \htwo\ column density peaks without CO emission. 

\subsubsection{Estimating Uncertainties}
\label{subsubsection:h2_unc}

The possible sources of systematic uncertainty in our \htwo\ estimate using \CII\ include the PACS spectrometer calibration, the uncertainty on the carbon abundance measurement, the possibility of minor \OI\ absorption and \CII\ optical depth, and the uncertainty in the assumptions in the low metallicity version of the PDR model. In Table \ref{table:h2_estimates_unc} we summarize the amount of uncertainty associated with each factor, which we explain how each factor is estimated in detail in Appendix \ref{appendix:uncertainty}. Each of the main factors tend to contribute $\leq{30\%}$ in uncertainty: $\sim\pm30\%$ from the PACS spectrometer calibration, $\sim\pm30\%$ from the carbon abundance measurements, and $\sim\pm20\%$ from the combination of the effects of \OI\ absorption and \CII\ optical depth. Adding the results in quadrature produces a total uncertainty estimate of $\sim\pm50\%$. Although formally adding the uncertainties in quadrature is not necessarily correct, this gives us a reasonable estimate of the possible effect of the combined systematics. This uncertainty neglects the factor of $\sim{2}$ difference in the \htwo\ masses measured using the higher metallicity PDR models from \citet{kau06}. The difference in the PDR models is driven by the different assumptions about dust grains compared to the Milky Way. One factor is the PAH abundance, which the SMC PDR models assume is $1/7.7$ times the Milky Way value based on the results from modeling of the mid-infrared spectra PAH band ratios using the \citet{dra07} models in the SMC from \citet{san10}. However, there is a large amount of scatter in the estimated PAH abundance in the SMC ranging from $1/2$ to $1/10$ times the Milky Way value and the regions we mapped in the Southwest Bar show the highest PAH abundances. A higher PAH abundance in our regions would cause the estimated \htwo\ masses to be higher and closer to the ones produced using the higher metallicity PDR models. In addition to the PAH abundance, there are a number of other assumptions used for the SMC PDR models that are uncertain and the high metallicity PDR model results that are a factor of $\sim{2}$ higher likely offer an upper limit on the \htwo\ mass estimates. Taking this into consideration, we choose to adopt an uncertainty of a factor of $\sim{2}$ for the mass of \htwo. 

\begin{figure}[t]
\epsscale{1.2}
\plotone{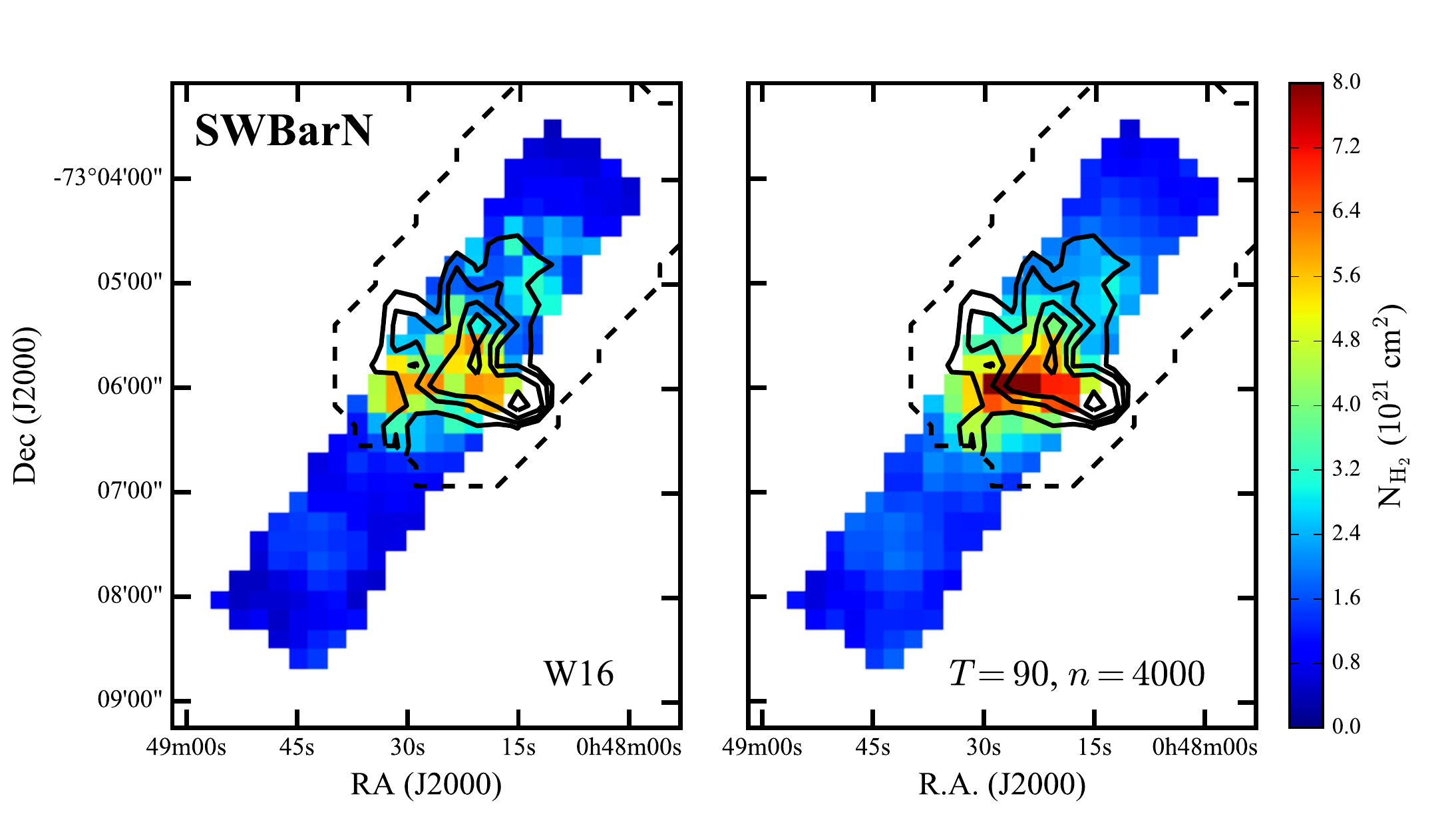}
\caption{Images of the estimate of \Nhtwo\ from the \CII\ emission for the SWBarN region using Method 2 using Wolfire et al. (2017, in preparation) models (left) and Method 1 with a fixed $T=90$ K and $n=4000$ cm$^{3}$ (right), shown at the same scale. The black contours show the estimate of \Nhtwo\ from \CO\ (using $X_{CO}=2\times10^{20}$) at levels of 0.4, 0.8, 1.4, and $2.4\times{10^{21}}$ cm$^{-2}$ and the black dashed line shows the coverage of the \CO\ map. Both \Nhtwo\ estimates from \CII\ produce similar average estimates of \Nhtwo, but the structures are different. When taking into account the local $T$ and $n$ using the Wolfire et al. (2017, in preparation) results, there is less \htwo\ needed to explain the \CII\ emission at the center of the cloud (the peak of the \CO\ emission). The estimated \Nhtwo\ peaks away from the \CO\ peak, which is consistent with the \CII\ emission tracing the CO-faint \htwo\ in the photodissociation region, suggesting that the \htwo\ estimates using the new models are more accurate.
\label{fig:H2_compare}}
\end{figure}

\begin{figure*}[t]
\plotone{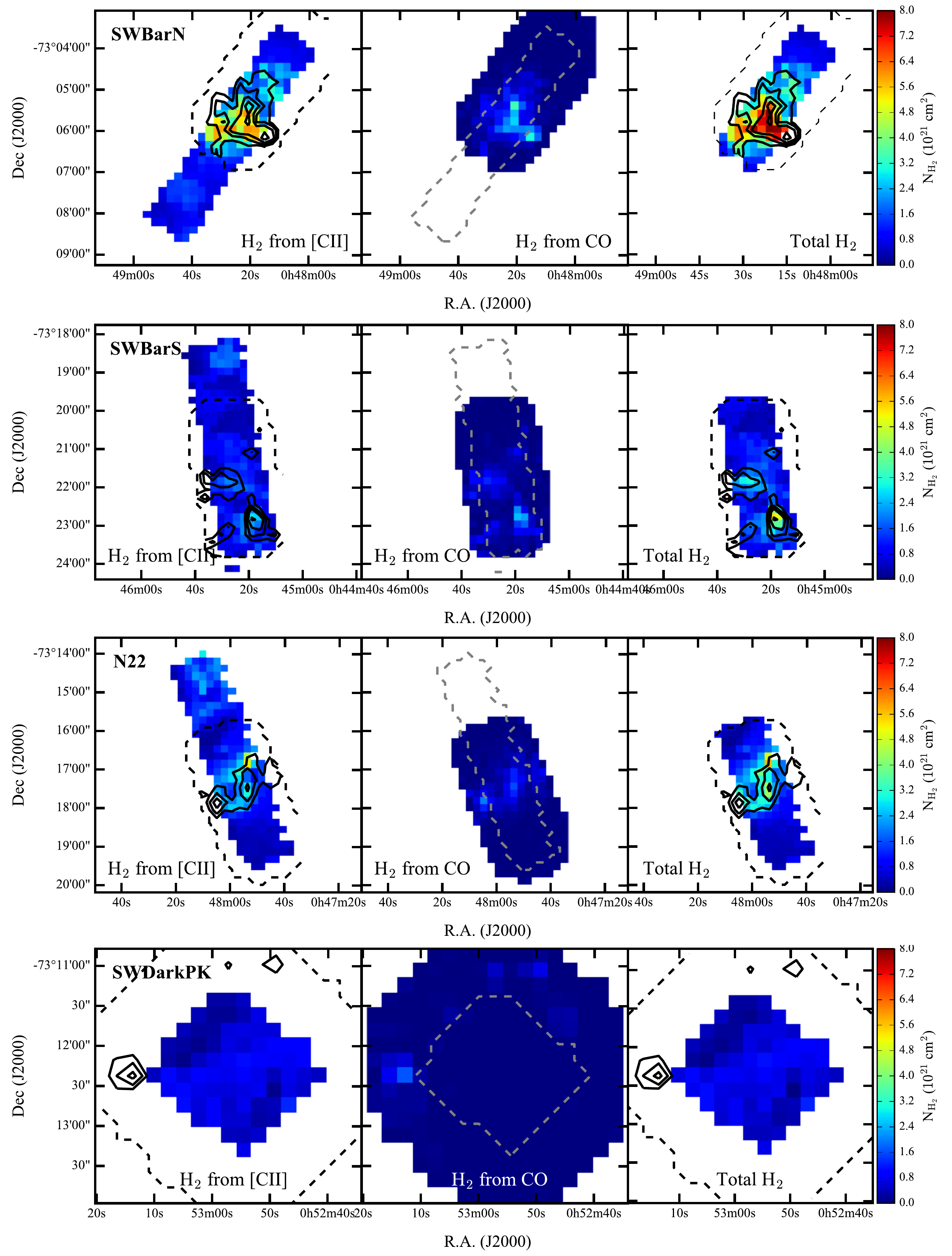}
\caption{Images of \Nhtwo\ for each of the regions with $Herschel$ and ALMA data. The images on the left shows the estimate of \Nhtwo\ from the \CII\ emission using Method 2 (W16 $n,~G_{0}$) described in Section \ref{subsubsection:method2}. The middle images shows \Nhtwo\ traced by \CO, with the ALMA $(2-1)$ converted to $(1-0)$ assuming thermalized emission, and assuming a Galactic $X_{CO}$ of $2\times10^{20}$ cm$^{-2}$ (K km s$^{-1}$)$^{-1}$. The images on the right show total \Nhtwo\ found by combining the estimate from \CII\ with the estimate from \CO. The black contours show the estimate of \Nhtwo\ from \CO\ at levels of 0.4, 0.8, 1.4, and $2.4\times{10^{21}}$ cm$^{-2}$ and the black dashed line shows the coverage of the \CO\ map. The grey dashed line shows the HS$^{3}$ (\CII\ and \OI) coverage. All maps are shown at the same color scale.
\label{fig:H2}}
\end{figure*}

\section{Discussion}
\label{section:discussion}

We have presented the results of our new far-IR emission line data, focusing primarily on the \CII\ 158 \micron\ and \OI\ 63 \micron\ lines from $Herschel$ PACS observations, and ALMA ACA CO $(2-1)$ data for a sample of star-forming regions in the SMC. Using the \CII\ and \CO\ line emission, we estimated the total amount of molecular gas in a low metallicity environment and found that the bulk of the molecular gas is traced by \CII. Here, we compare the \CII-based molecular gas estimates to previous dust-based estimates, discuss the implications of \CII\ molecular gas estimates and how it relates to CO, and use the \CII-based estimates to estimate the CO-to-\htwo\ conversion factor ($X_{CO}$) and compare to simulations and models of molecular clouds at low metallicity.

\input{table5a.tex}

\subsection{Comparison to Dust-based \htwo\ Estimates}

Another method that can trace ``CO-faint'' molecular gas is to use dust emission and a self-consistently determined gas-to-dust ratio in atomic regions using \hi\ emission \citep{dam01,isr97,ler09,bol11}. We compare our molecular gas estimates using \CII\ and CO to the recent dust-based molecular gas estimates for the SMC presented in \citet{jam16}. Table \ref{table:gas_masses} shows the \CII+CO and dust-based molecular gas estimates over the area with both \CII\ and ALMA \CO\ coverage, where we make a cut at the RMS level before calculating the total mass from CO.

The \CII+CO estimates are not inconsistent with the dust-based estimates from \citet{jam16} in the N22, SWBarN, and possibly the SWBarS regions, given the factor of $\sim{2}$ uncertainty in either methods. The \CII+CO estimates for the SWDarkPK and N83 regions, on the other hand, are significantly lower than the result of the dust method. All of the \CII+CO estimates are systematically lower than the dust-based \htwo\ mass estimates. The systematically lower \CII+CO estimates can be the result of our \CII\ methodology underestimating the amount of \htwo, or the dust methodology overestimating the total molecular gas mass. 

When estimating the amount of \htwo\ associated with \CII, we have been conservative in our estimates of the contribution to \CII\ from the ionized and neutral gas and the conditions of the gas ($n$ and $G_{0}$), which may cause us to underestimate the total \htwo. The dust-based method has a number of systematic uncertainties \citep[for more details, see][]{ler07,ler09,ler11,bol11,san13,jam16}, with the two main uncertainties being the assumptions surrounding the gas-to-dust ratio and the far-IR dust emissivity. In particular, the method from \citet{jam16} relies on self-consistently calibrating the gas-to-dust ratio in the atomic gas assuming that it only varies smoothly on large scales (500 pc). This methodology could overestimate the molecular gas if the gas-to-dust ratio is an overestimate of the actual gas-to-dust ratio, which could be the case if the gas-to-dust ratio decreases in molecular regions. Similarly, the dust properties may vary from the diffuse to dense gas, which could produce a similar effect as a decrease in the gas-to-dust. 

There is also a systematic uncertainty of $\gtrsim0.3$~dex in the dust-based molecular gas estimate. \citet{bot04} compared $IRAS$ observations of the far-IR dust emissivity in the diffuse gas to extinction studies in the SMC and found evidence for a factor of $2-3$ decrease in the gas-to-dust ratio in dense gas. \citet{rom14} found that at the resolution of the NANTEN observation ($2.6\arcmin$), the degeneracy between the gas-to-dust ratio, effect of dust emissivity, and $X_{CO}$ are too large to provide definitive evidence of a change in the gas-to-dust ratio between the diffuse and molecular gas in the SMC, but allow for up to a factor of $\sim{2-3}$ decrease in the gas-to-dust ratio between the diffuse and dense gas. Using depletion measurements based on far-UV spectra to estimate the gas-to-dust ratio in the SMC, \citet{tch15} find evidence for up to a factor of $\sim{5}$ decrease in the gas-to-dust ratio from the diffuse to dense gas. However, these results are based on difficult measurements and rely on a number of assumptions. \citet{jam16} found that scaling down the gas-to-dust ratio by a factor of 2 in the dense gas leads to a similar factor of $\sim{2}$ decrease in the total molecular gas mass estimate. If the dust-based molecular gas estimates were lower by a factor of $\sim{2}$, which is a possibility given the existing measurements, then all of the dust estimates would be consistent with the \CII+CO estimates, and there would be one region (SWBarN) where the \CII+CO molecular gas estimate would be higher than the dust-based estimate. Similarly, if we used the density and radiation field estimates from the higher metallicity PDR models, the \CII+CO estimates would be $\sim{2\times}$ higher and would no longer be systematically below the dust-based estimates.

\subsection{\CII\ as a Tracer of \htwo\ and Star Formation Rate at Low Metallicity}

On larger scales and in high-metallicity environments, \CII\ acts as a calorimeter as it is the main coolant of the ISM. The heating comes from FUV photons causing electrons to be ejected from dust grains due to the photoelectric effect. Most recently \citet{del14} and \citet{her15} have shown that \CII\ emission correlates well with other star formation rate tracers, specifically \ha\ and 24 \micron, on kiloparsec scales in nearby galaxies. On smaller scales \citet{kap15} found that the correlation between \CII\ and star formation rate holds, although much of the \CII\ emission in M31 comes from outside of star-forming regions, as the FUV photons that heat the gas travel long distances diffusing throughout the gas disk. When the M31 observations are averaged over kiloparsec scales, \citet{kap15} recovers the relationship between \CII\ and star formation rate observed in other galaxies. 

\input{table6.tex}

Both \citet{del14} and \citet{her15} show that the scatter in the correlation between \CII\ and star formation rate (SFR) increases as the metallicity decreases. The breakdown in the tight \CII-SFR relationship can be understood in terms of a decrease in the photoelectric heating efficiency and/or an increase in the FUV photon escape fraction. In a low-metallicity environment there is less dust, allowing FUV photons to propagate farther through the gas and potentially escape the region or even the galaxy. \citet{cor15} finds evidence of an increased UV escape fraction based on the observed $\OIII/L_{\rm{TIR}}$ for the DGS. The increase in the propagation distance causes the \CII\ emission to be less co-localized with the recent massive star formation. If the FUV propagation distance is longer than the scale on which the \CII\ emission is compared to other star formation indicators, that increases the scatter in the \CII-SFR relationship. When the propagation distance is larger than the scale of the gas disk a fraction of the FUV photons escape the galaxy. The \CII\ emission will then underestimate the star formation rate even when considering integrated properties. This is consistent with  the lower \CII\ emission per unit star formation rate (estimated combined from FUV and 24 \micron\ emission) observed at lower metallicities by \citet{del14}.

On small scales, \CII\ reflects the cooling of the gas and can trace the \htwo\ by accounting for the amount of emission attributed to collisions with \htwo\ molecules. This requires removing the contribution to the emission due to the cooling of the atomic gas, and understanding the conditions (temperature and density) that account for the \CII\ excitation. In this context, the star formation rate is only important in that it is the source of FUV photons that ionize the carbon and heats the PDR. Because this takes place on the scale of individual \hii\ regions, the fraction of FUV photons that ultimately escape the galaxy does not enter in the estimates. The fact that at lower metallicity the PDR region at $A_V\lesssim1$ is physically very extended makes \CII\ an excellent tracer of ``CO-faint'' molecular gas. The excellent correspondence between the \htwo\ S(0) line emission and \CII\ emission (see Figure \ref{fig:H2_S0}), and the general correspondence between the \CII\ and \CO\ emission (see Figure \ref{fig:CII_CO_images}) demonstrates that \CII\ traces the structure of the molecular gas at low metallicity.  

\subsection{Comparing \CII-bright and CO-bright \htwo}
\label{subsection:cii_co_h2}

\begin{figure}[t]
\epsscale{1.2}
\plotone{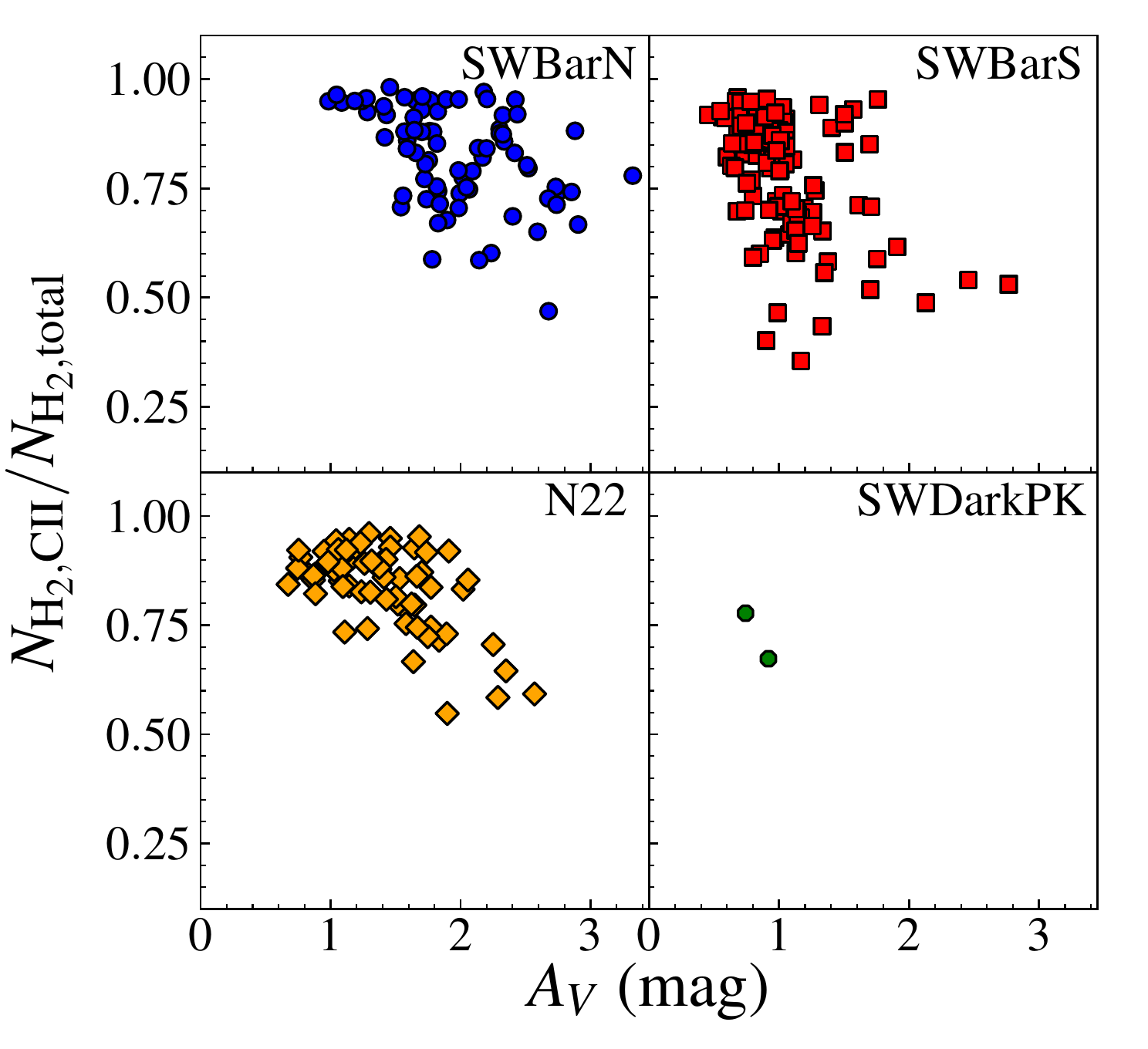}
\caption{Relationship between the ratio of $N_{\rm{\htwo, \CII}}$ to $N_{\rm{\htwo, total}}$ for the ALMA regions where $N_{\rm{\htwo, total}}$ is the combined estimate of \htwo\ from \CII\ and \htwo\ from CO, the latter using a Galactic CO-to-\htwo\ conversion factor $X_{CO} = 2\times10^{20}$ cm$^{-2}$ (K km s$^{-1}$)$^{-1}$. The colored symbols show independent measurements with $I_{\textnormal{\CO}}>3\sigma$. 
\label{fig:fCII}}
\end{figure}

We estimated the amount of molecular gas using \CII\ to trace the ``CO-faint'' molecular gas at low \Av\ (``\CII-bright'' molecular gas) and \CO\ to trace the molecular gas at high \Av\ using a Galactic $X_{CO}$ to convert the observed emission to molecular gas (``CO-bright'' molecular gas). Figure \ref{fig:fCII} shows the fraction of the total molecular gas coming from \CII\ as a function of $A_{V}$ estimated from dust. We see that at the low metallicity of our mapped SMC regions most of the molecular gas is traced by \CII, not CO, with $\sim{70\%}$ of the molecular gas coming from \CII-bright regions on average. The data also show the expected trend of a higher fraction of the total molecular gas being traced by CO as \Av\ increases, although with a lot of scatter. The scatter is in part due to the fact that our $A_V$ estimate uses data with lower resolution than the \CII\ or the CO observations, and in part due to our limitations at estimating the physical conditions from the available data. But more fundamentally, much of the scatter must arise from the fact that the relevant $A_V$ for photodissociation is not the line-of-sight extinction that we measure, but the $A_V$ toward the dominant sources of radiation. Note that given the methodology the molecular gas estimate from \CII\ is more likely to be underestimated rather than overestimated (see Section \ref{subsubsection:h2_unc}), suggesting that the ``\CII-bright'' molecular fractions may be preferentially higher.

In the SWBarN and N22 regions CO traces at most $\sim{50\%}$ of the molecular gas at our maximum \Av. The trend in the SWBarS region is considerably steeper, and around $A_V\sim1$ it reaches ``CO-bright'' fractions of $\sim{70\%}$. This is consistent with the fact that this region hosts the brightest \CO\ and \thirteenCO\ emission. The range of observed ``\CII-bright'' molecular gas fractions are consistent with the estimated fractions for regions in the SMC with velocity-resolved \CII\ spectra by \citet{req16}, who find $50\%$ to $\sim{90}\%$ of the molecular gas is traced by \CII\ emission. The fractions of ``CO-faint'' molecular gas in the SMC are much higher than fractions found in local Galactic clouds \citep[$f_{\rm{CO-faint}}=0.3$;][]{gre05}, and fractions in dense molecular clouds estimated from \CII\ emission \citep[$f_{\rm{CO-faint}}=0.2$;][]{lan14}. The fractions in the SMC are more similar to those found using \CII\ emission in diffuse Galactic clouds \citep[$f_{\rm{CO-faint}}=0.75$ for clouds with no detectable \CO\ emission;][]{lan14}. Overall, the high fractions of \CII-bright gas suggest that most of the \htwo\ gas in the SMC is found within PDRs or in regions with PDR-like conditions.

\begin{figure}[t]
\epsscale{1.2}
\plotone{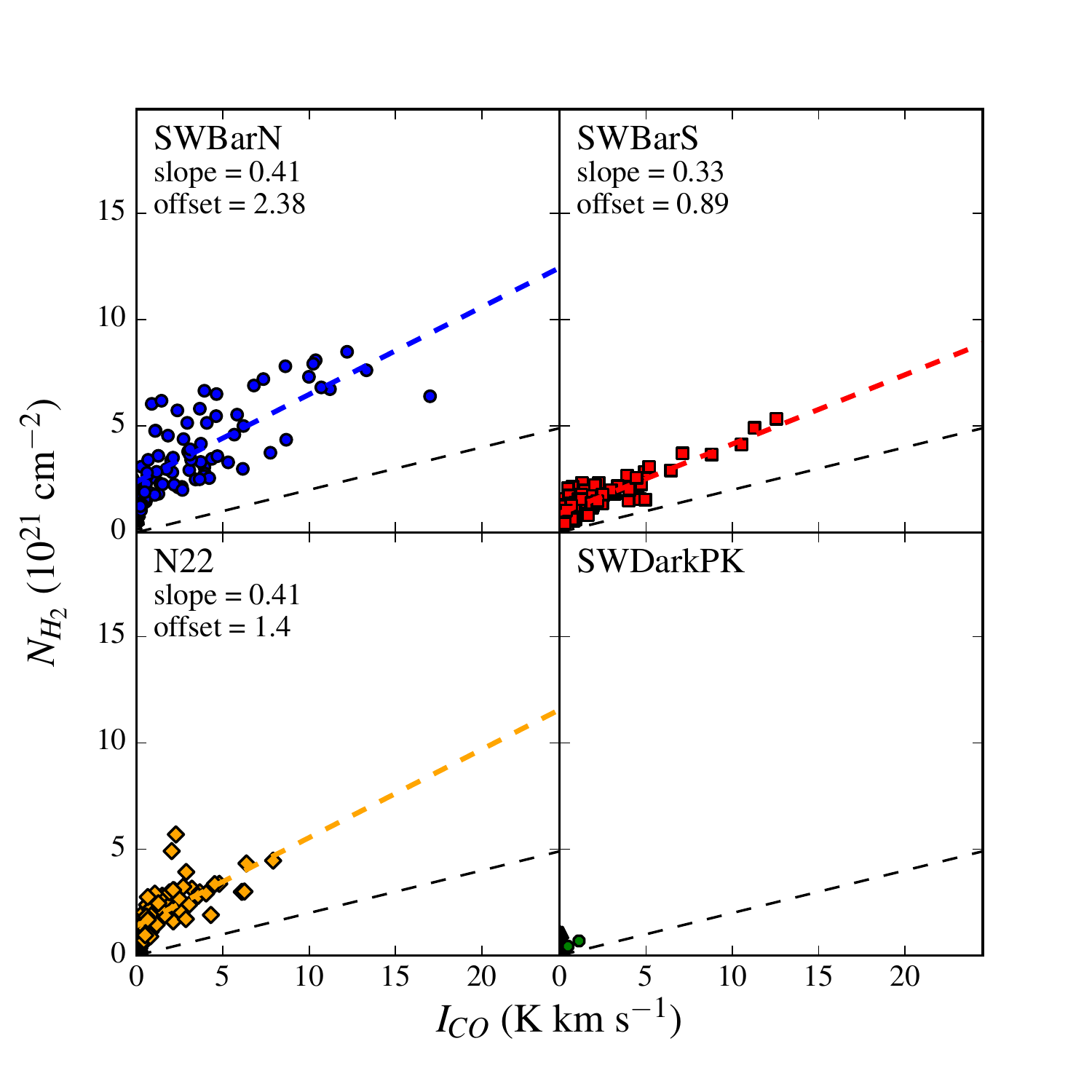}
\caption{Relationship between $I_{CO}$ and \Nhtwo\ for the ALMA regions where \Nhtwo\ is the combined estimate of \htwo\ from \CII\ and from CO using the Galactic CO-to-\htwo\ conversion factor $X_{CO} = 2\times10^{20}$ cm$^{-2}$ (K km s$^{-1}$)$^{-1}$. The colored symbols show independent measurements detected at $>3\sigma$ in \CO. There is good correlation between CO and \htwo, and the slope of the relationship is $X_{CO}$. The thick dashed colored lines show the linear fits to the data; the steeper slopes correspond to higher values of $X_{CO}$. The over plotted black dashed lines shows the Galactic conversion factor ($X_{CO}$) appropriate for the Milky Way and resolved measurements \citep{bol13}. 
\label{fig:Nh2_Ico}}
\end{figure}

Figure \ref{fig:Nh2_Ico} shows that, despite the fact that most molecular gas is related to \CII\ rather than CO emission, our estimates for \Nhtwo\ are correlated with the \CO\ integrated intensity. The black dashed line in Figure \ref{fig:Nh2_Ico} shows the estimated amount of \Nhtwo\ associated with CO using a Milky Way conversion factor of $X_{CO}=2\times10^{20}$ cm$^{-2}$ (K km s$^{-1}$)$^{-1}$. There is a clear offset between the results from assuming a Galactic $X_{CO}$ and our estimate for \Nhtwo. This offset is due to the presence of molecular gas traced primarily by \CII\ emission (at low $A_V$). Interestingly, we see that the relations are steeper than that in the Milky Way. The steeper slopes indicate that molecular gas builds up faster with CO intensity than it does in the Milky Way. This is due to the presence of ``\CII-bright'' \htwo\ along the line-of-sight toward CO emission: simply the result of the fact that ``CO-bright'' regions still have substantial \CII\ emission associated with them at the resolution of our measurements. 

\subsection{The CO-to-\htwo\ Conversion Factor in the SMC}

\begin{figure}[t]
\epsscale{1.2}
\plotone{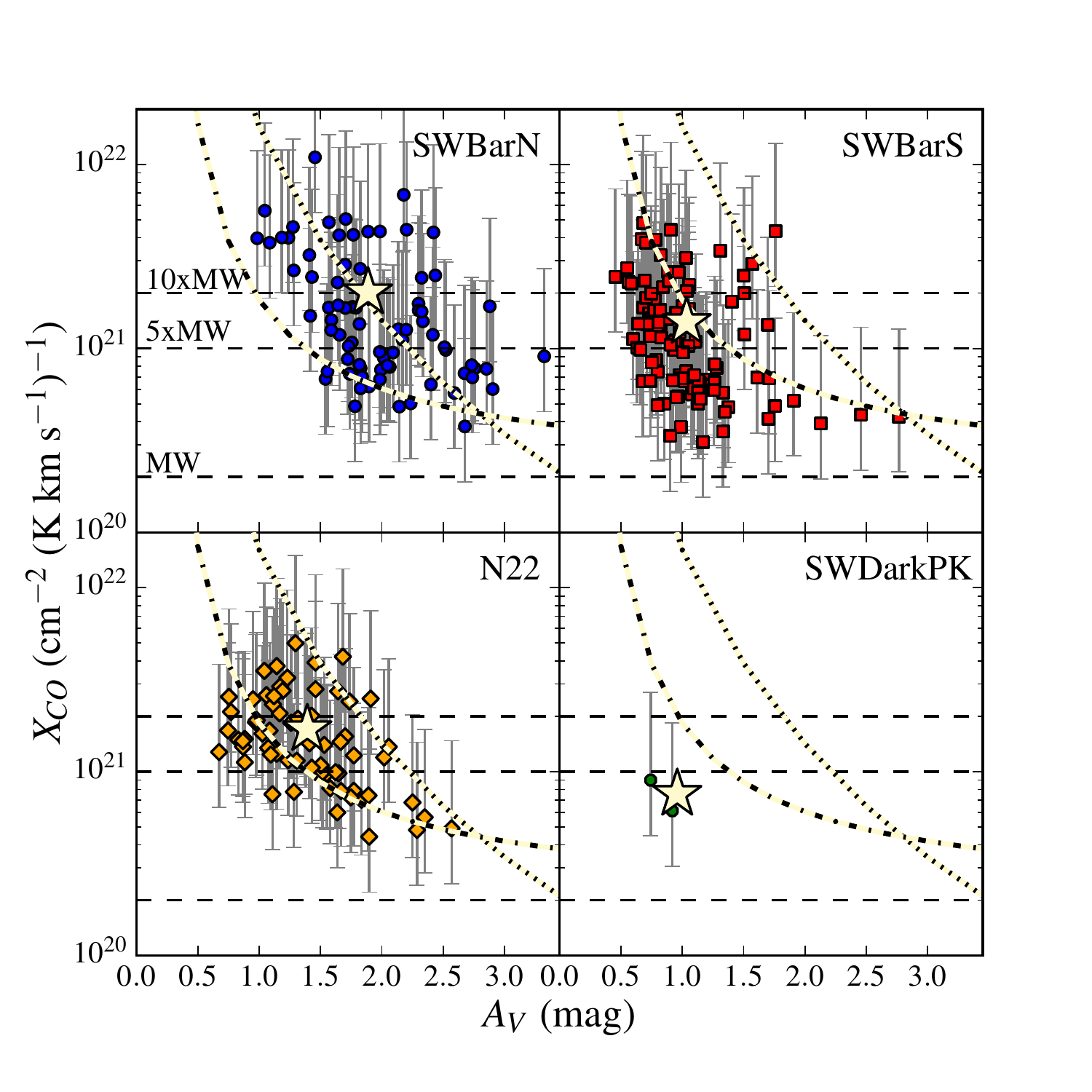}
\caption{The CO-to-\htwo\ conversion factor ($X_{CO}$) as a function of $A_{V}$ for the ALMA regions where the \htwo\ used is the combined estimate of \htwo\ from \CII\ and \htwo\ from CO using a Galactic value of $X_{CO} = 2\times10^{20}$ cm$^{-2}$ (K km s$^{-1}$)$^{-1}$. The colored symbols show independent measurements detected at $>3\sigma$ in \CO\ with the error bars showing the factor of 2 uncertainty in \Nhtwo, with the star symbols showing the mean \Av\ and $X_{CO}$ for each region. The over plotted dashed lines show scalings of the Galactic conversion factor appropriate for the Milky Way and resolved measurements \citep{bol13}. The dotted line shows the estimate of $X_{CO}$-based average CO luminosity, the average \Nhtwo, and the average $A_{V}$ over the entire cloud for the simulations of molecular clouds by \citet{glo11}. The dash-dotted line shows similar cloud-averaged estimate of $X_{CO}$ as a function of $A_{V}$ from the PDR models from \citet{wol10} for the SMC metallicity and radiation field strength of 40 $G_{0}$, the approximate mean radiation field found in the regions (see Section \ref{subsubsection:method2}). 
\label{fig:Xco}}
\end{figure}

We use our \Nhtwo\ determination to estimate the values of $X_{CO}$ throughout the mapped regions, accounting for both ``\CII-bright'' and ``CO-bright'' molecular gas. Figure \ref{fig:Xco} shows our estimates of $X_{CO}$ as a function of $A_V$, where there is a trend of decreasing $X_{CO}$ with increasing $A_{V}$ albeit with significant scatter. The trend in $X_{CO}$ matches the expectation of extended envelopes of ``\CII-bright'' molecular gas, and it compares well with similar trends seen in low metallicity molecular cloud simulations \citep{she11,szu16}. The scatter can be explained as due to the ``breakdown'' of a single value of $X_{CO}$ on small scales in simulations and even observed clouds, due to the clumpy nature of molecular clouds \citep{glo11,she11,bol13}. 

The values we find for our regions have on average a conversion factor of $X_{CO}\sim5\,X_{CO,\text{MW}}$. For the entire SMC, and using the dust-based molecular gas estimates, \citet{jam16} find a CO-to-\htwo\ conversion factor $\sim17$ times higher than the average Milky Way value. The difference between these two results can be mostly ascribed to the bias in our chosen HS$^3$ fields. Our survey selected actively star-forming regions with bright CO emission, whereas most of the SMC is faint in CO: in fact our regions contain some of the peaks of CO emission in the SMC. But even within our observed regions many lines of sight have high fractions of \htwo\ not traced by ``bright-CO'' emission (see Figure \ref{fig:fCII}). The higher global measurement is simply a statement of the fact that the extinction in a typical molecular line-of-sight in the SMC is probably $A_V\sim0.5-1$. Accounting for the dust-to-gas ratio of $1/5-1/7$ the Galactic value, the same line-of-sight would have $A_V\sim2.5-7$ in the Milky Way and emit brightly in CO. Even in our highly biased regions bright in CO we barely reach $A_V\sim2.5$ in the SMC (Figure \ref{fig:Xco}). \citet{sch17} also found variations in the CO-to-\htwo\ conversion factor using dust-based \htwo\ estimates in different regions of the low metallicity dwarf galaxy NGC 6822 with the lowest values (closest to the Milky Way value) in the regions with bright CO emission. Using galaxy-integrated measurements of \CII\ and \CO\ and isolates the \CII\ contributions from molecular gas using the results of models of \CII\ emission throughout a galaxy, \citet{acc17} estimated the conversion factor and found the values to range from $\sim{12.5-17}\times$ the Milky Way value for the metallicity of the SMC, which is within the range of our estimates of $X_{CO}$. 

We can compare our estimates of $X_{CO}$ versus $A_{V}$ with theoretical studies of the relation between CO and \htwo\ at low metallicity. \citet{wol10} present a model of molecular clouds in order to understand the fraction of \htwo\ not traced by bright CO emission as a function of metallicity. They assume spherical clouds with $r^{-1}$ density profile, predicting the fraction of ``CO-faint'' gas based on their mean volume density, external radiation field, mean $A_{V}$ ($\langle A_{V} \rangle$), and metallicity relative to Galactic ($Z^\prime$). \citet{glo11} and more recently \citet{szu16} create simulations of molecular clouds with varying conditions, including metallicity, finding an empirical relationship for the mean $X_{CO}$ value as a function of the mean $A_{V}$ for the simulated clouds. The trend predicted by the \citet{wol10} calculations and formulated by \citet{bol13} is
\begin{equation}
X_{CO} = X_{CO,0}e^{\left(4 \Delta A_{V}/\langle A_{V} \rangle \right)} e^{\left(-4 \Delta A_{V}/(\langle A_{V} \rangle/Z')\right)}
\end{equation}
where 
\begin{equation}
\Delta A_{V} = 0.53 - 0.045\log{\left(\frac{G_{0}}{n}\right)} - 0.097\log{\left(Z'\right)},
\end{equation}
\noindent while \citet{glo11} find
\begin{equation}
X_{CO} = \begin{cases}
		X_{CO,0}\left( \langle A_{V} \rangle/3.5 \right) ^{-3.5} & \text{if }\langle A_{V} \rangle<3.5\text{ mag}  \\
		X_{CO,0} & \text{if }\langle A_{V} \rangle\geq{3.5}\text{ mag}
		\end{cases}
\end{equation}
where $X_{CO,0}$ is the mean value for the Milky Way.

\begin{figure*}[t]
\epsscale{1.1}
\plotone{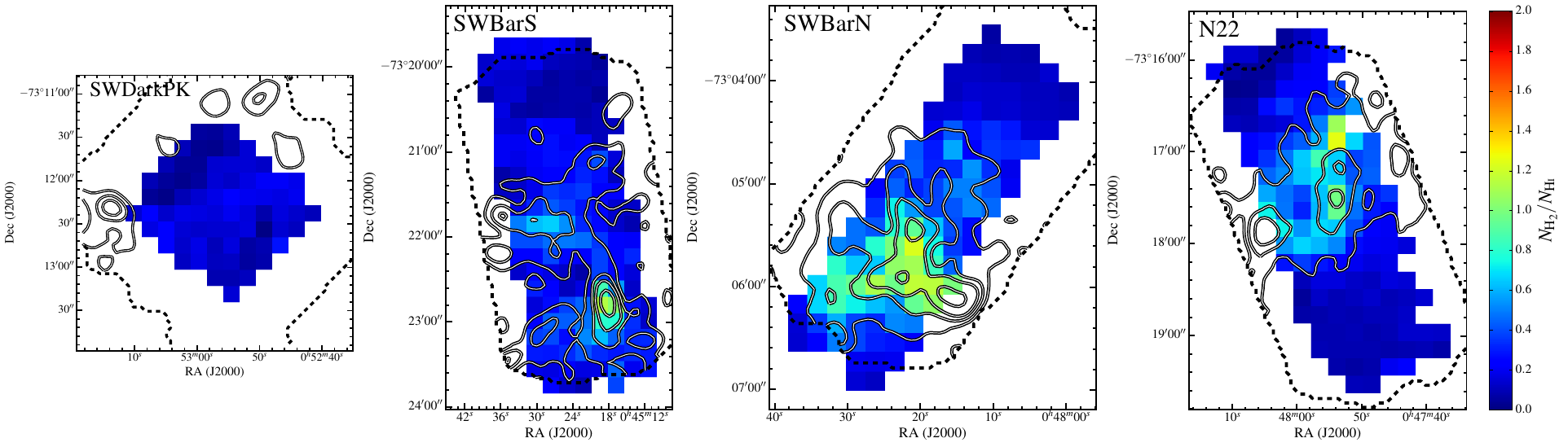}
\caption{Ratio of the estimate of the total molecular gas column (using \CII\ and \CO) to \Nhi\ estimated to be associated with the molecular cloud (see Section \ref{subsection:CII_from_HI}). The black contours show the ALMA \CO\ ($2-1$) integrated intensity (convolved to \CII\ resolution of $12\arcsec$) at levels of 1.6 ($\sim{3\sigma}$), 8, 16, 32, and 48 K km s$^{-1}$. We see that faint CO emission extends to $\Nhtwo/\Nhi\sim{0.5}$, which is beyond the estimated molecular to atomic transition.
\label{fig:mol2atomic}}
\end{figure*}

In Figure \ref{fig:Xco} we overplot the predicted $X_{CO}$ trend from \citet{wol10} using the metallicity of the SMC ($Z'=0.2$), an average $n=5000$ cm$^{-3}$, and $G_{0}=40$ based on the PDR modeling we did (see Section \ref{subsubsection:method2}). We also overplot the empirical fit from the simulations by \citet{glo11}. Both predictions are close to the data for both individual lines of sight and for the averages within a region (indicated by the star), although \citet{wol10} seems to do a better job overall. As both the \citet{wol10} model and \citet{glo11} simulations suggest, the main effect driving the relationship between $X_{CO}$ and $A_{V}$ is not metallicity itself, but how metallicity (and environment) affect the FUV shielding in the molecular gas. In particular, the critical element is the dust-to-gas ratio in the medium. This is also consistent with the findings of \citet{ler09} and \cite{lee15} that showed that the amount of \CO\ emission at a given value of \Av\ in the SMC resembles that in the Milky Way.

\subsection{\CO\ and the Molecular to Atomic Transition}

Our new measurements present an opportunity to look at the molecular-to-atomic transition at low metallicity. We estimate the molecular to atomic ratio using our total molecular gas estimate, \Nhtwo, and the \hi\ column density map associated with the molecular cloud. The \Nhi\ map is the same that we used to determine the amount of \CII\ emission associated with \hi\ gas (see Section \ref{subsection:CII_from_HI}), which was produced by only integrating the \hi\ emission over a velocity range that is centered on the \CO\ line velocity, but $\pm25\%$ wider than the range of velocities with observed \CO\ emission throughout  each the region. Figure \ref{fig:mol2atomic} shows the maps of the \Nhtwo/\Nhi\ ratio with overlaid ALMA \CO\ contours. As expected for the atomic-dominated SMC, we observe peak molecular-to-atomic ratios of $\Nhtwo/\Nhi\sim{1.5}$ that show that the amount of molecular gas does not greatly exceed the amount of \hi\ gas along the line-of-sight due to the large amount of atomic gas. We also detect \CO\ emission out to $\Nhtwo/\Nhi\sim{0.5}$, which is beyond the region of the cloud where we estimate the molecular gas to dominate over the atomic gas based on the measured column densities. The observation of \CO\ emission out to the region where we estimate the molecular-to-atomic transition occurs indicates that the molecular gas is truly ``CO-faint'' and not ``CO-dark.'' We note, however, that trying to use CO to estimate molecular masses at low metallicities rapidly runs into the problem of the fast growth in the conversion factor and its steep dependence on local conditions, nominally \Av\ or column density. 

\section{Summary and Conclusions}
\label{section:conclusions}

We present results from the $Herschel$ Spectroscopic Survey of the SMC (HS$^{3}$), a survey that mapped far-IR cooling lines in five star-forming regions. These data are complemented by APEX CO observations in the SMC Wing region N83, ALMA ACA observations of the Southwest Bar of the SMC that mapped four of the five HS$^{3}$ regions in \CO, \thirteenCO, and \CeighteenO, and by SOFIA GREAT observations that will be presented fully in a forthcoming paper. The main results from these new observations are:
\begin{enumerate}
\item The \CII\ 158 \micron\ line is detected throughout the entirety of the regions. The \OI\ 63 \micron\ emission is also very widespread, and detected throughout large portions of the regions including faint diffuse areas (\S \ref{subsection:CII_OI}, Figure \ref{fig:HS3_RGB}).
\item The \OI/\CII\ ratio is fairly uniform throughout all the regions with an average value of $\OI/\CII\sim{0.3}$ (\S \ref{subsection:CII_OI}, Figure \ref{fig:OI_CII}).
\item We do not detect the \NII\ 122 \micron\ line, but we do detect the \NII\ 205 \micron\ line in every region with FTS spectroscopy. Using the upper limit on the \NII\ 122 \micron\ observations, the observed \NII\ 122 \micron/205 \micron\ ratio is consistent with an electron density $n_e\lesssim{20}$ cm$^{-3}$ (\S \ref{subsection:NII}, Figure \ref{fig:NII_CII}). 
\item We find \OIII/\CII\ ratios in the \hii\ regions that are high compared to more massive, higher metallicity galaxies (\S \ref{subsection:OIII}, Figure \ref{fig:OIII_CII}). 
\item Our ALMA ACA \CO\ maps include the total power correction and show small (few pc), bright structures with bright \thirteenCO\ surrounded by more diffuse, faint emission. We do not detect \CeighteenO\ in any of the regions. We find $\CO/\thirteenCO\sim{7-13}$, which translates to \CO\ optical depths of $\tau_{\CO}\sim{4.5-11}$ assuming the \thirteenCO\ emission is optically thin (\S \ref{subsection:12CO_13CO}, Figures \ref{fig:ALMA_CO} and \ref{fig:12COto13CO}).
\item The \CO/\CII\ ratios are $\sim{1/5}$ the average ratio found for the Milky Way and increase with \Av, reaching the Milky Way ratio or higher. The low \CO/\CII\ ratios at low \Av\ suggests that there is a layer of molecular gas not traced by bright \CO\ emission, which is confirmed by our modeling (\S \ref{subsection:CO_CII}, Figures \ref{fig:CII_CO_images} and \ref{fig:CO_CII_OICII}).
\end{enumerate}

We see evidence that the \CII\ emission traces molecular gas in regions where the CO is expected to be photo-dissociated from the low \CO/\CII\ ratios and good correspondence between the \CII\ structure and the \htwo\ S(0) and \CO\ emission. We use the \CII\ and \CO\ emission to estimate the total molecular gas assuming that the bright \CII\ emission traces molecular gas at low \Av\ and \CO\ traces gas at high \Av. We first remove the possible \CII\ emission from the ionized and atomic gas (both CNM and WNM), and assume the remaining \CII\ emission arises from molecular gas. We convert this emission to \Nhtwo\ using two different methods: the first assumes a fixed temperature and density of $T=90$ K and $n=4000$ cm$^{-3}$, and the second uses the $n$ and $G_{0}$ from new SMC PDR models (see Appendix \ref{appendix:smc_pdr_models}) for the \CII, \OI, and total-IR combined with new \CII\ cooling rates from Wolfire et al. (2017, in preparation). The two methods produce similar results, but we choose to use the second method to account for variations in the conditions throughout the regions. We convert the \CO\ emission to a molecular gas estimate applicable in the high $A_V$ regions of the clouds using a Milky Way conversion factor of $X_{CO} = 2\times10^{20}$ cm$^{-2}$ (K km s$^{-1}$)$^{-1}$. We estimate a factor of 2 uncertainty in our molecular gas estimates, but note that given our assumptions for the carbon abundance and PDR models that we are more likely to underestimate \Nhtwo\ traced by \CII.

Using our estimates of \Nhtwo\ using \CII\ and \CO, we find that:
\begin{enumerate}
\item Our \htwo\ column density estimated from \CII\ and \CO\ is mostly consistent with (although systematically lower than) larger-scale dust-based estimates from \citet{jam16}.
\item We find average fractions of the molecular gas traced by \CII\ of $\sim{70\%}$ and show that most of the molecular gas in the SMC is not traced by bright \CO\ emission, which is expected for low metallicity environments where \htwo\ is able to self-shield while CO is photodissociated.
\item We use the \Nhtwo\ estimated from \CII\ and \CO\ to evaluate the CO-to-\htwo\ conversion factor ($X_{CO}$), and find $X_{CO}$ decreases with increasing \Av. The relationship between $X_{CO}$ and \Av\ is consistent with models from \citet{wol10} and simulations from \citet{glo11} and \citet{szu16}, which suggests that the main effect of lowering the metallicity is to decrease \Av\ and the amount of shielding for a given gas column density.
\item Despite the fact that most of the molecular gas in the SMC is not traced by bright CO emission, we do observe faint \CO\ emission out to the estimated location of the molecular-to-atomic transition. Using the aggregate CO emission to obtain total molecular mass in a low metallicity environment suffers from the problem that it is very sensitive to the local \Av\ or column density.  
\end{enumerate}




\acknowledgments

We would like to thank the anonymous referee and Prof. Adam Leroy for useful comments that improved the manuscript. This paper makes use of the following ALMA data: \dataset[ADS/JAO.ALMA\#2015.1.00581.S]{https://almascience.nrao.edu/aq/?project_code=2015.1.00581.S} and \dataset[ADS/JAO.ALMA\#2013.1.00652.S]{https://almascience.nrao.edu/aq/?project_code=2013.1.00652.S}. ALMA is a partnership of ESO (representing its member states), NSF (USA) and NINS (Japan), together with NRC (Canada), NSC and ASIAA (Taiwan), and KASI (Republic of Korea), in cooperation with the Republic of Chile. The Joint ALMA Observatory is operated by ESO, AUI/NRAO and NAOJ. The National Radio Astronomy Observatory is a facility of the National Science Foundation operated under cooperative agreement by Associated Universities, Inc. We used APEX data from project C-093.F-9711-2014. This work is based in part on observations made with the NASA/DLR Stratospheric Observatory for Infrared Astronomy (SOFIA). SOFIA is jointly operated by the Universities Space Research Association, Inc. (USRA), under NASA contract NAS2-97001, and the Deutsches SOFIA Institut (DSI) under DLR contract 50 OK 0901 to the University of Stuttgart. Financial support for this work was provided by the NASA-USRA grants SOF030120 and SOF040151. A.D.B. and K.E.J. acknowledge the partial support from a CAREER grant NSF-AST0955836, from NSF-AST1139998, from NASA-JPL 1454733, and from USRA-SOF030120. K.E.J. acknowledges support from the NRAO Student Observer Support program grant 347065 and  partial support from CONICYT (Chile) through FONDECYT grant No. 1140839. M.R. wishes to acknowledge support from CONICYT (CHILE) through FONDECYT grant No. 1140839 and partial support from project BASAL PFB-06.



\facilities{ALMA, Herschel, APEX, SOFIA, Spitzer}.

\software{HIPE (v 12.0.2765, \citealt{ott10}), CASA \citep{mcm07}, NumPy \citep{van11}, SciPy \citep{jon01}, Astropy \citep{ast13}, Matplotlib \citep{hun07}, NADA \citep{lee17}}


 
\clearpage

\appendix

\section{SMC PDR Models}
\label{appendix:smc_pdr_models}

\begin{figure*}[t]
\plotone{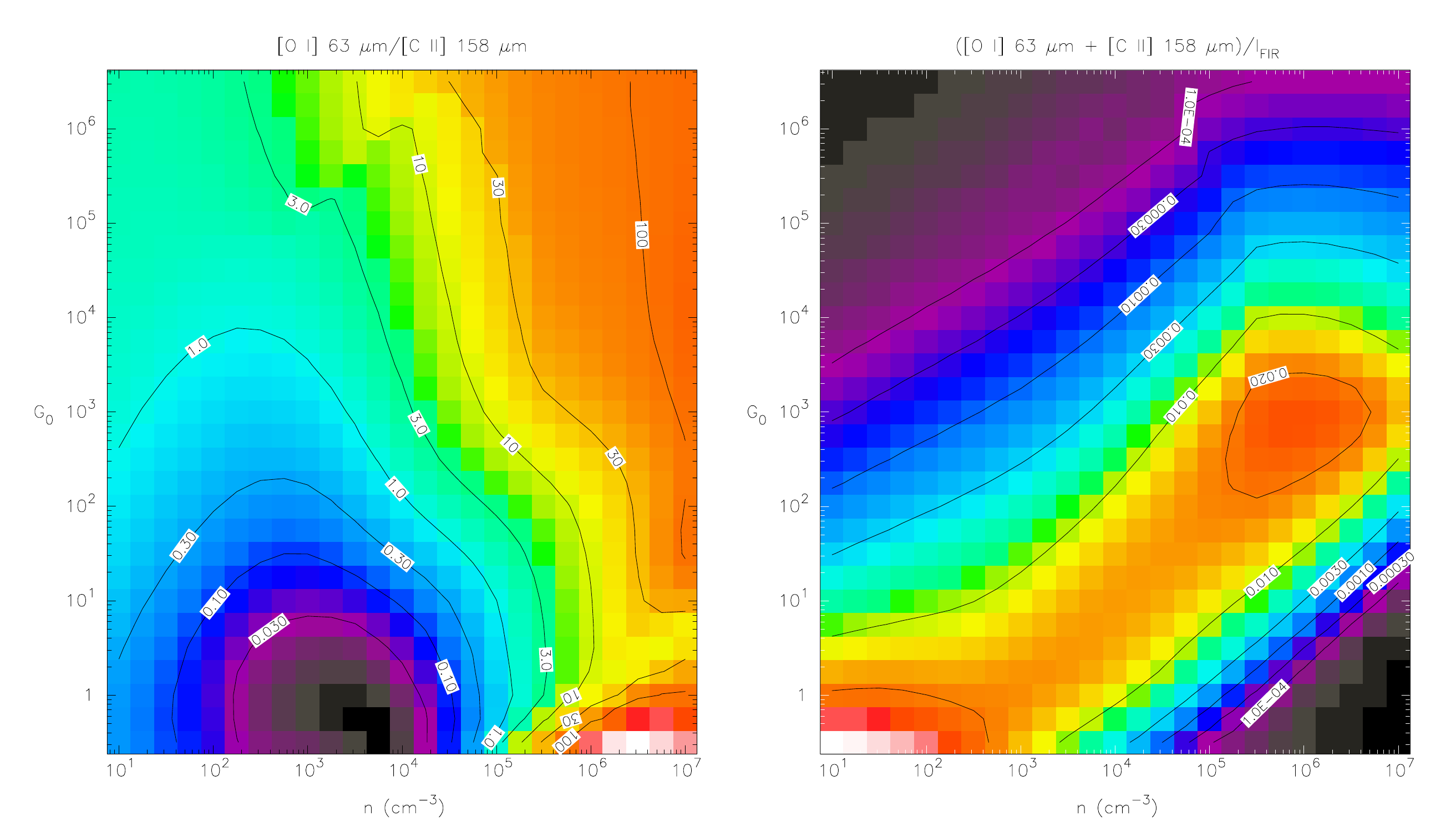}
\caption{New SMC PDR model contour plots for the \OI\ 63 \micron/\CII\ 158 \micron\ (left) and (\OI\ 63 \micron\ + \CII\ 158 \micron)/FIR ratios as a function of $n$ and $G_{0}$. 
\label{fig:SMC_PDR_models_contour}}
\end{figure*}

The PDR model we use is based on that of \citet{kau06}, \citet{wol10}, and \citet{hol12} but with a modified grain extinction, grain abundances, and gas phase abundances as appropriate for the SMC. For the gas phase metal abundances we use $1/5$ of the Galactic values. We use the conversion between optical depth and column density for the average ``Bar Sample'' from \citet{gor03}, $N/A_{V} = 1.32\times{10^{22}}$ cm$^{-2}$. For the small grain abundance, which scales the photoelectric heating rate, we use an abundance $1/7.7$ of the Galactic value from \citet{san10}. The FUV rise of the dust extinction is known to be steeper than the Galactic case \citep{gor03} which decreases the penetration of FUV photons into the PDR. The general form of the photo-chemical rates is $G_{0} \alpha e^{(-\beta A_{V})}$, where $\alpha$ scales the overall rate and $\beta$ accounts for the dust attenuation relative to visible wavelengths. We assume that the beta coefficients for the SMC are a factor of 2 higher than for local Galactic grains. This assumption was checked using the Meudon PDR code \citep{lep06}. First we added an SMC extinction curve to the code using the fits in \citet{gor03} to the \citet{fit90} parameterized extinction curves. Second we fit the resulting beta dependence as a function of Av to several key rates calculated by the Meudon PDR code. We find the factor of 2 results in less than $10\%$ difference in $\alpha$ compared to the Meudon code for the rates that we tested.

\begin{figure*}[t]
\plotone{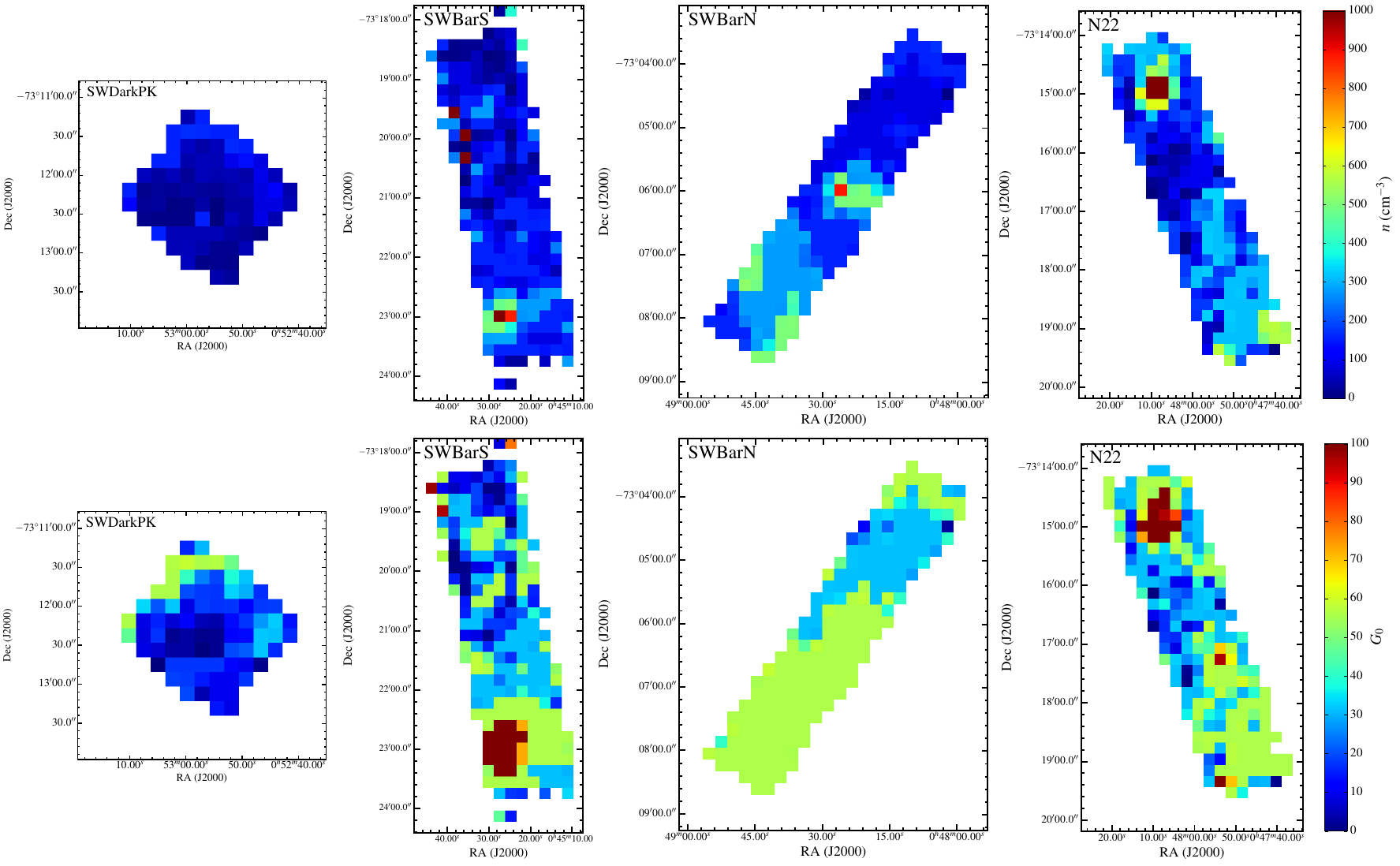}
\caption{Images of the results using the previous PDR models for Galactic conditions from \citet{kau06} for volume density ($n$; top row), and FUV radiation field strength ($G_{0}$; bottom row) for each of the regions using the same \CII, \OI, and TIR data as Figure \ref{fig:PDRT}. These models indicate similar radiation field strengths, but the densities are a factor of $\sim{10}$ lower than those found using the SMC PDR models.
\label{fig:PDRT_highZ}}
\end{figure*}

Resulting contour plots for \OI\ 63 \micron/\CII\ 158 \micron\ and (\OI\ 63 \micron\ + \CII\ 158 \micron)/FIR as a function of $G_0$ and $n$ are shown in Figure \ref{fig:SMC_PDR_models_contour}. For reference, Figure \ref{fig:PDRT_highZ} shows the PDR model results using the models for Galactic conditions from \citet{kau06}. Compared to the Galactic case \citep{kau99,kau06} at fixed $G_{0}$, the contours of ([OI] 63 + [CII] 158)/FIR are generally shifted to higher density or lower $G_{0}/n$ ratio. This line-to-continuum ratio can be interpreted as the heating efficiency or the fraction of energy that goes into gas heating compared to the total FUV photon energy (note that the total FIR is approximately twice the FUV energy due to dust heating by optical photons).  Theoretical models predict that in low metallicity environments, the reduced abundance of electrons from metals, leads to a higher grain charge and lower heating efficiencies \citep[e.g.,][]{rol06}. Increasing $n$, or lowering $G_{0}/n$, leads to more neutral grains and higher heating efficiencies. For low $G_{0}$ (approx less than 10) and low densities $n$ (approx less $10^{3}$) the \OI\ 63 \micron/\CII\ 158 \micron\ ratio for the SMC deviates from the Galactic curves due to the strong temperature dependence in this region and the effects of metallicity on the heating efficiency. Lower heating efficiency leads to a lower temperature, low \OI\ 63 \micron\ line intensity, and low \OI\ 63 \micron/\CII\ 158 \micron\ ratio (we note that at very low $G_{0}$ and high density the \CII\ line intensity rapidly drops leading to a high ratio of \OI\ 63 \micron/\CII\ 158 \micron). At higher density and $G_{0}$ the curves are similar to the Galactic case.

\section{\htwo\ Estimate from \CII\ Uncertainties}
\label{appendix:uncertainty}

In this section we explore the effects of different potential sources of uncertainty that could affect the \htwo\ column density estimates. We explore in detail how \OI\ absorption and \CII\ optical depth could affect our adopted values of $G_{0}$ and $n$ from the models, as well as the difference between the results using the SMC PDR models and the higher metallicity models. In general, increasing $n$ and/or $G_{0}$ will increase the \CII\ cooling rate, which decreases the amount of \htwo\ needed to account for the observed \CII\ emission. The amount of increase in \CII\ cooling depends on the specific value of $n$ since it is not a linear function of $n$ (see Figure \ref{fig:CII_cooling}), whereas the \CII\ cooling scales linearly with $G_{0}$. 

The values of $G_{0}$ appear to be robust to the choice of PDR model since the results using the SMC PDR models are similar to those found using the high metallicity models from \citet{kau06} (see Appendix \ref{appendix:smc_pdr_models}). The $G_{0}$ values are also consistent with the values found by \citet{isr11} for the N22 and SWBarN (N27) regions using Lyman continuum and dust emission and by \citet{san10} from modeling the mid-IR spectral energy distribution (SED). The results for $n$ using the new SMC PDR models are a factor of $\sim{10}$ higher than what would have been obtained using a high metallicity PDR model \citep[e.g.,][]{kau99}. For a given $G_{0}$ and $n$ the grains will be more charged at low metallicity due to fewer free electrons (primarily from the lower carbon abundance). The increase in grain charge will decrease the photoelectric heating efficiency and a higher density is required to produce the observed line intensity ratios. See Appendix \ref{appendix:smc_pdr_models} for a full description of the changes in the new PDR models. This increase in density translates to an increase in the total \htwo\ mass from \CII\ ($M_{\htwo,\CII}$) by factors of $\sim{1.5-3.0}$. 

There are uncertainties associated with the interpretation of the available observations; in particular, the possibility of \OI\ self-absorption and optically thick \CII\ emission. We already discussed the likelihood of \OI\ absorption or self-absorption in Section \ref{subsubsection:OI_abs}, and found no evidence that there is significant absorption present in our observations. To explore how much a low level of absorption would affect our results, we assume $30\%$ of the integrated \OI\ intensity is absorbed leading to a 30\% underestimate in the real \OI\ flux. We find that if we allow for 30\% larger \OI\ flux, the model density increases by a factor of $\sim{3}$ while the radiation field $G_{0}$ decreases by a similar factor of $\sim{3}$. The increase in density increases the cooling rate, which decreases the amount of \htwo\ needed to explain the \CII\ emission. Conversely, decreasing $G_{0}$ also decreases the cooling rate, which increases the amount of \htwo. At the moderate densities in the regions ($\sim{10^{3}-10^{4}}$ cm$^{-3}$), the cooling rate does not increase much with $n$, whereas the cooling rate will scale linearly with $G_{0}$. As a result, the effect of increasing the amount of \htwo\ from the change in $G_{0}$ dominates over the decrease due to the change in $n$ and the estimated $M_{\htwo,\CII}$ decreases by $\sim{30\%}$.

We have no evidence that the \CII\ 158 \micron\ emission is optically thick in our SMC regions. Measurements of \thirteenCII\ and \twelveCII\ in dense, Galactic PDRs find optical depths $\tau_{\rm [CII]}\approx{0-3}$ \citep{oss13}. A moderate optical depth of $\tau_{\rm [CII]}=1$ seen in Galactic regions scaled to account for the lower carbon abundance in the SMC gives $\tau_{\rm [CII]}=0.2$. This level of optical depth is also possible in the higher \Nhtwo\ regions; using the equation for $\tau_{\rm [CII]}$ from \citet{cra85} with $\Nhtwo=10^{22}$ cm$^{-2}$ (values found in the peaks of our \Nhtwo\ estimates), $T=100$ K, and $n=5000$ cm$^{-3}$ produces $\tau_{\rm [CII]}=0.2$. To explore how this level of moderate optical depth would affect our \htwo\ estimates by correcting the \CII\ emission by a factor of $\tau/(1-\exp(-\tau))=1.1$. We find that this has little effect on the model values of $n$ and $G_{0}$, and only increasing $n$ by a factor of $\lesssim{2}$ in some of the denser ($n\sim{5000-10000}$ cm$^{-3}$) regions and decreasing $G_{0}$ by the same factor. Since $n$ and $G_{0}$ remain largely the same, the \CII\ cooling rate is the same as for the uncorrected, lower \CII\ intensity while the optical depth correction increases amount of \CII\ emission, which increases the amount of \htwo. The correction for $\tau_{\rm [CII]}=0.2$ increases the mass estimates by $\sim{10\%}$.

Additionally sources of uncertainty include the carbon abundance, absolute flux calibration of the PACS spectrometer, and the \CII\ intensity attributed to ionized and neutral gas. The measurements of the gas-phase carbon abundance in the SMC range from C/H$\sim 1.4\times{10^{-5}}-4.2\times{10^{-5}}$ \citep{kur99,tch15}. This introduces a factor of $\sim{30\%}$ uncertainty on our assumed value of $2.8\times10^{-5}$, which translates into a factor of $\sim{30\%}$ uncertainty in the \htwo\ estimate as it scales linearly with carbon abundance. The absolute flux calibration of the PACS spectrometer is an additional source of uncertainty on the order of $30\%$ \citep{pog10}. Finally, there is uncertainty involved in our estimates of the \CII\ intensity attributed to neutral and ionized gas. Our estimate of \CII\ emission from ionized gas would likely only overestimate the actual contribution, and the ultimate contribution is small ($\lesssim{10\%}$) in the regions in and around the molecular cloud and does not significantly affect the total \htwo\ mass estimate. The contribution from neutral gas is more uncertain. Changing the amount of \hi\ associated with the WNM has little effect on the estimated \CII\ intensity due to the high temperature and low density (compared to the critical density) of the gas, but the high temperature causes the \hi\ emission from the WNM to dominate over the CNM. Changing the amount of \hi\ in the CNM has a stronger effect on the attributed \CII\ intensity due to the density being closer to the critical density (see Equation \ref{eq:I_CII}). Varying the fraction of the \hi\ associated with the CNM from our assumed value of $50\%$ to $25\%$ and $75\%$ changes the fraction of the \CII\ intensity from neutral gas by $\pm2\%$, which would ultimately have a similar effect on the mass as the optical depth correction. Since a correction to $I_{\CII}$ of $+10\%$ had an effect of $\sim{+10\%}$ on the mass, a $\sim{2\%}$ correction to $I_{\CII}$ will cause a $\sim{2\%}$ increase of the \htwo\ mass, which is negligible compared to the other sources of uncertainty discussed.



\begin{thebibliography}{}

\bibitem[Accurso et al.(2017)]{acc17} Accurso, G., Saintonge, A., Catinella, B., et al.\ 2017, \mnras, 470, 4750 
\bibitem[Aniano et al.(2011)]{ani11} Aniano, G., Draine, B.~T., Gordon, K.~D., \& Sandstrom, K.\ 2011, \pasp, 123, 1218
\bibitem[Astropy Collaboration et al.(2013)]{ast13} Astropy Collaboration, Robitaille, T.~P., Tollerud, E.~J., et al.\ 2013, \aap, 558, A33 
\bibitem[Bakes \& Tielens(1994)]{bak94} Bakes, E.~L.~O., \& Tielens, A.~G.~G.~M.\ 1994, \apj, 427, 822 
\bibitem[Barinovs et al.(2005)]{bar05} Barinovs, {\u G}., van Hemert, M.~C., Krems, R., \& Dalgarno, A.\ 2005, \apj, 620, 537 
\bibitem[Bolatto et al.(1999)]{bol99} Bolatto, A.~D., Jackson, J.~M., \& Ingalls, J.~G.\ 1999, \apj, 513, 275 
\bibitem[Bolatto et al.(2007)]{bol07} Bolatto, A.~D., Simon, J.~D., Stanimirovi{\'c}, S., et al.\ 2007, \apj, 655, 212 
\bibitem[Bolatto et al.(2008)]{bol08} Bolatto, A.~D., Leroy, A.~K., Rosolowsky, E., Walter, F., \& Blitz, L.\ 2008, \apj, 686, 948 
\bibitem[Bolatto et al.(2011)]{bol11} Bolatto, A.~D., Leroy, A.~K., Jameson, K., et al.\ 2011, \apj, 741, 12 
\bibitem[Bolatto et al.(2013)]{bol13} Bolatto, A.~D., Wolfire, M., \& Leroy, A.~K.\ 2013, \araa, 51, 207 
\bibitem[Bot et al.(2004)]{bot04} Bot, C., Boulanger, F., Lagache, G., Cambr{\'e}sy, L., \& Egret, D.\ 2004, \aap, 423, 567 
\bibitem[Bourke et al.(1997)]{bou97} Bourke, T.~L., Garay, G., Lehtinen, K.~K., et al.\ 1997, \apj, 476, 781 
\bibitem[Brand \& Wouterloot(1995)]{bra95} Brand, J., \& Wouterloot, J.~G.~A.\ 1995, \aap, 303, 851 
\bibitem[Brauher et al.(2008)]{bra08} Brauher, J.~R., Dale, D.~A., \& Helou, G.\ 2008, \apjs, 178, 280-301 
\bibitem[Chevance et al.(2016)]{che16} Chevance, M., Madden, S.~C., Lebouteiller, V., et al.\ 2016, \aap, 590, A36 
\bibitem[Cohen et al.(1988)]{coh88} Cohen, R.~S., Dame, T.~M., Garay, G., et al.\ 1988, \apjl, 331, L95
\bibitem[Cormier et al.(2015)]{cor15} Cormier, D., Madden, S.~C., Lebouteiller, V., et al.\ 2015, \aap, 578, A53
\bibitem[Crawford et al.(1985)]{cra85} Crawford, M.~K., Genzel, R., Townes, C.~H., \& Watson, D.~M.\ 1985, \apj, 291, 755 
\bibitem[Croxall et al.(2012)]{cro12} Croxall, K.~V., Smith, J.~D., Wolfire, M.~G., et al.\ 2012, \apj, 747, 81
\bibitem[Croxall et al.(2017)]{cro17} Croxall, K.~V., Smith, J.~D., Pellegrini, E., et al.\ 2017, \apj, 845, 96 
\bibitem[Dame et al.(2001)]{dam01} Dame, T.~M., Hartmann, D., \& Thaddeus, P.\ 2001, \apj, 547, 792
\bibitem[Dalgarno \& Black(1976)]{dal76} Dalgarno, A., \& Black, J.~H.\ 1976, Reports on Progress in Physics, 39, 573
\bibitem[Davies et al.(1976)]{dav76} Davies, R.~D., Elliott, K.~H., \& Meaburn, J.\ 1976, \memras, 81, 89
\bibitem[de Blok et al.(2016)]{deb16} de Blok, W.~J.~G., Walter, F., Smith, J.-D.~T., et al.\ 2016, \aj, 152, 51 
\bibitem[De Looze et al.(2014)]{del14} De Looze, I., Cormier, D., Lebouteiller, V., et al.\ 2014, \aap, 568, A62 
\bibitem[Dickey et al.(2000)]{dic00} Dickey, J.~M., Mebold, U., Stanimirovic, S., \& Staveley-Smith, L.\ 2000, \apj, 536, 756
\bibitem[Dickman(1978)]{dic78} Dickman, R.~L.\ 1978, \apjs, 37, 407 
\bibitem[Draine \& Li(2001)]{dra01} Draine, B.~T., \& Li, A.\ 2001, \apj, 551, 807
\bibitem[Draine \& Li(2007)]{dra07} Draine, B.~T., \& Li, A.\ 2007, \apj, 657, 810 
\bibitem[Draine(2011)]{dra11} Draine, B.~T.\ 2011, Physics of the Interstellar and Intergalactic Medium by Bruce T.~Draine.~Princeton University Press, 2011.~ISBN: 978-0-691-12214-4 
\bibitem[Dufour(1984)]{duf84} Dufour, R. J. 1984, in IAU Symp. 108, Structure and Evolution of the Magellanic
Clouds, ed. S. van den Bergh \& K. S. D. de Boer (Dordrecht: Reidel), 353
\bibitem[Fitzpatrick \& Massa(1990)]{fit90} Fitzpatrick, E.~L., \& Massa, D.\ 1990, \apjs, 72, 163
\bibitem[Galametz et al.(2013)]{gal13} Galametz, M., Kennicutt, R.~C., Calzetti, D., et al.\ 2013, \mnras, 431, 1956 
\bibitem[Garden et al.(1991)]{gar91} Garden, R.~P., Hayashi, M., Hasegawa, T., Gatley, I., \& Kaifu, N.\ 1991, \apj, 374, 540 
\bibitem[Glover \& Mac Low(2011)]{glo11} Glover, S.~C.~O., \& Mac Low, M.-M.\ 2011, \mnras, 412, 337 
\bibitem[Glover et al.(2015)]{glo15} Glover, S.~C.~O., Clark, P.~C., Micic, M., \& Molina, F.\ 2015, \mnras, 448, 1607 
\bibitem[Goldsmith et al.(2012)]{gol12} Goldsmith, P.~F., Langer, W.~D., Pineda, J.~L., \& Velusamy, T.\ 2012, \apjs, 203, 13
\bibitem[Gordon et al.(2003)]{gor03} Gordon, K.~D., Clayton, G.~C., Misselt, K.~A., Landolt, A.~U., \& Wolff, M.~J.\ 2003, \apj, 594, 279
\bibitem[Gordon et al.(2011)]{gor11} Gordon, K.~D., Meixner, M., Meade, M.~R., et al.\ 2011, \aj, 142, 102 
\bibitem[Grenier et al.(2005)]{gre05} Grenier, I.~A., Casandjian, J.-M., \& Terrier, R.\ 2005, Science, 307, 1292 
\bibitem[Griffin et al.(2010)]{gri10} Griffin, M.~J., Abergel, A., Abreu, A., et al.\ 2010, \aap, 518, L3
\bibitem[Helsel(2005)]{hel05} Helsel, D.~R.\ 2005, Nondetects and Data Analysis: Statistics for Censored Environmental Data (Wiley; New York)
\bibitem[Henize(1956)]{hen56} Henize, K.~G.\ 1956, \apjs, 2, 315
\bibitem[Herrera-Camus et al.(2015)]{her15} Herrera-Camus, R., Bolatto, A.~D., Wolfire, M.~G., et al.\ 2015, \apj, 800, 1 
\bibitem[Herrera-Camus et al.(2016)]{her16} Herrera-Camus, R., Bolatto, A., Smith, J.~D., et al.\ 2016, \apj, 826, 175 
\bibitem[Heyminck et al.(2012)]{hey12} Heyminck, S., Graf, U.~U., G{\"u}sten, R., et al.\ 2012, \aap, 542, L1 
\bibitem[Heyer et al.(2001)]{hey01} Heyer, M.~H., Carpenter, J.~M., \& Snell, R.~L.\ 2001, \apj, 551, 852 
\bibitem[Heyer et al.(2009)]{hey09} Heyer, M., Krawczyk, C., Duval, J., \& Jackson, J.~M.\ 2009, \apj, 699, 1092 
\bibitem[Hollenbach et al.(1991)]{hol91} Hollenbach, D.~J., Takahashi, T., \& Tielens, A.~G.~G.~M.\ 1991, \apj, 377, 192
\bibitem[Hollenbach \& Tielens(1999)]{hol99} Hollenbach, D.~J., \& Tielens, A.~G.~G.~M.\ 1999, Reviews of Modern Physics, 71, 173 
\bibitem[Hollenbach et al.(2012)]{hol12} Hollenbach, D., Kaufman, M.~J., Neufeld, D., Wolfire, M., \& Goicoechea, J.~R.\ 2012, \apj, 754, 105
\bibitem[Hughes et al.(2010)]{hug10} Hughes, A., Wong, T., Ott, J., et al.\ 2010, \mnras, 406, 2065 
\bibitem[Hunter(2007)]{hun07} Hunter, J.~D.\ 2007, Computing in Science and Engineering, 9, 90 
\bibitem[Indebetouw et al.(2013)]{ind13} Indebetouw, R., Brogan, C., Chen, C.-H.~R., et al.\ 2013, \apj, 774, 73 
\bibitem[Inoue et al.(2016)]{ino16} Inoue, A.~K., Tamura, Y., Matsuo, H., et al.\ 2016, Science, 352, 1559
\bibitem[Israel et al.(1993)]{isr93} Israel, F.~P., Johansson, L.~E.~B., Lequeux, J., et al.\ 1993, \aap, 276, 25 
\bibitem[Israel et al.(1996)]{isr96} Israel, F.~P., Maloney, P.~R., Geis, N., et al.\ 1996, \apj, 465, 738 
\bibitem[Israel(1997)]{isr97} Israel, F.~P.\ 1997, \aap, 328, 471 
\bibitem[Israel et al.(2003)]{isr03} Israel, F.~P., Johansson, L.~E.~B., Rubio, M., et al.\ 2003, \aap, 406, 817 
\bibitem[Israel \& Maloney(2011)]{isr11} Israel, F.~P., \& Maloney, P.~R.\ 2011, \aap, 531, A19 
\bibitem[Jameson et al.(2016)]{jam16} Jameson, K.~E., Bolatto, A.~D., Leroy, A.~K., et al.\ 2016, \apj, 825, 12
\bibitem[Jones et al.(2001)]{jon01} Jones, E., Oliphant, E., Peterson, P., et al.\ 2001--, SciPy: Open source scientific tools for Python, http://www.scipy.org/  
\bibitem[Kapala et al.(2015)]{kap15} Kapala, M.~J., Sandstrom, K., Groves, B., et al.\ 2015, \apj, 798, 24  
\bibitem[Kaufman et al.(1999)]{kau99} Kaufman, M.~J., Wolfire, M.~G., Hollenbach, D.~J., \& Luhman, M.~L.\ 1999, \apj, 527, 795 
\bibitem[Kaufman et al.(2006)]{kau06} Kaufman, M.~J., Wolfire, M.~G., \& Hollenbach, D.~J.\ 2006, \apj, 644, 283   
\bibitem[Krips et al.(2010)]{kri10} Krips, M., Crocker, A.~F., Bureau, M., Combes, F., \& Young, L.~M.\ 2010, \mnras, 407, 2261 
\bibitem[Kurt \& Dufour(1998)]{kur98} Kurt, C.~M., \& Dufour, R.~J.\ 1998, Revista Mexicana de Astronomia y Astrofisica Conference Series, 7, 202 
\bibitem[Kurt et al.(1999)]{kur99} Kurt, C.~M., Dufour, R.~J., Garnett, D.~R., et al.\ 1999, \apj, 518, 246 
\bibitem[Langer et al.(2010)]{lan10} Langer, W.~D., Velusamy, T., Pineda, J.~L., et al.\ 2010, \aap, 521, L17
\bibitem[Langer et al.(2014)]{lan14} Langer, W.~D., Velusamy, T., Pineda, J.~L., Willacy, K., \& Goldsmith, P.~F.\ 2014, \aap, 561, A122 
\bibitem[Lee et al.(2015)]{lee15} Lee, C., Leroy, A.~K., Schnee, S., et al.\ 2015, \mnras, 450, 2708 
\bibitem[Lee (2017)]{lee17} Lee, L.\ 2017, "NADA: Nondetects and Data Analysis for Environmental Data", R package version 1.6-1, https://CRAN.R-project.org/package=NADA
\bibitem[Le Petit et al.(2006)]{lep06} Le Petit, F., Nehm{\'e}, C., Le Bourlot, J., \& Roueff, E.\ 2006, \apjs, 164, 506
\bibitem[Leroy et al.(2007)]{ler07} Leroy, A., Bolatto, A., Stanimirovi{\'c}, S., et al.\ 2007, \apj, 658, 
\bibitem[Leroy et al.(2009)]{ler09} Leroy, A.~K., Bolatto, A., Bot, C., et al.\ 2009, \apj, 702, 352 
\bibitem[Leroy et al.(2011)]{ler11} Leroy, A.~K., Bolatto, A., Gordon, K., et al.\ 2011, \apj, 737, 12 
\bibitem[Leurini et al.(2015)]{leu15} Leurini, S., Wyrowski, F., Wiesemeyer, H., et al.\ 2015, \aap, 584, A70 
\bibitem[Madden et al.(1997)]{mad97} Madden, S.~C., Poglitsch, A., Geis, N., Stacey, G.~J., \& Townes, C.~H.\ 1997, \apj, 483, 200 
\bibitem[Madden et al.(2013)]{mad13} Madden, S.~C., R{\'e}my-Ruyer, A., Galametz, M., et al.\ 2013, \pasp, 125, 600
\bibitem[Malhotra et al.(2001)]{mal01} Malhotra, S., Kaufman, M.~J., Hollenbach, D., et al.\ 2001, \apj, 561, 766  
\bibitem[McMullin et al.(2007)]{mcm07} McMullin, J.~P., Waters, B., Schiebel, D., Young, W., \& Golap, K.\ 2007, Astronomical Data Analysis Software and Systems XVI, 376, 127 
\bibitem[Meixner et al.(2013)]{mei13} Meixner, M., Panuzzo, P., Roman-Duval, J., et al.\ 2013, \aj, 146, 62 
\bibitem[Mizuno et al.(2001)]{miz01} Mizuno, N., Yamaguchi, R., Mizuno, A., et al.\ 2001, \pasj, 53, 971
\bibitem[Muller et al.(2010)]{mul10} Muller, E., Ott, J., Hughes, A., et al.\ 2010, \apj, 712, 1248 
\bibitem[Muraoka et al.(2017)]{mur17} Muraoka, K., Homma, A., Onishi, T., et al.\ 2017, \apj, 844, 98 
\bibitem[Nikoli{\'c} et al.(2007)]{nik07} Nikoli{\'c}, S., Garay, G., Rubio, M., \& Johansson, L.~E.~B.\ 2007, \aap, 471, 561 
\bibitem[Nordon \& Sternberg(2016)]{nor16} Nordon, R., \& Sternberg, A.\ 2016, \mnras, 462, 2804 
\bibitem[Okada et al.(2015)]{oka15} Okada, Y., Requena-Torres, M.~A., G{\"u}sten, R., et al.\ 2015, \aap, 580, A54 
\bibitem[Ossenkopf et al.(2013)]{oss13} Ossenkopf, V., R{\"o}llig, M., Neufeld, D.~A., et al.\ 2013, \aap, 550, A57  
\bibitem[Ott(2010)]{ott10} Ott, S.\ 2010, Astronomical Data Analysis Software and Systems XIX, 434, 139 
\bibitem[Padoan et al.(2000)]{pad00} Padoan, P., Juvela, M., Bally, J., \& Nordlund, {\AA}.\ 2000, \apj, 529, 259 
\bibitem[Paglione et al.(2001)]{pag01} Paglione, T.~A.~D., Wall, W.~F., Young, J.~S., et al.\ 2001, \apjs, 135, 183
\bibitem[Pak et al.(1998)]{pak98} Pak, S., Jaffe, D.~T., van Dishoeck, E.~F., Johansson, L.~E.~B., \& Booth, R.~S.\ 1998, \apj, 498, 735 
\bibitem[Pagel(2003)]{pag03} Pagel, B.~E.~J.\ 2003, CNO in the Universe, 304, 187 
\bibitem[Paradis et al.(2011)]{par11} Paradis, D., Paladini, R., Noriega-Crespo, A., et al.\ 2011, \apj, 735, 6
\bibitem[Penzias et al.(1971)]{pen71} Penzias, A.~A., Jefferts, K.~B., \& Wilson, R.~W.\ 1971, \apj, 165, 229 
\bibitem[Pilbratt et al.(2010)]{pil10} Pilbratt, G.~L., Riedinger, J.~R., Passvogel, T., et al.\ 2010, \aap, 518, L1 
\bibitem[Pineda et al.(2013)]{pin13} Pineda, J.~L., Langer, W.~D., Velusamy, T., \& Goldsmith, P.~F.\ 2013, \aap, 554, A103
\bibitem[Pineda et al.(2017)]{pin17} Pineda, J.~L., Langer, W.~D., Goldsmith, P.~F., et al.\ 2017, \apj, 839, 107
\bibitem[Planck Collaboration et al.(2011)]{pla11} Planck Collaboration, Ade, P.~A.~R., Aghanim, N., et al.\ 2011, \aap, 536, A17 
\bibitem[Poglitsch et al.(1995)]{pog95} Poglitsch, A., Krabbe, A., Madden, S.~C., et al.\ 1995, \apj, 454, 293 
\bibitem[Poglitsch et al.(1996)]{pog96} Poglitsch, A., Herrmann, F., Genzel, R., et al.\ 1996, \apjl, 462, L43 
\bibitem[Poglitsch et al.(2010)]{pog10} Poglitsch, A., Waelkens, C., Geis, N., et al.\ 2010, \aap, 518, L2 
\bibitem[Polk et al.(1988)]{pol88} Polk, K.~S., Knapp, G.~R., Stark, A.~A., \& Wilson, R.~W.\ 1988, \apj, 332, 432 
\bibitem[Pound \& Wolfire(2008)]{pou08} Pound, M.~W., \& Wolfire, M.~G.\ 2008, Astronomical Data Analysis Software and Systems XVII, 394, 654 
\bibitem[Requena-Torres et al.(2016)]{req16} Requena-Torres, M.~A., Israel, F.~P., Okada, Y., et al.\ 2016, \aap, 589, A28
\bibitem[Reynolds(1991)]{rey91} Reynolds, R.~J.\ 1991, \apjl, 372, L17 
\bibitem[R{\"o}llig et al.(2006)]{rol06} R{\"o}llig, M., Ossenkopf, V., Jeyakumar, S., Stutzki, J., \& Sternberg, A.\ 2006, \aap, 451, 917
\bibitem[Roman-Duval et al.(2010)]{rom10} Roman-Duval, J., Israel, F.~P., Bolatto, A., et al.\ 2010, \aap, 518, L74  
\bibitem[Roman-Duval et al.(2014)]{rom14} Roman-Duval, J., Gordon, K.~D., Meixner, M., et al.\ 2014, \apj, 797, 86 
\bibitem[Rubin et al.(2009)]{rub09} Rubin, D., Hony, S., Madden, S.~C., et al.\ 2009, \aap, 494, 647 
\bibitem[Rubio et al.(1991)]{rub91} Rubio, M., Garay, G., Montani, J., \& Thaddeus, P.\ 1991, \apj, 368, 173
\bibitem[Rubio et al.(1993a)]{rub93a} Rubio, M., Lequeux, J., Boulanger, F., et al.\ 1993a, \aap, 271, 1 
\bibitem[Rubio et al.(1993b)]{rub93} Rubio, M., Lequeux, J., \& Boulanger, F.\ 1993b, \aap, 271, 9 
\bibitem[Rubio et al.(1996)]{rub96} Rubio, M., Lequeux, J., Boulanger, F., et al.\ 1996, \aaps, 118, 263
\bibitem[Rubio et al.(2015)]{rub15} Rubio, M., Elmegreen, B.~G., Hunter, D.~A., et al.\ 2015, \nat, 525, 218 
\bibitem[Russell \& Dopita(1992)]{rus92} Russell, S.~C., \& Dopita, M.~A.\ 1992, \apj, 384, 508 
\bibitem[Sandstrom et al.(2010)]{san10} Sandstrom, K.~M., Bolatto, A.~D., Draine, B.~T., Bot, C., \& Stanimirovi{\'c}, S.\ 2010, \apj, 715, 701
\bibitem[Sandstrom et al.(2012)]{san12} Sandstrom, K.~M., Bolatto, A.~D., Bot, C., et al.\ 2012, \apj, 744, 20  
\bibitem[Sandstrom et al.(2013)]{san13} Sandstrom, K.~M., Leroy, A.~K., Walter, F., et al.\ 2013, \apj, 777, 5 
\bibitem[Schruba et al.(2012)]{sch12} Schruba, A., Leroy, A.~K., Walter, F., et al.\ 2012, \aj, 143, 138 
\bibitem[Schruba et al.(2017)]{sch17} Schruba, A., Leroy, A.~K., Kruijssen, J.~M.~D., et al.\ 2017, \apj, 835, 278 
\bibitem[Scowcroft et al.(2016)]{sco16} Scowcroft, V., Freedman, W.~L., Madore, B.~F., et al.\ 2016, \apj, 816, 49 
\bibitem[Shetty et al.(2011)]{she11} Shetty, R., Glover, S.~C., Dullemond, C.~P., et al.\ 2011, \mnras, 415, 3253 
\bibitem[Smith \& MCELS Team(1999)]{smi99} Smith, R.~C., \& MCELS Team 1999, New Views of the Magellanic Clouds, 190, 28
\bibitem[Smith et al.(2017)]{smi17} Smith, J.~D.~T., Croxall, K., Draine, B., et al.\ 2017, \apj, 834, 5 
\bibitem[Sofia et al.(1997)]{sof97} Sofia, U.~J., Cardelli, J.~A., Guerin, K.~P., \& Meyer, D.~M.\ 1997, \apjl, 482, L105
\bibitem[Solomon et al.(1979)]{sol79} Solomon, P.~M., Sanders, D.~B., \& Scoville, N.~Z.\ 1979, \apjl, 232, L89
\bibitem[Stacey et al.(1983)]{sta83} Stacey, G.~J., Smyers, S.~D., Kurtz, N.~T., \& Harwit, M.\ 1983, \apjl, 265, L7 
\bibitem[Stacey et al.(1991)]{sta91} Stacey, G.~J., Geis, N., Genzel, R., et al.\ 1991, \apj, 373, 423 
\bibitem[Stanimirovi\'{c} et al.(1999)]{sta99} Stanimirovi\'{c}, S., Staveley-Smith, L., Dickey, J.~M., Sault, R.~J., \& Snowden, S.~L.\ 1999, \mnras, 302, 417 
\bibitem[Stanimirovi{\'c} et al.(2004)]{sta04} Stanimirovi{\'c}, S., Staveley-Smith, L., \& Jones, P.~A.\ 2004, \apj, 604, 176 
\bibitem[Sternberg \& Dalgarno(1995)]{ste95} Sternberg, A., \& Dalgarno, A.\ 1995, \apjs, 99, 565 
\bibitem[Subramanian \& Subramaniam(2012)]{sub12} Subramanian, S., \& Subramaniam, A.\ 2012, \apj, 744, 128
\bibitem[Sz{\H u}cs et al.(2016)]{szu16} Sz{\H u}cs, L., Glover, S.~C.~O., \& Klessen, R.~S.\ 2016, \mnras, 460, 82
\bibitem[Tayal(2008)]{tay08} Tayal, S.~S.\ 2008, \aap, 486, 629
\bibitem[Tayal(2011)]{tay11} Tayal, S.~S.\ 2011, \apjs, 195, 12
\bibitem[Tchernyshyov et al.(2015)]{tch15} Tchernyshyov, K., Meixner, M., Seale, J., et al.\ 2015, \apj, 811, 78 
\bibitem[Temi et al.(2014)]{tem14} Temi, P., Marcum, P.~M., Young, E., et al.\ 2014, \apjs, 212, 24 
\bibitem[Tielens \& Hollenbach(1985)]{tie85} Tielens, A.~G.~G.~M., \& Hollenbach, D.\ 1985, \apj, 291, 722 
 \bibitem[Van Der Walt et al.(2011)]{van11} Van Der Walt, S., Colbert, S.~C., \& Varoquaux, G.\ 2011, arXiv:1102.1523 
\bibitem[van Loon et al.(2010)]{van10} van Loon, J.~T., Oliveira, J.~M., Gordon, K.~D., Sloan, G.~C., \& Engelbracht, C.~W.\ 2010, \aj, 139, 1553
\bibitem[Velusamy et al.(2012)]{vel12} Velusamy, T., Langer, W.~D., Pineda, J.~L., \& Goldsmith, P.~F.\ 2012, \aap, 541, L10 
\bibitem[Wiesenfeld \& Goldsmith(2014)]{wie14} Wiesenfeld, L., \& Goldsmith, P.~F.\ 2014, \apj, 780, 183 
\bibitem[Wilson \& Rood(1994)]{wil94} Wilson, T.~L., \& Rood, R.\ 1994, \araa, 32, 191
\bibitem[Wolfire et al.(1989)]{wol89} Wolfire, M.~G., Hollenbach, D., \& Tielens, A.~G.~G.~M.\ 1989, \apj, 344, 770 
\bibitem[Wolfire et al.(1990)]{wol90} Wolfire, M.~G., Tielens, A.~G.~G.~M., \& Hollenbach, D.\ 1990, \apj, 358, 116
\bibitem[Wolfire et al.(1995)]{wol95} Wolfire, M.~G., Hollenbach, D., McKee, C.~F., Tielens, A.~G.~G.~M., \& Bakes, E.~L.~O.\ 1995, \apj, 443, 152 
\bibitem[Wolfire et al.(2003)]{wol03} Wolfire, M.~G., McKee, C.~F., Hollenbach, D., \& Tielens, A.~G.~G.~M.\ 2003, \apj, 587, 278
\bibitem[Wolfire et al.(2010)]{wol10} Wolfire, M.~G., Hollenbach, D., \& McKee, C.~F.\ 2010, \apj, 716, 1191
\bibitem[Wong et al.(2017)]{won17} Wong, T., Hughes, A., Tokuda, K., et al.\ 2017, \apj, 850, 139
\bibitem[Wu et al.(2013)]{wu13} Wu, R., Polehampton, E.~T., Etxaluze, M., et al.\ 2013, \aap, 556, A116 


\end{thebibliography}
\end{document}

